\documentclass{article}
\usepackage[english]{babel}
\usepackage[T1]{fontenc}
\usepackage{fullpage}
\usepackage[utf8]{inputenc}
\usepackage[normalem]{ulem}
\usepackage{mathrsfs} 
\usepackage{newlfont}
\usepackage{fancyhdr}
\usepackage{mhchem}
\usepackage{indentfirst}
\usepackage{graphicx}
\usepackage{amssymb}
\usepackage{amsmath}
\usepackage{latexsym}
\usepackage{amsthm}
\usepackage{bm}
\usepackage{bbold}
\usepackage{xcolor}
\usepackage{cancel}
\usepackage{listings}
\usepackage[hidelinks]{hyperref}
\usepackage[capitalise]{cleveref}
\usepackage{subfigure}
\hypersetup{
    colorlinks=true,
    linkcolor=black,
    citecolor=blue,
    filecolor=black,
    urlcolor=blue,
    pdfauthor={M. Cadoni, M. Oi and A. P. Sanna},
    pdftitle={Evaporation and information puzzle for $2$D nonsingular asymptotically flat black holes}
}
\usepackage{cite}

\crefname{plural}{Eqs.}{Eqs.}
\Crefname{plural}{Eqs.}{Eqs.}

\crefformat{plural}{#2eqs.~(#1)#3}
\Crefformat{plural}{#2Eqs.~(#1)#3}

\newcommand{\D}{\mathscr{D}}
\newcommand{\K}{\mathscr{K}}
\newcommand{\V}{\mathscr{V}}

\newcommand{\M}{\mathcal{M}}
\newcommand{\A}{\mathscr{A}}
\newcommand{\B}{\mathscr{B}}
\newcommand{\Th}{T_\text{H}}
\newcommand{\rH}{r_\text{H}}
\newcommand{\I}{\mathcal{I}}
\newcommand{\J}{\mathcal{J}}
\newcommand{\Ib}{\mathcal{I}_{\text{b}}}
\newcommand{\IGHY}{\mathcal{I}_{\text{GHY}}}

\newcommand{\rext}{r_\text{ext}}
\newcommand{\rst}{r_\ast}
\newcommand{\Mext}{\mathcal{M}_\text{ext}}
\newcommand{\JGS}{\J_\text{GS}}
\newcommand{\ii}{\text{i}}
\newcommand{\phiGS}{\phi_\text{GS}}
\newcommand{\kappaH}{\kappa_\text{H}}
\newcommand{\dd}{\text{d}}

\title{\bf Evaporation and  information  puzzle for 2D nonsingular  asymptotically flat  black holes}
\author{M.~Cadoni${}^{ab}$\thanks{E-mail: mariano.cadoni@ca.infn.it}, \, M.~Oi${}^{abc}$\thanks{E-mail: mauro.oi@ca.infn.it}, \, A.~P.~Sanna${}^{ab}$\thanks{E-mail: asanna@dsf.unica.it} \ 
\\
${}^a$\emph{Dipartimento di Fisica, Universit\`a di Cagliari}
\\
{\em Cittadella Universitaria, 09042 Monserrato, Italy}
\\
\\
${}^b$\emph{I.N.F.N, Sezione di Cagliari}
\\
{\em  Cittadella Universitaria, 09042 Monserrato, Italy}
\\
\\
${}^c$\emph{CENTRA, Departamento de F\'isica, Instituto Superior T\'ecnico -- IST,}
\\
{\em  Universidade de Lisboa -- UL, Avenida Rovisco Pais 1, 1049 Lisboa, Portugal}
\\
\\}

\date{}

\begin{document}
\maketitle
\begin{abstract}
We investigate the thermodynamics and the classical and semiclassical dynamics of two-dimensional ($2\text{D}$), asymptotically flat, nonsingular dilatonic black holes. They are characterized by a de Sitter core, allowing for the smearing of the classical singularity, and by the presence of two horizons with a related extremal configuration. For concreteness, we focus on a  $2\text{D}$ version of the Hayward black hole. We find a second order thermodynamic phase transition, separating large unstable black holes from stable configurations close to extremality. We first describe the black-hole evaporation process using a quasistatic approximation and we show that it ends in the extremal configuration in an infinite amount of time. We go beyond the quasistatic approximation by numerically integrating the field equations for $2\text{D}$ dilaton gravity coupled to $N$ massless scalar fields, describing the radiation. We find that the inclusion of large backreaction effects ($N \gg 1$) allows for an end-point extremal configuration after a finite evaporation time. Finally, we evaluate the entanglement entropy (EE) of the radiation in the quasistatic approximation and construct the relative Page curve. We find that the EE initially grows, reaches a maximum and then goes down towards zero, in agreement with previous results in the literature. Despite the  breakdown of the semiclassical approximation prevents the description of the evaporation process near extremality, we have a clear indication that the end point of the evaporation is a regular, extremal state with vanishing EE of the radiation. This suggests that the nonunitary evolution, which commonly characterizes the evaporation of singular black holes, could be traced back to the presence of the singularity. 
\end{abstract}
%


\tableofcontents

\clearpage

\section{Introduction}
\label{sec:Introduction}

Despite the huge recent progress on the observational side, achieved through gravitational-wave detection \cite{LIGOScientific:2016aoc,LIGOScientific:2017vwq} and imaging \cite{EventHorizonTelescope:2019dse,EventHorizonTelescope:2022wkp}, black holes are still a source of challenges for theoretical fundamental physics. The usual black-hole solutions of general relativity (GR) harbour, shielded behind event horizons, spacetime singularities, i.e. regions where the classical and semiclassical descriptions break down \cite{Penrose:1964wq,Hawking:1970zqf}. As discovered more than 50 years ago, black holes behave as thermodynamic systems, whose microscopic description remains, however, still mysterious \cite{Strominger:1996sh,Ashtekar:1997yu,Carlip:2002be,Ryu:2006b,Padmanabhan:2009vy,Dvali:2011aa,McGough:2013gka,Cadoni:2021jer}. They emit thermal radiation, but the description of the information flow during the evaporation has led to the information paradox, which most embodies the apparent incompatibility between quantum mechanics and GR \cite{Hawking:1976ra,Page:1993wv,Mathur:2009hf}.

A possible solution to the information puzzle, which has been pursued in the literature, is linking it to the singularity problem \cite{Hasslacher:1980hd,Horowitz:2003he,Ashtekar:2005qt,Ashtekar:2008jd,Hossenfelder:2009xq,Cadoni:2022chn}.
The presence of a spacetime singularity makes the very notion of a global quantum state for matter fields in the black-hole background ill-defined. The loss of unitarity in the evolution of quantum states could be, therefore, traced back to the bad definition of the latter.
The main objection to this argument is that the ``unitarity crisis'' shows up also for macroscopic black holes, i.e. those with masses hierachically larger than the Planck mass  $m_\text{p} \simeq 10^{19} \, \text{GeV}$, that is at energy scales where the singularity cannot play any role. Also the possibility to shift the solution of the problem to the late stages of the evaporation, for instance through the formation of Planck-scale remnants, seems untenable owing to difficulty of storing/recovering the huge amount of information at these small scales \cite{Giddings:1993km,Chen:2014jwq}.
 
The recent reformulation and proposal for  a solution of the information puzzle \cite{Penington:2019npb,Penington:2019kki,Almheiri:2019qdq,Almheiri:2020cfm} seem to bring further arguments against a close relationship between the singularity and the information problem. This is because this novel approach is focused \emph{only} on reconstructing the correlations between early and late Hawking radiation and, thus, refers mainly to near-horizon physics. 

There is, however, an important feature of black-hole solutions, which could change drastically the debate about the relationship between the singularity problem and the information paradox. The most commonly used spacetime setup is that of a black-hole solution with a single event horizon. Black holes with two (an inner and an outer) horizons introduce a new ingredient, which drastically changes the rules of the game. First of all, these black holes typically admit a ground state (GS) represented by an extremal configuration, in which the inner and outer horizons merge in a single one. Moreover, the radius of the extremal black hole  could be hierachically larger than the Planck scale \cite{Cadoni:2022chn,Cadoni:2022vsn}. In the near-horizon region and in the extremal regime, the geometry factorizes as a two-dimensional (2D) anti de Sitter ($\text{AdS}_2$) spacetime times a $2\text{D}$ sphere of constant radius. This opens the way to the intriguing possibility that the information issue could be solved in the final stages of the evaporation process using properties of $\text{AdS}_2$ quantum gravity, e.g., by reconstructing correlations between the two disconnected parts of $\text{AdS}_2$ spacetime \cite{Maldacena:2001kr} or by the topological properties of the fragmented GS \cite{Maldacena:1998uz}. Moreover, there is some evidence that, for black holes with two-horizons and without a central singularity \cite{Hayward:2005gi,Frolov:2014jva,Sueto:2023ztw}, the evaporation process could  be unitary. Specifically, the presence of the inner Cauchy horizon could act as a trapping region for high energy modes, which could be responsible for the release of information at late times, when the two horizons are about to merge \cite{Bianchi:2014bma,Frolov:2017rjz}. 

The most natural candidates for testing these ideas are four-dimensional ($4\text{D}$) nonsingular black holes with a de Sitter (dS) core \cite{bardeen1968proceedings,Dymnikova:1992ux,Hayward:2005gi,Nicolini:2008aj,Modesto:2010uh,Frolov:2016pav,Fan:2016hvf,Cadoni:2022chn,Cadoni:2023nrm}. They appear as static solutions of Einstein's equations sourced by an anisotropic fluid. The corresponding spacetime is asymptotically flat (AF) and at great distances is indistinguishable from the Schwarzschild solution, whereas the  singularity at the origin of the radial coordinate is regularized due to inner dS behavior. The latter also produces an additional hair $\ell$, which could have interesting observational signatures in the geodesics motion of massless and massive particles, quasinormal modes spectrum and gravitational waves (see, e.g., Refs.~\cite{Cadoni:2022chn,Cadoni:2022vsn,Hu:2018nhg,Lamy:2018zvj,Guo:2021bhr,DellaMonica:2021fdr} and references therein). Another consequence of the dS core is the presence, depending on the value of $\ell$, of two horizons and an extremal solution. From the thermodynamic point of view, these models are characterized by a second order phase transition: the spectrum has a branch of large unstable configurations and a stable branch of near-extremal solutions.

There are two main obstructions that prevent the direct use of such $4\text{D}$ models to address the information paradox. Firstly, we do not have a microscopic model describing the sources of the solutions. We can just give a coarse-grained description in terms of an anisotropic fluid, with equation of state $p=-\rho$, and a given profile for the energy density $\rho$. Secondly, there is the usual difficulty of describing semiclassical dynamics, including backreaction effects on the geometry, of quantum Hawking radiation in the  4D classical black-hole background.

In this paper, we show that both issues can be addressed by considering $2\text{D}$ dilaton gravity models of AF, nonsingular black holes with a dS core. As we shall show, these models can be formulated at Lagrangian level and describe, in a simplified setting, the $S$-wave sector (radial modes) of their $4\text{D}$ nonsingular counterparts. This will allow us to retain the qualitative features of the higher-dimensional models, keeping, however, under control their dynamics. For concreteness, our investigations will be focused on a particular, but quite relevant, case, namely the $2\text{D}$ Hayward black hole.   
     
We will be able to capture the main thermodynamic features of $4\text{D}$ regular models and, at the same time, to describe their evaporation process and to solve the classical and semiclassical dynamics, including the backreaction of Hawking radiation on the geometry. Having under control the latter will allow us to partially answer some important questions regarding the end point of the evaporation process, the time evolution of the entanglement entropy and the shape of the related Page curve \cite{Page:1993wv,Page:2013dx}. The main limitation of this approach is obviously represented by the limit of validity of the semiclassical approximation. Nonetheless, our results, together with some known features of $\text{AdS}_2$ quantum gravity, will allow us to have clear indications about the fate of information during the evaporation of nonsingular black holes with a dS core.

The structure of this paper is as follows.

In \cref{sec:Twodimensionalregularmodels}, we review some general properties of 2D dilaton gravity models and we present our class of $2\text{D}$, nonsingular, AF solutions with a dS core. 

In \cref{sec:HaywardSol}, we introduce the prototype-model we will use throughout our paper, namely the $2\text{D}$ Hayward black hole.  

\cref{sec:FirstLawThermo} is devoted to the investigation of the thermodynamic properties of our 2D models.

In \cref{subsec:semiclassicalevaporation}, we discuss black-hole evaporation using a quasistatic approximation.

The coupling with conformal matter, in the form of  $N$ massless scalars, is introduced in \cref{sec:CouplingConformalMatter}. We consider, in particular, classical solutions corresponding to a shock wave.

In \cref{sec:blackholeevaporationandbackreaction}, we discuss the evaporation process by quantizing matter in the classical gravitational background and by including backreaction effects. The semiclassical dynamics cannot be solved analytically, so we resort to numerical integration. 

The entanglement entropy of the Hawking radiation is computed in \cref{sec:EEandPagecurve} and its Page curve is presented.

Finally, we draw our conclusions in \cref{sec:conclusions}.

Some details of the calculations concerning the absence of divergences for the stress-energy tensor in the GS and the boundary conditions used for the numerical integration of the field equations are presented in \cref{sec:nodivergencesTmunuGS} and
\cref{sec:boundaryconditionsnumerical}, respectively.

\section{Two-dimensional regular  dilatonic black holes}
\label{sec:Twodimensionalregularmodels}

The simplest black-hole models can be constructed in a 2D spacetime. However, the pure 2D Einstein-Hilbert action is a topological invariant and a metric theory of gravity has to be built by coupling the Ricci scalar with a scalar field $\phi$, the dilaton. $2$D dilaton gravity is generally described by the action (see Ref.~\cite{Grumiller:2002nm} for a review; for a generalization, see Ref.~\cite{Grumiller:2021cwg})
\begin{equation}\label{LagrangianGeneral}
\mathscr{L} = \sqrt{-g}\left[\D(\phi) R + \K(\phi) g^{\mu\nu} \partial_\mu \phi \partial_\nu \phi + \V(\phi)\right] \, ,
\end{equation}
where $\D$, $\K$ and $\V$ are functions of the dilaton, representing, respectively, the coupling with the Ricci scalar $R$ (the dimensionless inverse Newton constant), the kinetic term of the scalar field and the potential. Using a Weyl transformation of the metric
\begin{equation}\label{Weyltransf}
g_{\mu\nu} = e^{\mathscr{P}(\phi)}\tilde g_{\mu\nu}\, , \qquad \mathscr{P} = -\int^\phi \dd\psi \, \frac{\K(\psi)}{\D(\psi)}\, ,
\end{equation}
together with a field redefinition
\begin{equation}\label{FieldRedef}
\D(\phi) \to \phi\, ,
\end{equation}
it is always possible to set $\K = 0$ and to recast the lagrangian into the simpler form  \cite{Banks:1990mk,Cavaglia:2000uw}
\begin{equation}\label{actionpresentpaper}
\mathscr{L} = \sqrt{-g} \left[\phi R + \V(\phi) \right]\, . 
\end{equation} 
This choice of the conformal (Weyl) frame is typically used when dealing with asymptotically AdS black holes, while for AF configurations (which is the focus of the present paper), a conformal frame with $\K \neq 0$ is generally considered more appropriate \cite{Callan:1992rs, Bogojevic:1998ma,Cadoni:2005ej, Frolov:2021kcv,Fitkevich:2022ior}. 
This is particularly true  when $2\text{D}$ dilatonic black holes are used to describe the $S$-wave sector of higher dimensional models (but see, e.g., Refs.~\cite{Trodden:1993dm, Banks:1992xs, Lowe:1993zw,Cadoni:1995dd,Ai:2020nyt}).
However, as we shall see below, the description of also AF black holes is much simpler in the conformal frame \eqref{actionpresentpaper} than  in a frame with $\K \neq 0$.  Moreover, the lagrangian \eqref{actionpresentpaper} is fully characterized by the dilaton potential $\V(\phi)$; this allows for a simple classification of regular black-hole models in terms of the properties of $\V$. For these reasons, in the following, we will work in this frame, although this choice will introduce some difficulties concerning  the physical interpretation of the parameters we use to describe black holes.

The equations of motion stemming from \cref{actionpresentpaper}, in the absence of matter fields, read
\begin{subequations}\label[plural]{EqnsMotion}
\begin{align}
&R + \frac{\dd\V}{\dd\phi} = 0 \, ; \label{ScalarFieldEqMotion}\\
&\left(g_{\mu\nu}\Box - \nabla_\mu \nabla_\nu \right)\phi -\frac{1}{2}g_{\mu\nu} \V =0\, . \label{MetricEqMotion}
\end{align}
\end{subequations}
\subsection{Linear dilaton solution}
\label{subsec:LDS}
Let us now first  consider static solutions of \cref{MetricEqMotion}. In this case, the dilaton can be used as a spacelike ``radial'' coordinate of the 2D spacetime, 
\begin{equation}\label{LDsolution}
\phi = \lambda r,
\end{equation}
where $\lambda$ is a constant, with dimensions of the inverse of a length, characterizing the potential $\V$. This parametrization of the dilaton is particularly useful when the 2D theory is used to describe the $S$-wave sector of $4\text{D}$ black holes. In this case, the dilaton is proportional to the radius of the transverse two-sphere. Notice also that the dilaton represents the inverse of the $2\text{D}$ Newton constant. This means that the region $r\ll\lambda^{-1}$ is in a strong coupling regime, whereas $r\gg\lambda^{-1}$ is a weak-coupling region. To be consistent, both interpretations of $\phi$ require to limit the range of variation of the radial coordinate to $r\in [0,\infty)$.  

\cref{LDsolution} allows us  to write the most general static solution of \cref{MetricEqMotion,ScalarFieldEqMotion} as the linear dilaton solution (LDS)
\begin{equation}\label{LDsolution1}
ds^2 = -f(r) \, \dd t^2 + \frac{\dd r^2}{f(r)}\, , \qquad   f =  c_1 + \frac{1}{\lambda}\int \dd r \, \V \,  
\end{equation}
where $c_1$ is a dimensionless integration constant, which can be written in terms of the covariant mass $\mathcal{M}$, which can be defined for a  generic $2\text{D}$ dilaton  gravity theory \cite{Mann:1992yv}. For a static spacetime, $\M$ is the conserved charge associated with the Killing vector $\chi^\mu = F_0 \epsilon^{\mu\nu}\partial_\nu \phi$, generating time translations \footnote{In this case  $\M$ corresponds to the ADM mass of the solution \cite{Cadoni:1995mi,Mignemi:1994wg}.}. $F_0$ is a constant, which is  
 fixed by the normalization of $\chi^\mu$. As we will see in the following, to make contact with four dimensional models, a convenient choice\footnote{Note that this normalization differs only in the sign with respect to the expression of Ref.~\cite{Mann:1992yv}.} is $F_0 = -1/\lambda$.

In our Weyl frame, $\M$ reads
\begin{equation}\begin{split}\label{Mgeneral}
\mathcal{M}& = \frac{F_0}{2} \left[\int^\phi \dd\phi \, \V - g^{\mu\nu}\partial_\mu \phi \partial_\nu \phi \right]= -\frac{F_0}{2}c_1 \lambda^2 \, .
\end{split}\end{equation}
Choosing appropriately the form of  $\V$ allows to generate different solutions. In particular, we focus on dilaton gravity models allowing for AF nonsingular black holes. 

A useful information for classifying different classes of models can be obtained from the existence of solutions characterized by a constant dilaton, the so-called constant dilaton vacua (CDV). Owing to the $r\text{-dependent}$ parametrization of the dilaton \eqref{LDsolution}, these solutions cannot be obtained as particular LDSs given by \cref{LDsolution1} and must be discussed separately.

\subsection{Constant dilaton vacua}
\label{subsec:CDS}

The CDV solutions of \Cref{EqnsMotion} can be obtained by setting $\phi = \text{constant} \equiv \phi_0$. According to \cref{MetricEqMotion}, these vacuum configurations must correspond to zeroes of the potential, $\V(\phi_0)=0$. On the other hand, \cref{ScalarFieldEqMotion} tells us that they correspond to $2\text{D}$ spacetimes with constant curvature and can be classified according to the sign of $\dd\V/\dd\phi |_{\phi = \phi_0}$. We have three possible cases:

\begin{enumerate}
\item If $\dd\V/\dd\phi |_{\phi_0}<0$, $R> 0$ and we have a dS spacetime;
\item If $\dd\V/\dd\phi |_{\phi_0} >0$, $R< 0$ and we have an AdS spacetime;
\item If $\dd\V/\dd\phi |_{\phi_0} =0$, $R=0$ and we have a flat, Minkowski spacetime.
\end{enumerate}

Notice that the condition for having a flat CDV is rather strong. It requires $\phi_0$ to be both a zero and an extremum of $\V$. 

Asymptotic flatness, however, also implies that we always have $\V(\infty)=\dd\V/\dd\phi |_{\infty}=0$. In this case, formally, we can consider $\phi=\infty$, which corresponds to a decoupled configuration (the 2D Newton constant vanishes), as a flat CDV.  
  
In the first case above, instead, we can define
\begin{equation}\label{dsL}
\frac{\dd\V}{\dd\phi}\biggr|_{\phi_{0}} \equiv -\frac{1}{L^2_\text{dS}}\, .
\end{equation} 
From this, using \cref{ScalarFieldEqMotion}, we have
\begin{equation}\label{dS2CDV}
 f(r) =  1 - \frac{r^2}{L^2_\text{dS}}\, ,
\end{equation}
which describes  two-dimensional dS (dS$_2$) spacetime, with an associated dS length $L_\text{dS}$.

In the second case, we define instead
\begin{equation}\label{AdsL}
\frac{\dd\V}{\dd\phi}\biggr|_{\phi_{0}} \equiv \frac{1}{L^2_\text{AdS}}\, .
\end{equation} 
\Cref{ScalarFieldEqMotion} now yields
\begin{equation}\label{AdS2LengthCDV}
f(r) = \frac{r^2}{L^2_\text{AdS}}\,  ,
\end{equation}
which describes the $\text{AdS}_2$ spacetime, with an associated AdS length $L_\text{AdS}$.

\subsection{General class of 2D nonsingular, asymptotically-flat black holes with a de Sitter core}
\label{subsec:Nonsingularconditions}

We are interested in $2\text{D}$ dilatonic black holes, which mimic the behavior of $4\text{D}$ regular black holes. Therefore we have to choose the form of the potential $\V$ such that: (a) the spacetime has one or at most two horizons; (b) the spacetime curvature remains everywhere finite, in particular the usual $4\text{D}$ curvature singularity at $r=0$ is removed; (c) the spacetime is AF, with a Schwarzschild subleading behavior at $r\to\infty$, i.e., $f \sim 1 - c/r$, with $c$ a constant.
 
Condition (a) requires $f(r)$ to have at least one zero and to stay positive outside the (outer) horizon, i.e.,
\begin{equation}\label{cond1}
\int_{\phi_\text{H}}^\phi \dd\psi \, \V(\psi)\ge 0\, .
\end{equation}
Condition (b) can be reformulated as a condition on the first derivative of $\V$.
In two dimensions, the Ricci scalar is the only curvature invariant. Therefore, bounding it is sufficient to generate regular spacetime metrics. According to \cref{ScalarFieldEqMotion}, this translates into requiring regularity of $\dd\V/\dd\phi$. Regularity at $r=0$ can be achieved in different ways. The simplest and more physical one, which has also been used for $4\text{D}$ models \cite{Cadoni:2022chn,bardeen1968proceedings,Dymnikova:1992ux,Hayward:2005gi,Nicolini:2008aj,Modesto:2010uh,Frolov:2016pav,Fan:2016hvf}, is to impose $\dd\V/\dd\phi |_{0}$ to be finite and 
\begin{equation}\label{cond2}
\V(0) = 0,\quad  \frac{\dd\V}{\dd\phi}\biggr|_0<0 \, .
\end{equation}
According to the general discussion of \cref{subsec:CDS}, this implies that our model must allow for a dS$_2$ CDV at $\phi=0$, given by \cref{dS2CDV}, which will therefore describe the inner core of our black-hole solutions. Using \cref{dS2CDV}, one can easily find the form of the  potential in the  $\phi\sim 0$ region  is,
\begin{equation}\label{deSittercorecondition}
 \V \sim -2\frac{\phi}{\lambda \hat L^2}\, .
\end{equation}
Condition (c), i.e., asymptotic flatness and a Schwarzschild subleading behavior, can be implemented by fixing the asymptotic behavior for $\phi\to \infty$ 
\begin{equation}\label{cond3}
\V \sim \frac{\lambda^2}{\phi^2}\, .
\end{equation}

\begin{figure}[h!]
\centering
\includegraphics[width= 7 cm, height = 7 cm,keepaspectratio]{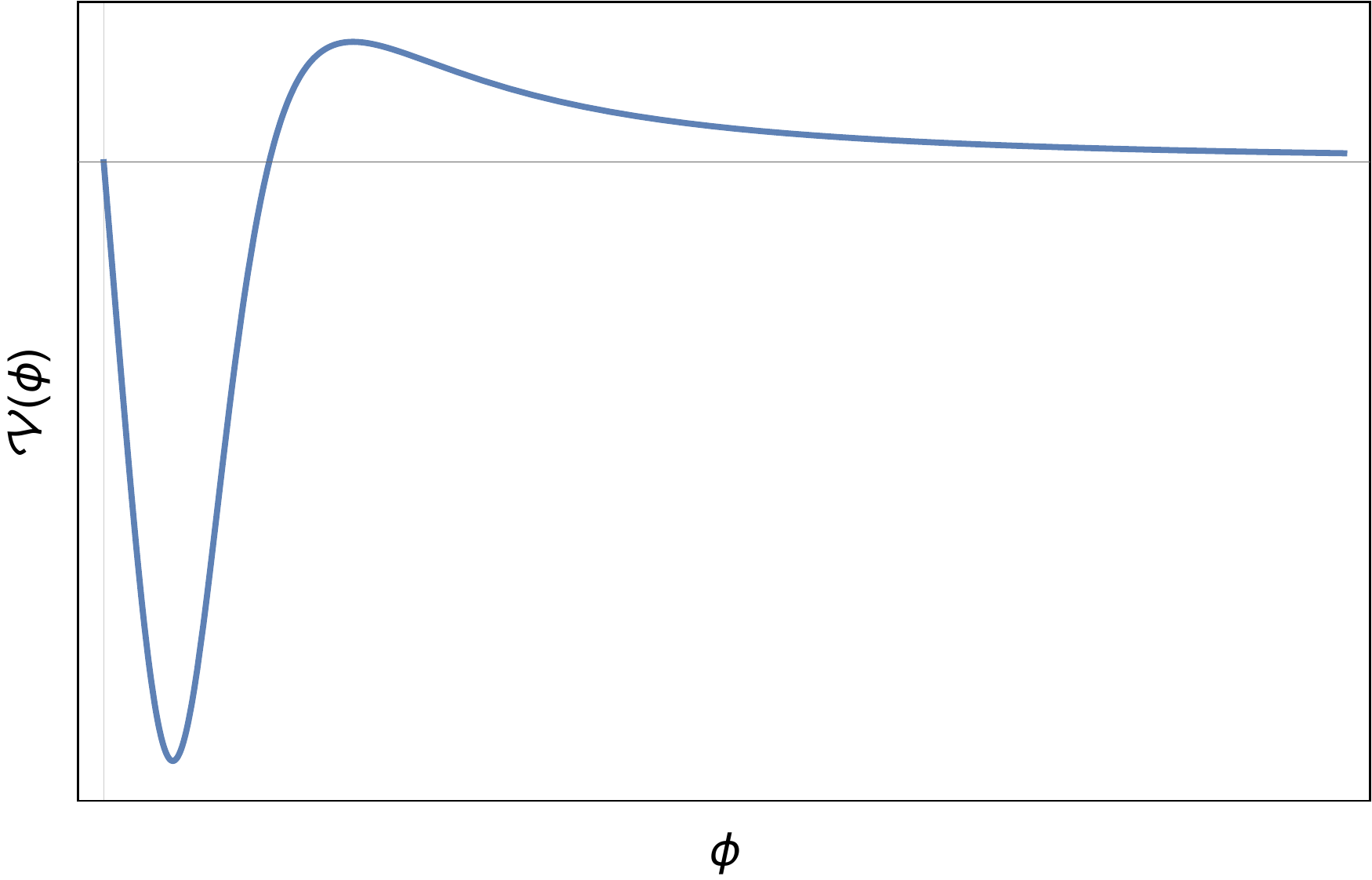}
\caption{Qualitative behavior of the potential, characterizing the broad class of the regular models satisfying the conditions $(1)$, $(2)$, $(3)$. We restricted ourselves to the case of a single maximum and a single minimum. The horizontal  and vertical lines correspond to $\V = 0$  and $\phi=0$, respectively.}
	\label{Potential}
\end{figure}

\Cref{cond2} implies the dilaton potential to be zero at $\phi=0$, and to become negative and decrease near $\phi \sim 0$. However, it has to grow again, cross the $\phi\text{-axis}$ at a finite value $\phi=\phi_1$, develop at least one minimum and one maximum to guarantee the positive fall of $\V$ at asymptotic infinity, implied by \cref{cond3}. The model, therefore, must allow for three different CDV solutions at $\phi = 0$, $\phi=\phi_1$ and $\phi=\infty$, describing, respectively, a dS$_2$, $\text{AdS}_2$ and flat spacetimes.  

In principle, the potential could show any number of oscillations, but for simplicity in the following we restrict ourselves to potentials with a single maximum and a single minimum. The qualitative behavior of the general form of our potential $\V$ is shown in \cref{Potential}. Quantitatively, the potential will depend on some dimensional parameters. The most natural, minimal choice is a potential depending on two parameters. Since $\phi$ is dimensionless, the parameter $\lambda$, introduced above, is needed to give the right dimensions to $\V$. At least a second parameter, however, which we will call $\ell$, is needed if we want to express the CDV $\phi_0$ in terms of parameters of the model. The values of such parameters will impact on the behavior of the metric function. 
Indeed, since $\V=\lambda \, \dd f/\dd r$ from \cref{LDsolution1}, we see that the presence of two zeros for $\V$ implies the existence of a minimum for $f(r)$. Depending on the value of $\ell$, this minimum can be above, below or exactly at the $r\text{-axis}$, producing horizonless, two-horizon or extremal configurations, respectively. The parameter $\lambda$, instead, does not affect the presence of extrema in $f(r)$, nor their location in the radial direction.
Notice that the $\text{AdS}_2$ CDV describes the near-horizon behavior of the extremal black hole \cite{Navarro-Salas:1999zer,Cadoni:2017dma}.  

One can easily construct dilatonic potentials $\V(\phi)$ behaving as in \cref{Potential}. Basically, for every spherically-symmetric, regular $4\text{D}$ black hole, characterized by a single metric function $f$,
one can easily construct the corresponding $2\text{D}$ dilaton gravity theory by solving \cref{LDsolution1}, determining in this way the form of $\V$. For instance, notable models are those which can be obtained from the Hayward black hole \cite{Hayward:2005gi}, Gaussian-core black hole \cite{Nicolini:2008aj}, the Fan-Wang model \cite{Fan:2016hvf} or the Bardeen solution \cite{bardeen1968proceedings}.   

For the sake of concreteness, in the following we will focus on a $2\text{D}$ dilaton gravity model reproducing the Hayward black hole. However, all the considerations of the next sections can be extended to the general class of models described in this section.  

\section{Two-dimensional Hayward black hole}
\label{sec:HaywardSol}

One of the simplest cases of potentials behaving as shown in \cref{Potential} is given by
\begin{equation}\label{potentialHayward}
\V(\phi) = \lambda^2 \frac{\phi^4-2\ell^3\lambda^3 \phi}{(\phi^3+\lambda^3\ell^3)^2}\, ,
\end{equation}
where $\ell$ is a parameter with dimensions of length. The potential has a zero at $\phi=0$, which gives the dS CDV with a related dS length \eqref{dsL}, given in this case by $L^2_\text{dS}  = \lambda \, \ell^3$, and goes to zero for $\phi\to\infty$. The other zero is at
\begin{equation}\label{CDVsol}
\left(\frac{\phi}{\lambda} \right)^3 = 2 \ell^3 \quad \Rightarrow \quad \phi_0 = \sqrt[3]{2}\, \lambda\, \ell \, ,
\end{equation}
which gives the $\text{AdS}_2$ CDV and, as we shall see below, describes extremal black holes in the near-horizon regime. The associated AdS length (see \cref{AdsL}) is $L^2_\text{AdS} = 3 \lambda \, \ell^3$.

With the potential \eqref{potentialHayward}, solving \cref{LDsolution1} yields
\begin{equation}\label{solgen}
f = \frac{2\mathcal{M}}{\lambda} - \frac{1}{\lambda}\frac{r^2}{r^3 + \ell^3}\, ,
\end{equation}
which interpolates between the Schwarzschild spacetime at great distances and the dS one at $r \sim 0$, modulo a rescaling of the coordinates $t$ and $r$ by the constant quantity $2\M/\lambda$. This peculiar behavior, in which the mass term in the line element dominates at great distances, was analyzed in 2$D$ very recently in Ref. \cite{Bagchi:2014ava} and termed ``mass-dominated'' dilaton gravity.

The $4\text{D}$ Hayward black hole \cite{Hayward:2005gi} is described by the metric element $\dd s_4^2=-f_\text{H}(r)\,\dd t^2+ f_\text{H}^{-1} (r)\,\dd r^2+ r^2 \dd\Omega_2$, with the metric function given by
\begin{equation}\label{Hawyard4D}
f_\text{H} (r) = 1-\frac{2Gm r^2}{r^3 + \ell^3}\, ,
\end{equation}
where $m$ is the $4\text{D}$ ADM mass. 

One can easily check that the (constant) Weyl rescaling of the 2D metric, together with a rescaling of the time coordinate
\begin{equation}\label{Weylrescaling}
g_{\mu\nu} \to \frac{\lambda}{2\mathcal{M}} g_{\mu\nu},\,
\quad t \to   \frac{\lambda}{2\M} \, t
\end{equation}
brings the metric into the form
\begin{equation}\label{tdme}
\dd s^2_2 = -\left(1-\frac{1}{2\mathcal{M}}\frac{r^2}{r^3 + \ell^3} \right)\dd t^2  + \left(1-\frac{1}{2\mathcal{M}}\frac{r^2}{r^3 + \ell^3} \right)^{-1}\dd r^2\, .
\end{equation}
This transformation leaves the $2\text{D}$ dilaton gravity action invariant up to a constant factor, which does not alter the equations of motion.

If we now write the covariant mass of the 2D solution in terms the mass $m$ of the $4\text{D}$ black-hole solution \eqref{Hawyard4D} we get 
\begin{equation}\label{masses2D4D}
\frac{1}{2\M} \equiv \frac{2m}{\lambda^2}\, .
\end{equation}
The $4\text{D}$ metric element of the Hayward black hole can be simply written in terms of the $2\text{D}$ one \eqref{tdme} and the dilaton as $\dd s^2_4=\dd s^2_2+ (\phi/\lambda)^2 \dd\Omega_2$.
 
The peculiar relation \eqref{masses2D4D} can be seen as a consequence of both the specific conformal frame chosen here (see the discussion at the beginning of \cref{sec:Twodimensionalregularmodels}), which is particularly suited for asymptotically AdS spacetimes, and of the normalization of the Killing vector $F_0$ adopted in \cref{subsec:LDS}. As we shall see, the minus sign in $F_0$ implies that an asymptotic observer in $2\text{D}$ spacetime  measures the energy of the system with the opposite sign with respect to the asymptotic observer in the $4\text{D}$  spacetime. Hence, when the $2\text{D}$ mass  becomes bigger, the corresponding  $4\text{D}$ mass   decreases and viceversa, which is reflected in the inverse relation \eqref{masses2D4D}. We will further confirm this below, when studying the thermodynamic properties of the 2D model.

In the remainder of the paper, we will consider the metric function $f$ in the form \eqref{solgen}. It has a minimum at 
\begin{equation}\label{extremalradius}
r_\text{min} = \sqrt[3]{2}\, \ell\, .
\end{equation}
If $f(r_\text{min})<0$, the metric has two horizons, solutions of $f(r) = 0$, while if $f(r_\text{min})>0$, it has no horizons. If $f(r_\text{min})=0$, the two horizons merge, become degenerate and the configuration becomes extremal, with an event horizon located at $r_\text{min}\equiv \rext$. Using the latter and setting $f(r_\text{ext})=0$ yields the critical value of $\ell$ at extremality
\begin{equation}\label{ellcritico}
\ell_c = \frac{1}{3 \sqrt[3]{2}\,  \M}\, .
\end{equation}
Thus, for $\ell < \ell_c$ the black hole has  two horizons; for  $\ell = \ell_c$  the two horizon merge in a single one; whereas 
for $\ell > \ell_c$ the spacetime has no horizons.

Interestingly, the value of
$r_\text{ext}$ in \cref{extremalradius} is the same at which the potential \eqref{potentialHayward} changes sign (see \cref{CDVsol}). Indeed, as it is usually the case for two-horizon models \cite{Cadoni:2022chn,Giddings:1992kn,Bardeen:1999px}, the extremal, near-horizon metric is that of $\text{AdS}_2$ spacetime. 

\section{Black-hole thermodynamics }
\label{sec:FirstLawThermo}

\subsection{Thermodynamic potentials and the first principle}
\label{subsec:potentials_firstprinciple}
\begin{figure}[!h]
\centering
\includegraphics[width= 7 cm, height = 7 cm,keepaspectratio]{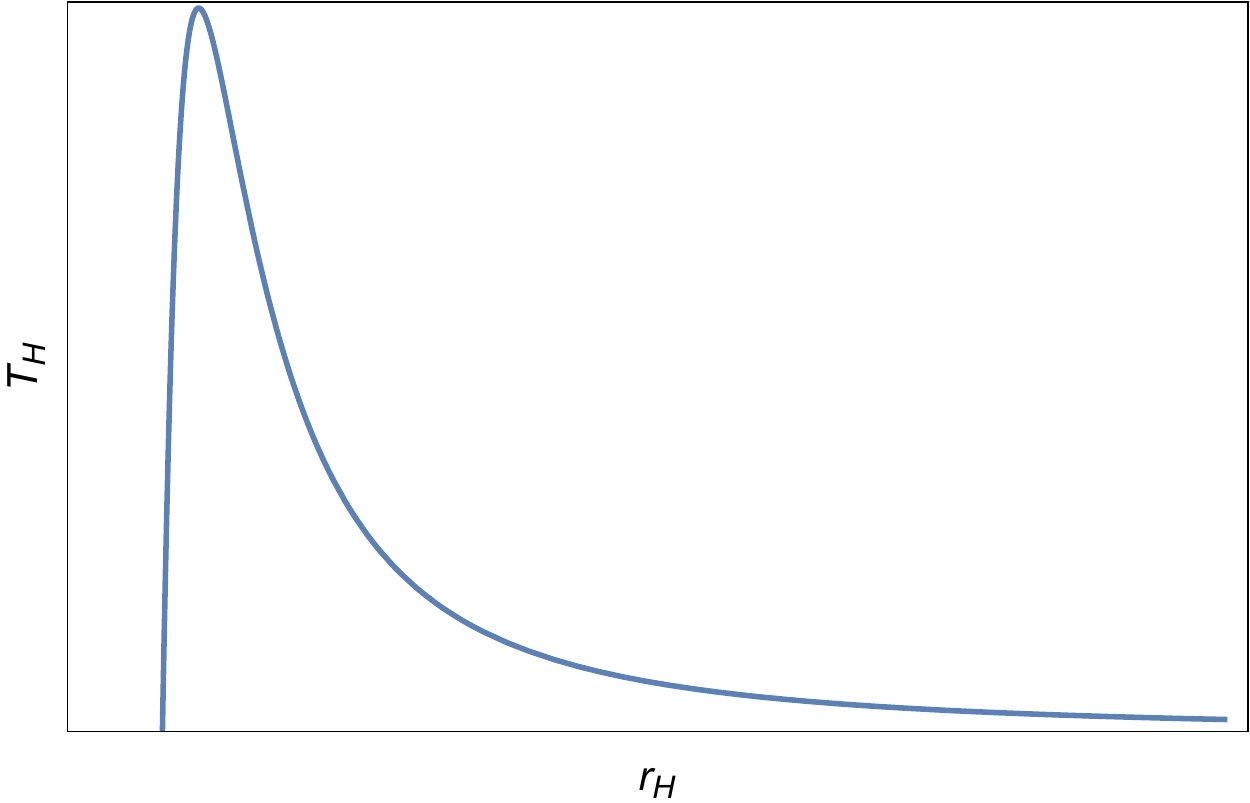}
\caption{Qualitative behavior of the temperature of the 2D black hole, according to \cref{HawkingTemp}. }
	\label{Temperaturegeneral}
\end{figure}

Our $2\text{D}$ black-hole solutions can be considered as thermodynamic systems, characterized by a Hawking tempertaure $\Th$, an internal energy $E$ and an entropy $S$.  
The Hawking temperature is given by the standard formula
\begin{equation}\label{HawkingTemp}
\Th = \frac{f'(\rH)}{4\pi} = \frac{\V(\phi_\text{H})}{4\pi}= \frac{\rH^4-2 \ell^3 \rH}{4 \pi  \lambda  \left(\rH^3+\ell^3\right)^2}\, ,
\end{equation}
where $\rH$ is the radius of the (outer) event horizon. The temperature becomes zero both at extremality, i.e., for $\rH = \rext =  \sqrt[3]{2}\, \ell$, and for $\rH\to \infty$, whereas it reaches a maximum at 
\begin{equation}\label{Temppeak}
r_\text{H, peak} = \frac{\ell}{\sqrt[3]{2}}\sqrt[3]{7 + \sqrt{45}}\, .
\end{equation}
The qualitative behavior of $\Th$ is plotted in \cref{Temperaturegeneral}.

The internal energy is usually identified with the black-hole mass. However, in our case, we  have chosen a negative normalization of the Killing vector generating time translation,  opposite to the usual positive one. Consistency with this normalization requires $E=-\M$. Using \cref{solgen}, we can express $\M$ as a function of the outer event horizon radius
\begin{equation}\label{masshorizon}
E = -\M= -\frac{1}{2}\frac{\rH^2}{\rH^3 + \ell^3}\, .
\end{equation}
A negative internal energy is somehow unusual for black holes, but it is perfectly consistent with their description (and normalizations) as thermodynamic systems. Indeed, we will confirm this below by using the euclidean action approach and proving the consistency of the first law of thermodynamics. The energy $E$, thus, is always negative and goes to its maximum value $E=0$ as $\rH \to \infty$, whereas it reaches its minimum value $E_\text{ext}= -(3\sqrt[3]{2}\, \ell)^{-1}$ for the extremal black hole, whose radius is given by \cref{extremalradius}\footnote{Notice that, according to \cref{masses2D4D}, $E=0$ corresponds to the $4\text{D}$ Hayward black hole with an infinite  mass $m$, whereas the $E_\text{ext}$ corresponds to the mass $m_\text{ext}$ of the extremal $4\text{D}$ black hole.}. Moreover, for $\rH \ge \rext$, as expected, $E(\rH)$ is a monotonic increasing function. 

Let us now calculate the entropy of the $2\text{D}$ Hayward black hole using the Euclidean action formalism, which allows us to calculate the partition function $\mathcal{Z}$ of the thermodynamic ensemble in terms of the Euclidean action $\I$, i.e., $\I= -\ln \mathcal{Z}$. By a Wick rotation of the time $t$, the action of the lagrangian \eqref{actionpresentpaper} becomes the euclidean bulk action $\Ib$. It has been shown that, in order to have a well-defined variational principle for ``mass-dominated'' dilaton gravity theories, the action must be supported by (one half) the usual Gibbons-Hawking-York (GHY) boundary term $\IGHY$ plus an additional one, containing the normal derivative of the dilaton \cite{Bagchi:2014ava}.

The action reads
\begin{equation}\label{fea}
\I = \Ib + \IGHY + \mathcal{I}_{\partial \mathcal{\phi}}= -\frac{1}{2}\int \dd^2 x \, \sqrt{-g} \left(\phi R +\V \right) -\frac{1}{2}\int_{r=r_\infty} \dd\tau\, \sqrt{h} \, \left[\phi \, \mathcal{K}-n^\mu \partial_\mu \phi \right]\, . 
\end{equation}
We  enclose the system into a hypersurface at constant $r=r_\infty$, where we define an induced, one-dimensional, metric $h$, whose extrinsic curvature is described by $\mathcal{K}_{\mu\nu}$ ($\mathcal{K}$ is its trace). $\tau$ is the euclidean time, which is periodic with period equal to the inverse of the temperature $\Th^{-1} \equiv \beta$. All quantities on the boundary will be evaluated at the cutoff $r = r_\infty$ and then we will let $r_\infty$ go to infinity.

Moreover, one could add a purely topological, Einstein-Hilbert term of the action $\I_\text{topo} \propto \phi_0 \int d^2 x \, \sqrt{-g} R$, with $\phi_0$ a constant, which only changes the value of the entropy by an additional constant $S_0$, depending on $\phi_0$. This constant value can be identified  as  the entropy of the extremal configuration.

Let us now evaluate the boundary term on the LDS. The induced metric with euclidean signature reads $h_{\mu\nu} = h_{00} = f$. The extrinsic curvature is defined in terms of the normal vector to the hypersurface $n_\mu$ as $\mathcal{K}_{\mu\nu} \equiv \frac{1}{2}\nabla_\mu n_\nu + \frac{1}{2}\nabla_\nu n_\mu$, where the normal vector reads, in this case, $n_\mu = f^{-1/2} \, \delta^r_\mu$. Therefore, when evaluated on the solution for the dilaton, we have $\IGHY=  -\frac{\lambda}{4}\beta \, r \, f' |_{r=r_\infty}$, which vanishes in the limit $r_\infty \to \infty$ when $f$ is given by \cref{solgen}. The remaining boundary term, instead, gives 
\begin{equation}
\mathcal{I}_{\partial \mathcal{\phi}} = \frac{\beta}{2} \, \lambda\,  f\biggr|_{r = r_\infty} = \beta \M \, .
\label{finalboundaryterm}
\end{equation}

Usually, one has to add a counterterm to the boundary action, needed to regularize divergences arising in the limit $r_\infty \to \infty$. This counterterm is written in terms of the extrinsic curvature of the boundary embedded in flat spacetime. In our case, such term is not needed because there are no divergences. Moreover, there is no contribution from flat spacetime ($f(r) = 1$) since \cref{fea} gives $\mathcal{S}_\text{flat} = 2\int dt \, dr \, \phi''$, which is zero for the LDS \eqref{LDsolution1}.

We now evaluate the bulk action. Using \cref{LDsolution1,LDsolution1}, we have
\begin{equation}\begin{split}\label{euclideanactionlinearfinal}
\Ib = -\frac{1}{2}\int \dd^2 x \, \sqrt{-g}\left(-\phi f'' + \V \right) =-\frac{\beta \lambda}{2} \left[-rf'+2f \right]^{r_\infty}_{\rH} =-2\beta \M - 2\pi \lambda \rH \, ,
\end{split}\end{equation}
where we used $f(\rH)=0$, $r_\infty f'(r_\infty) \to 0$ and $f(r_\infty) \to \, 2\M/\lambda$ for $r_\infty\to\infty$, and $f'(\rH) = 4\pi \Th = 4\pi/\beta$.

Combining \cref{finalboundaryterm,euclideanactionlinearfinal} yields $\I = -\beta \M - 2\pi \lambda \rH=-\ln \mathcal{Z}$, where $\mathcal{Z}$ is the partition function. The internal energy and entropy, thus, read
\begin{subequations}
\begin{align}
&E = -\partial_\beta \ln \mathcal{Z} = -\M \, ; \label{InternalenergyEuclidean}\\
&S = \beta \partial_\beta \ln \mathcal{Z} -  \ln \mathcal{Z} = 2\pi \lambda \rH = 2\pi \phi(\rH) \label{entropythermod}\, .
\end{align}
\end{subequations}

\Cref{InternalenergyEuclidean} confirms \cref{masshorizon}, as expected. 

The black-hole entropy, instead, scales as the dilaton, i.e., as the inverse $2\text{D}$ Newton constant, evaluated at the horizon. This is the usual formula for the entropy of $2\text{D}$ black holes \cite{Nappi:1992as,Kunstatter:1997my} and represents the extension to two spacetime dimensions of the usual area law in higher dimensions. This is also quite evident when the $2\text{D}$ black hole is derived from the dimensional reduction of the $4\text{D}$ one, with the dilaton playing the role of the radius of the transverse $S^2$ sphere, as it is here the case. 

Contrary to standard $4D$ regular black-hole solutions \cite{Cadoni:2022chn}, here  the entropy naturally follows the area law. This is because our $2\text{D}$ solutions do not require external matter sources and, therefore, are not coupled to a stress-energy tensor, which in the $4\text{D}$ case  describes an anistropic fluid. In general, for regular models, the latter is characterized by a density that depends on the ADM mass of the model, which introduces extra (bulk) terms altering the area-scaling of the entropy \cite{Cadoni:2022chn,Ma:2014qma}.

Adding the contribution $S_0$ of the topological action leads to
\begin{equation}\label{entropyfromaction}
S = S_0 + 2\pi \lambda \rH \, .
\end{equation}
Because of the minus sign in \cref{masshorizon}, we need to check the consistency of our derivation with the first principle of thermodynamics. By differentiating $E$ and $S$, given respectively by \cref{masshorizon,entropythermod}, with respect to $\rH$, and using \cref{HawkingTemp} for $\Th$, we can easily check that the identity $dE= \Th \dd S$ is satisfied.

The previous results allow us to compute the energy difference between two configurations, characterized by the two values $\phi_1$ and $\phi_2$ of the dilaton, in terms of the integral of the dilaton potential. This can be done in all generality by integrating the first law, considering \cref{HawkingTemp,entropythermod}
\begin{equation}\label{DeltaE12}
\Delta E_{1,2} \equiv E(\phi_1) - E(\phi_2) = \frac{1}{2}\int_{\phi_2}^{\phi_1}\V(\phi_\text{H})\, \dd\phi_\text{H}\, .
\end{equation}
Let us assume $\phi_2 < \phi_1$ and that $\phi_2$ represents a black-hole configuration. Then, for $\phi_2$, the condition \eqref{cond1} holds, whether or not $\phi_1$ represents a black hole. Therefore, $\Delta E_{1,2}>0$ and the black-hole energy increases monotonically, as already seen by analysing \cref{masshorizon}. The consequence of this is that the configuration retaining the least internal energy will be the one with the least dilaton, i.e., the extremal configuration. 

\subsection{Thermodynamic stability and second order phase transition}
\label{subsec:PhaseTransition}

Let us now investigate the thermodynamic stability of our regular black-hole solutions. This can be done by studying the specific heat and the free energy.    

\subsubsection{Specific Heat}
\label{subsubsec:SpecificHeat}

The specific heat of our solutions is given by 
\begin{equation}\label{sh}
C_\text{H} = \frac{\dd E}{\dd\Th} = \frac{\dd E}{\dd\phi_\text{H}} \left(\frac{\dd\Th}{\dd\phi_\text{H}} \right)^{-1} =4\pi  \frac{\dd E}{\dd\phi_\text{H}} \left(\frac{\dd\V}{\dd\phi_\text{H}} \right)^{-1}\, .
\end{equation}
In \cref{subsec:potentials_firstprinciple}, we showed that the internal energy of black-hole configurations is always increasing with $\phi_H$. Therefore, the sign of $C_\text{H}$ is determined by the sign of $\dd\V/\dd\phi_\text{H}$. As already discussed in \cref{sec:Twodimensionalregularmodels}, requiring the potential to satisfy the minimal requirements listed in \cref{subsec:Nonsingularconditions} implies $\V$ to be necessarily nonmonotonic. Moreover, imposing a dS-like behavior in the interior constrains the potential to have another zero at $\phi_\text{H}=\phi_0$, which further restricts the interval to $\phi_\text{H} \in [\phi_0,\infty)$ where we have at least an extremum. 
The specific heat, thus, shows a single maximum in this interval, located at $\phi_\text{peak}$ (see \cref{Potential}). Here, $\dd\V/\dd\phi_\text{H}$ changes sign (from positive to negative) as the potential falls monotonically at infinity. Therefore, we have 
\begin{itemize}
\item For $\phi_0 \leq \phi_\text{H} < \phi_\text{peak}$, $\dd\V/\dd\phi_\text{H} >0$ and therefore $C_\text{H}>0$. This correspods to a branch of thermodynamically stable configurations; 
\item For $\phi_\text{H}> \phi_\text{peak}$, $\dd\V/\dd\phi_\text{H} < 0$ and therefore $C_\text{H}<0$. This is, instead, the branch of thermodynamically unstable configurations;
\item For $\phi_\text{H} = \phi_\text{peak}$, $\dd\V/\dd\phi_\text{H} = 0$ and $C_\text{H} \to \infty$, which signals the onset of a second order phase transition. 
\end{itemize}
\begin{figure}[!h]
\centering
\includegraphics[width= 9 cm, height = 9 cm,keepaspectratio]{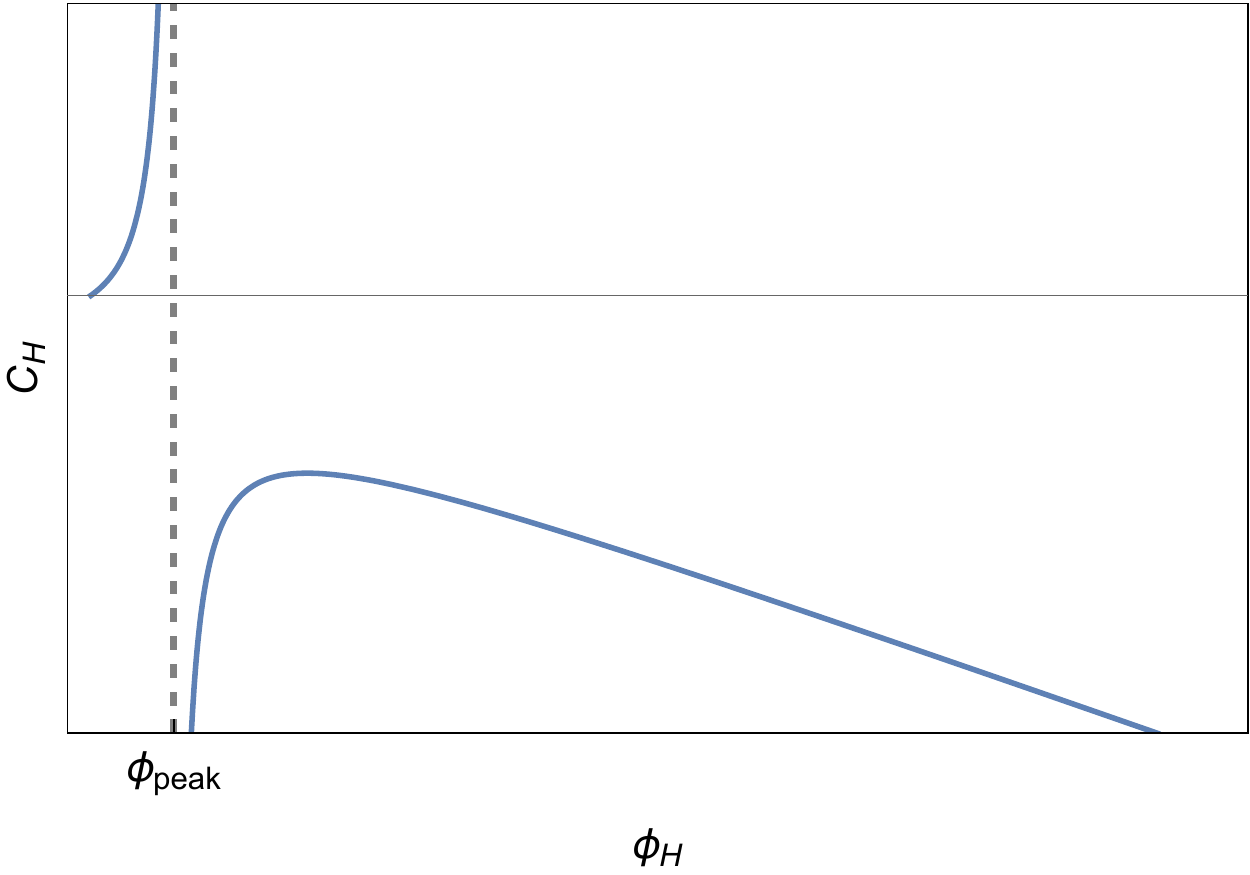}
\caption{Qualitative behavior of the specific heat of the 2D black hole. The vertical dashed line corresponds to the peak of the potential, $\phi_\text{peak}$.}
	\label{SpecificHeat}
\end{figure}

This general discussion also applies to the 2D Hayward black hole, described by the potential \eqref{potentialHayward}. Using \cref{{HawkingTemp},masshorizon}, \cref{sh} reads
\begin{equation}\label{specificheat}
C_\text{H} = \frac{\pi  \lambda  \left(\ell^3+\rH^3\right) \left(2 \ell^3 \rH-\rH^4\right)}{\ell^6-7 \ell^3 \rH^3+\rH^6}\, ,
\end{equation}
which diverges at the peak temperature \eqref{Temppeak}. In terms of the event horizon radius we have
\begin{itemize}
\item An unstable branch of large black holes ($\rH \gg r_\text{ext}$), with negative specific heat;
\item A stable branch of black holes close to extremality ($\rH \gtrsim r_\text{ext}$) with positive specific heat.
\end{itemize}
$C_\text{H}$ goes to zero at extremality and, near $\rH \sim \rext$, it behaves as
\begin{equation}
C_\text{H} \sim 2\pi \lambda (\rH -\rext) + \mathcal{O}\left[(\rH-\rext)^2\right]\, ,
\end{equation}
which scales linearly with $\rH$, as the specific heat of $\text{AdS}_2$ black holes \cite{Kumar:1994ve}. This is fully consistent with the $\text{AdS}_2$ behavior of the $4\text{D}$ LDS Hayward metric in the near-extremal, near-horizon regime. However, differently from its four dimensional counterpart, for which the $\text{AdS}_2$ spacetime represents only an approximate solution, the $\text{AdS}_2$ appears also as an exact CDV solution for the 2D model.
   
\subsubsection{Free energy}
\label{subsec:freeenergyanalysis}
The analysis of the specific heat allows us to distinguish between stable and unstable branches, but it is not sufficient to select the energetically preferred configurations. To this end, we need to analyze the difference in the free energy between different configurations  sharing the same temperature. We will  consider LDS on different thermodynamic branches, with the same temperature, but different dilaton (and, therefore, different horizon radius). In our analysis, we will not consider the CDV solutions, but only LDS. The inclusion of the CDV will be discussed in the next subsection.

\begin{figure}
\centering
\includegraphics[scale=0.38]{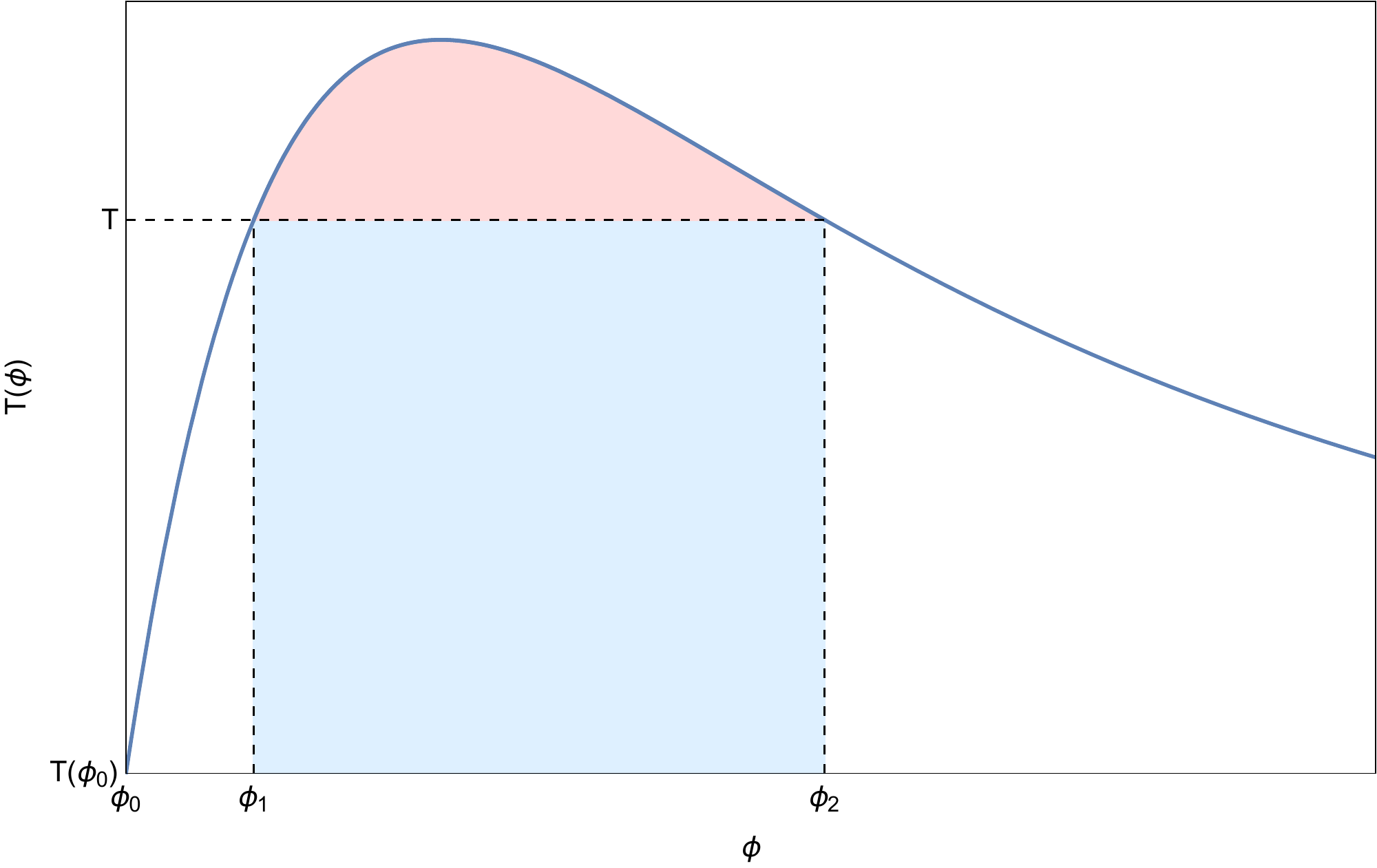} 
\caption{Temperature $T(\phi)$ as a function of $\phi$. Here, we highlighted the two configurations corresponding to $\phi_1$ and $\phi_2$, respectively in the stable and unstable branch and their common temperature $T=T(\phi_1)=T(\phi_2)$. The areas representing $(\phi_2 - \phi_1)T$ and the right-hand side of \cref{DF_stable_unstable}, are  coloured in  blue and  red,  respectively.}
\label{FreeEnergy}
\end{figure}

We consider the situation depicted in \cref{FreeEnergy}, where we present a  qualitative plot of the  temperature $T(\phi)$ of our $2\text{D}$ black holes. We focus on two different configurations, $\phi_1$ and $\phi_2$, in the stable and unstable branches, respectively, but with the same temperature $T(\phi_1) = T(\phi_2) \equiv T$. We now evaluate the free energy difference $\Delta F_{2,1} \equiv F(\phi_2) - F(\phi_1) = \Delta E_{2,1}-T \Delta S_{2,1}$. Using \cref{DeltaE12,entropythermod}, we have 
\begin{equation}\label{DF_stable_unstable}
\Delta F_{2,1} = 2\pi \int_{\phi_1}^{\phi_2} \left[T(\phi) - T \right]\, \dd\phi \, .
\end{equation}
From a geometric point of view, $\Delta F_{2,1}/2\pi$ represents the area limited by the graph of $T(\phi)$ (solid blue line in \cref{FreeEnergy}) and the straight line $T$ (dashed horizontal line in \cref{FreeEnergy}). Since $T(\phi)$ shows a single maximum located at $\phi = \phi_\text{peak}$ and $\phi_1 < \phi_\text{peak} < \phi_2$, then the relation $T(\phi) > T$ holds true in the interval $[\phi_1, \phi_2]$. This implies $\Delta F_{2,1}$ to be strictly positive, i.e., $F(\phi_2) > F(\phi_1)$.  Therefore, generic configurations in the stable branch are energetically preferred with respect to configurations in the stable branch.
These results confirm those obtained in \cref{subsubsec:SpecificHeat} and show that configurations in the stable branch retain the least free energy and are, thus, thermodynamically favoured.

\subsection{Including the constant dilaton vacuum}
So far, in our thermodynamic considerations, we have considered only the LDSs. We have already seen in \cref{sec:Twodimensionalregularmodels}, however, that our dilaton gravity model allows, in its spectrum, a solution with constant dilaton describing an $\text{AdS}_2$ spacetime, the CDV. This solution represents a GS of the theory, with different asymptotics not only with respect to the ``excited'' LDS, but also to the extremal one. In fact, the LDS, including the extremal one, are metrically AF and the dilaton depends linearly on $r$. Conversely, the CDV describes an $\text{AdS}_2$ spacetime and the dilaton is identically constant. Extremal LDSs, in the near-horizon approximation, are described by an $\text{AdS}_2$ spacetime, endowed, however, with a linear dilaton, the so called linear dilaton vacuum (LDV) \cite{Cadoni:2017dma}.

The situation described above is very common for $4\text{D}$ charged black holes \cite{Carroll:2009maa}, for which we have both an extremal, AF solution, described in the near-horizon approximation by $\text{AdS}_2$ with a linear dilaton, and an AdS$_2\times \text{S}^2$  solution (our CDV). 

Working in the context of $2\text{D}$ dilaton gravity, it has been shown that the $\text{AdS}_2$ CDV does not admit finite energy excitations, i.e., it is separated from the $\text{AdS}_2$ LDV by a mass gap \cite{Almheiri:2016fws}. Moreover, there is the additional difficulty that the two spacetimes have different asymptotics (linearly varying versus constant dilaton). The latter point makes it conceptually problematic to compare the free energies of the two configurations and to assess which one is thermodynamically favoured. These difficulties can be circumvented, and one can show, computing the free energy, that the $\text{AdS}_2$ CDV is energetically preferred with respect to the  $\text{AdS}_2$ LDV \cite{Cadoni:2017dma}.
Here, we will use a similar procedure for the solutions of the $2\text{D}$ Hayward model and  compare the free energy of the $\text{AdS}_2$ CDV with that of the extremal LDS.  

Let us first note that we can formally consider zero mass, thermal excitations of the CDV using a Rindler-like coordinate transformation, which generates a horizon with a related temperature
\begin{equation}\label{CDVnewT}
\dd s^2 = -\left(\frac{r^2}{L^2_\text{AdS}}-4\pi^2 \Th^2 L^2_\text{AdS} \right)\, \dd t^2 + \left(\frac{r^2}{L^2_\text{AdS}}-4\pi^2 \Th^2 L^2_\text{AdS} \right)^{-1} \, \dd r^2\, .
\end{equation}
We can now evaluate the bulk euclidean action of the CDV \eqref{CDVnewT}, considering that the dilaton is constant $\phi = \phi_\text{CDV}$ and the potential is zero when evaluated in the CDV. We have \footnote{We are implicitly considering also a renormalization contribution to the action, to renormalize the divergent contribution at infinity due to the AdS asymptotics. }
\begin{equation}\begin{split}
\Ib^{\text{CDV}} &= -\frac{1}{2}\int \dd^2 x \, \sqrt{-g} \, \left(-\phi_\text{CDV} \, f'' \right) =-\frac{\beta \phi_\text{CDV}}{2}4\pi \Th =-2\pi \phi_\text{CDV}\, .
\end{split}\end{equation}
The free energy $F^\text{CDV} = -T \, \ln \mathcal{Z}$ reads
\begin{equation}
F^{\text{CDV}} = -2\pi \phi_\text{CDV} \, \Th\, .
\end{equation}
We can now compute the difference in the free energy between a generic black-hole configuration of the LDS and the CDV. Using the result of \cref{entropythermod} for the entropy of the LDS and the equation $F= E-\Th S$ we obtain
\begin{equation}
\Delta F^{\text{BH}-\text{CDV}}\equiv F^{\text{BH}} - F^{\text{CDV}} = -\M - 2\pi \Th \left(\phi_\text{H} - \phi_\text{CDV} \right)\, .
\end{equation}
Since, for every black-hole solution, $\phi_\text{H} > \phi_\text{CDV}$, we have $F^{\text{BH}} < F^{\text{CDV}}$, and thus the black-hole configuration is always thermodynamically preferred. 

However, this is true whenever  $\Th \neq 0$. At $\Th=0$, we do not have thermal contributions to the free energy anymore and $\Delta F$ reduces to the difference of the masses contributions. On the other hand, at $\Th = 0$ the semiclassical approximation is broken, as signalized by the generation of the mass gap (see \cref{subsubsec:approachingextandsemiclassicalapprox}) \cite{Maldacena:1998uz,Almheiri:2016fws,Cadoni:2017dma}, and thus we cannot rely  on the euclidean action approach anymore. This does not allow to define a proper mass for the CDV, consistently with the fact that pure $\text{AdS}_2$ spacetime does not admit finite energy excitations. One could argue, following the argument of Ref.~\cite{Cadoni:2017dma}, that at $\Th=0$ the  only contribution to $\Delta F^{\text{BH}-\text{CDV}}$ comes from the mass difference, which, due the absence of finite energy excitations, should diverge, which makes the CDV energetically preferred with respect to  the extremal LDV.   

\section{Black-hole evaporation in the quasistatic approximation}
\label{subsec:semiclassicalevaporation}

In the following, we will describe the evaporation process of our regular $2$D black hole working in the quasistatic approximation and in the semiclassical regime, in which the mass is slowly varying with time so that it can be considered almost constant for each individual evaporation step.
In this way, the  backreaction of the geometry due to the radiation is not taken into account in a fully dynamic way, but it is described in a very simplified, rough manner. The dynamic character of this backreaction will be, instead, fully taken into account in the next section, where we will consider the coupling of gravity to the matter fields describing Hawking radiation. We expect our quasistatic approximation to hold for black holes very far from extremality and to break down in the near-extremal regime, where the semiclassical approximation is not capable of describing the dynamics.

Since our black holes behave as black bodies with a Planckian thermal spectrum, we use the Stefan-Boltzmann (SB) law to describe the time variation of the internal energy, which in arbitrary $d+1$ dimensions reads \cite{Cardoso:2005cd}
\begin{equation}
\frac{\dd E}{\dd t} = \sigma \mathcal{A}_{d-1} \Th^{d+1} \, , \qquad \sigma = \frac{d \, \Gamma\left(\frac{d}{2} \right) \zeta \left(d+1 \right)}{2\pi^{d/2+1}}\, , \qquad \mathcal{A}_{d-1} = \frac{2\pi^{d/2}}{\Gamma \left(\frac{d}{2} \right)} r^{d-1}\, ,
\end{equation}
where $\sigma$ is the SB constant, $\mathcal{A}_{d-1}$ is the $(d-1)$-dimensional emitting surface, $\Gamma(x)$ and $\zeta(s)$ are the gamma and Riemann zeta functions, respectively. In the present case, $d = 1$, and thus
\begin{equation}\label{2DSBLawGeneral}
\frac{\dd E}{\dd t} = -\frac{\dd\M}{\dd t} = -\frac{\pi}{6}\Th^2\, ,
\end{equation}
where we used the fact that $E = -\M$.

\subsection{Evaporation time}
\label{subsubsec:evaporationtime}

To compute the evaporation time, we express $\M$ as a function of the event horizon radius and use $\dd\M/\dd t = (\dd\M/\dd\rH)(\dd\rH/\dd t)$ together with  \cref{{masshorizon},HawkingTemp,2DSBLawGeneral}, to obtain
\begin{equation}\label{drHdtSB}
\frac{\dd\rH}{\dd t} = -\frac{1}{12\lambda}\Th\, .
\end{equation}
Inverting and integrating yields the evaporation time required to pass from an initial configuration with event horizon radius $r_{\text{H},0}\gg r_\text{ext}$ to a final one with radius $r_\text{H, final}$
\begin{equation}\label{Deltatgeneral}
\Delta t = -12\lambda \int_{r_{\text{H},0}}^{r_\text{H, final}} \frac{\dd\rH}{\Th}\, .
\end{equation}
From this expression, it is already evident that, as $\Th \to 0$, i.e., as we approach extremality, $\Delta t \to \infty$. In other words, reaching the extremal configuration, in the quasistatic semiclassical approximation, requires an infinite time, consistently with the thermodynamic stability analysis of the previous subsections. 

For the particular model under investigation described by \cref{solgen}, \cref{Deltatgeneral} reads
\begin{equation}\label{Deltatevaporation}
\Delta t = -48\pi \lambda^2 \int_{r_{\text{H},0}}^{r_\text{H, final}} \dd\rH \, \frac{(\rH^3+\ell^3)^2}{\rH^4 - 2\ell^3 \rH} = -8\pi \lambda^2 \left[2 \rH^3-3\ell^3 \ln \rH + 9 \ell^3 \, \ln \left(\rH^3 - 2\ell^3 \right) \right]_{r_{\text{H},0}}^{r_\text{H, final}}\, .
\end{equation}
In the extremal limit $r_\text{H, final} \to \sqrt[3]{2} \, \ell = r_\text{ext}$ we have a logarithmic divergence, as expected. This is consistent with the behavior of $4\text{D}$ regular models (see \cite{Akil:2022coa, Alesci:2011wn, Carballo-Rubio:2018pmi} and references therein).

\subsection{Time variation of mass and entropy}
\label{subsubsec:massevolution}

\begin{figure}[!h]
\centering
\subfigure[]{\includegraphics[width= 8 cm, height = 8 cm,keepaspectratio]{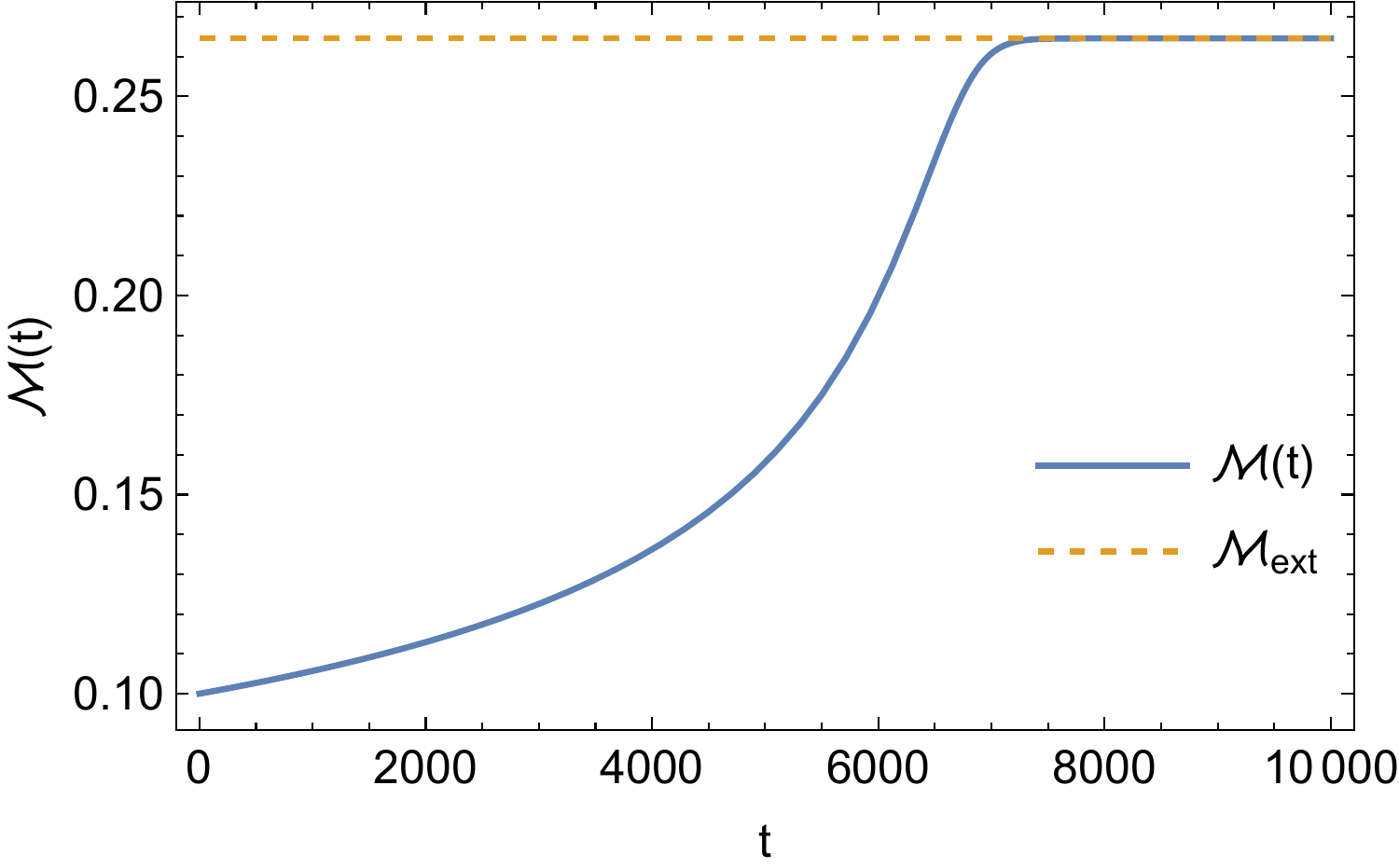} \label{Timevariationmass}}
\hfill
\subfigure[]{\includegraphics[width= 7.9 cm, height = 7.9 cm,keepaspectratio]{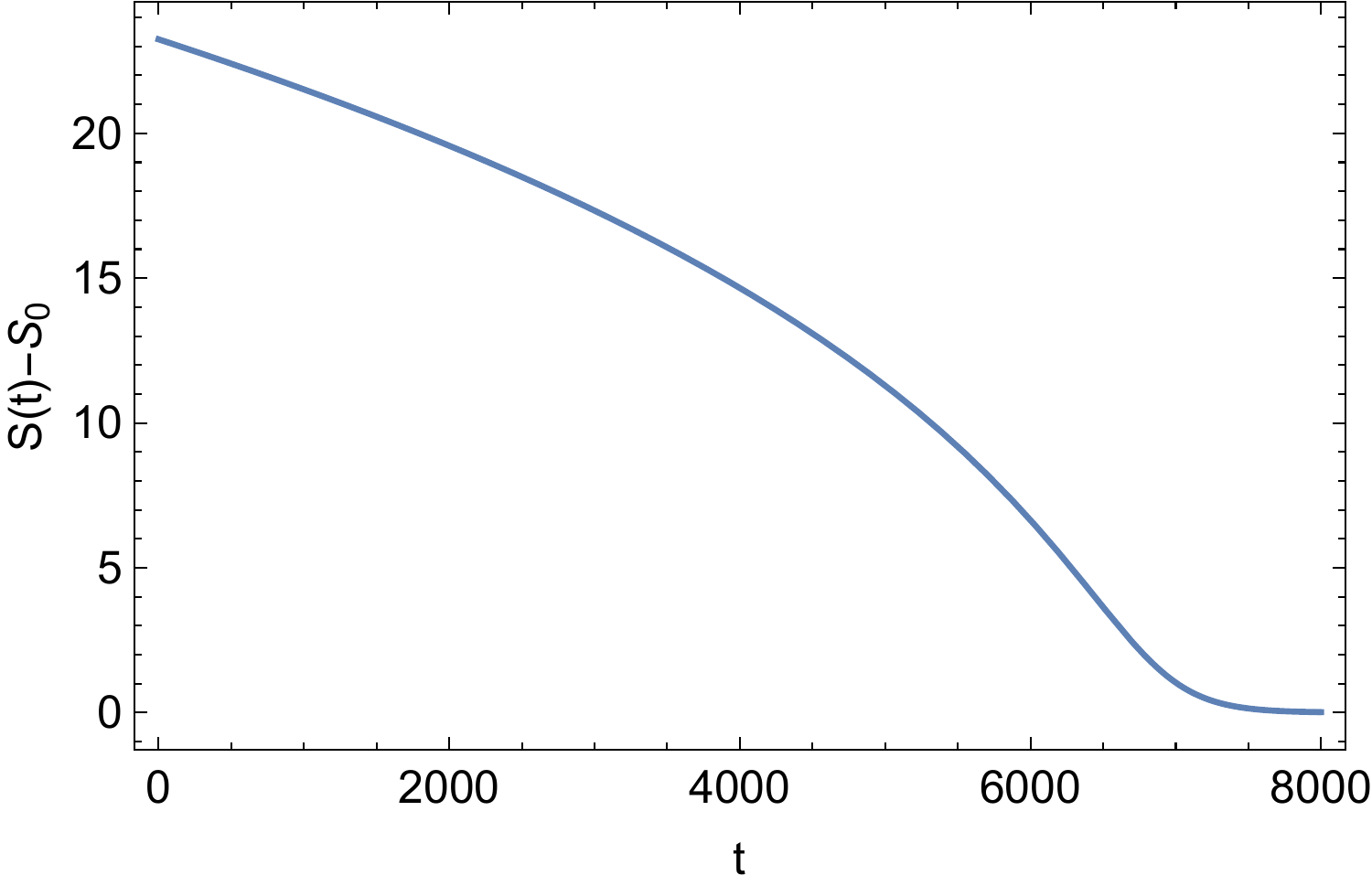} \label{Timevariationentropy}}
\caption{\textbf{Figure (a):} Numerical solution of \cref{2DSBLawGeneral} (blue solid line), which shows the time evolution of the mass of an evaporating black hole as a function of time, in the quasistatic approximation. We see that at large times, the numerical solution asymptotes the extremal mass (orange dashed line). For the numerical integration, we set $\M(t=0) = 0.1 \, \lambda$ as an initial condition. \textbf{Figure (b): }Time variation of the entropy of the black hole according to \cref{entropyfromaction}. At late times, the entropy goes to zero, after subtracting  the topological term $S_0$. In both figures, we set $\lambda = \ell = 1$.}
\end{figure}

Let us now compute how the mass of our $2\text{D}$ Hayward black hole evolves in time due to the emission of Hawking radiation, according to the SB law \eqref{2DSBLawGeneral}. This can be done by first inverting $\rH$ and $\Th$, to express  them as functions of $\M$ and  then by numerically solving \cref{2DSBLawGeneral} written in the form $\frac{\dd\M}{\dd t} = \frac{\pi}{6}\Th^2(\M)\,$. As a boundary condition, we impose $\M(t= 0) = 0.1\, \lambda$. We also set $\lambda = \ell = 1$. Using the expression for $\rH$ as a function of the mass obtained by inverting \cref{masshorizon}, for $\M = 0.1$, we have $\rH \simeq 4.96$, which is about twice as large as that pertaining to the temperature peak \eqref{Temppeak} ($r_\text{H, peak} \simeq 1.9$ with $\ell=1$), which confirms that the initial state belongs to the unstable branch. We have also checked  that the final results are independent of the value of the initial mass.

The result of the numerical integration is shown in \cref{Timevariationmass}. We see that, at large times, the mass asymptotes the extremal one. The mass increases during the evaporation due to the negative normalization of the Killing vector generating time translations. Indeed, the internal energy $E=-\M$ is decreasing, as it should be during the evaporation.  

We can also derive the time evolution of the black-hole entropy. This can be done by simply combining the solution of \cref{2DSBLawGeneral} together with the function $\rH(\M)$ (obtained by inverting \cref{masshorizon}) and \cref{entropyfromaction}. The result is reported in \cref{Timevariationentropy} and confirms the fact that, as the solution asymptotes the extremal one, the entropy reduces to zero. This is true only if we subtract the contribution $S_0$ of \cref{entropyfromaction}.  

These results will be extended in \cref{sec:blackholeevaporationandbackreaction}, where we will go beyond the quasistatic approximation and we will consider the full dynamics of the backreaction of Hawking radiation on the background geometry. We will show that the inclusion of the latter causes the evaporation process to take place in a \emph{finite} time. 

\subsection{Approaching extremality and  breakdown of the semiclassical approximation}
\label{subsubsec:approachingextandsemiclassicalapprox}

The divergence of the evaporation time for black holes approaching extremality, found in the previous section, signalizes the breakdown of the semiclassical approximation. Let us now study in details how an excited configuration approaches the extremal limit, by solving \cref{drHdtSB} at leading order around extremality, i.e., around $\rH \sim \sqrt[3]{2} \, \ell$. 

Near extremality, at leading order, the temperature varies linearly with $\rH$, 
\begin{equation}\label{Temperaturearoundextremality}
\Th \simeq \frac{\rH - \sqrt[3]{2}\, \ell}{6\pi \lambda \ell^3} + \mathcal{O}\left[\left(\rH - \sqrt[3]{2} \ell \right)^2 \right]\, ,
\end{equation}
which gives, after solving \cref{drHdtSB}, the time-dependence of $\rH$
\begin{equation}\label{rtapproachingextremality}
\rH(t) \simeq \sqrt[3]{2} \ell + \alpha_1 e^{-\frac{t}{72 \pi \lambda^2 \ell^3}}\, ,
\end{equation}
where $\alpha_1$ is an integration constant. 

Moreover, $\M(t)$ behaves, near extremality, as
\begin{equation}
\M \simeq \frac{\left(\sqrt[3]{2} \ell + \alpha_1 e^{-\frac{t}{72 \pi \lambda^2 \ell^3}} \right)^2}{2 \left[\ell^3 + \left(\sqrt[3]{2} \ell +  \alpha_1 e^{-\frac{t}{72 \pi \lambda^2 \ell^3}} \right)^3 \right]}\, ,
\end{equation}
which reduces to $\frac{1}{3 \sqrt[3]{2} \, \ell} = \Mext$ only for $t \to \infty$, confirming the numerical results of the previous subsection. The entropy, instead, approaches exponentially that of the extremal configuration at $t \to \infty$, according to \cref{rtapproachingextremality}.

However, one must question the validity of the semiclassical approximation near extremality. The latter breaks down when the energy of Hawking quanta, which is of  order  $\Th$, becomes comparable with the energy of the black-hole energy above extremality $\Delta E=\left|\M - \Mext \right|$ (see, e.g., Ref.~\cite{Maldacena:1998uz}). The energy scale  at which this breakdown occurs determines the mass gap separating the CDV from the continuous part of the spectrum (the LDS) \cite{Maldacena:1998uz, Almheiri:2016fws}.

\begin{figure}[h!]
\centering
\subfigure[]{\includegraphics[width= 8 cm, height = 8 cm,keepaspectratio]{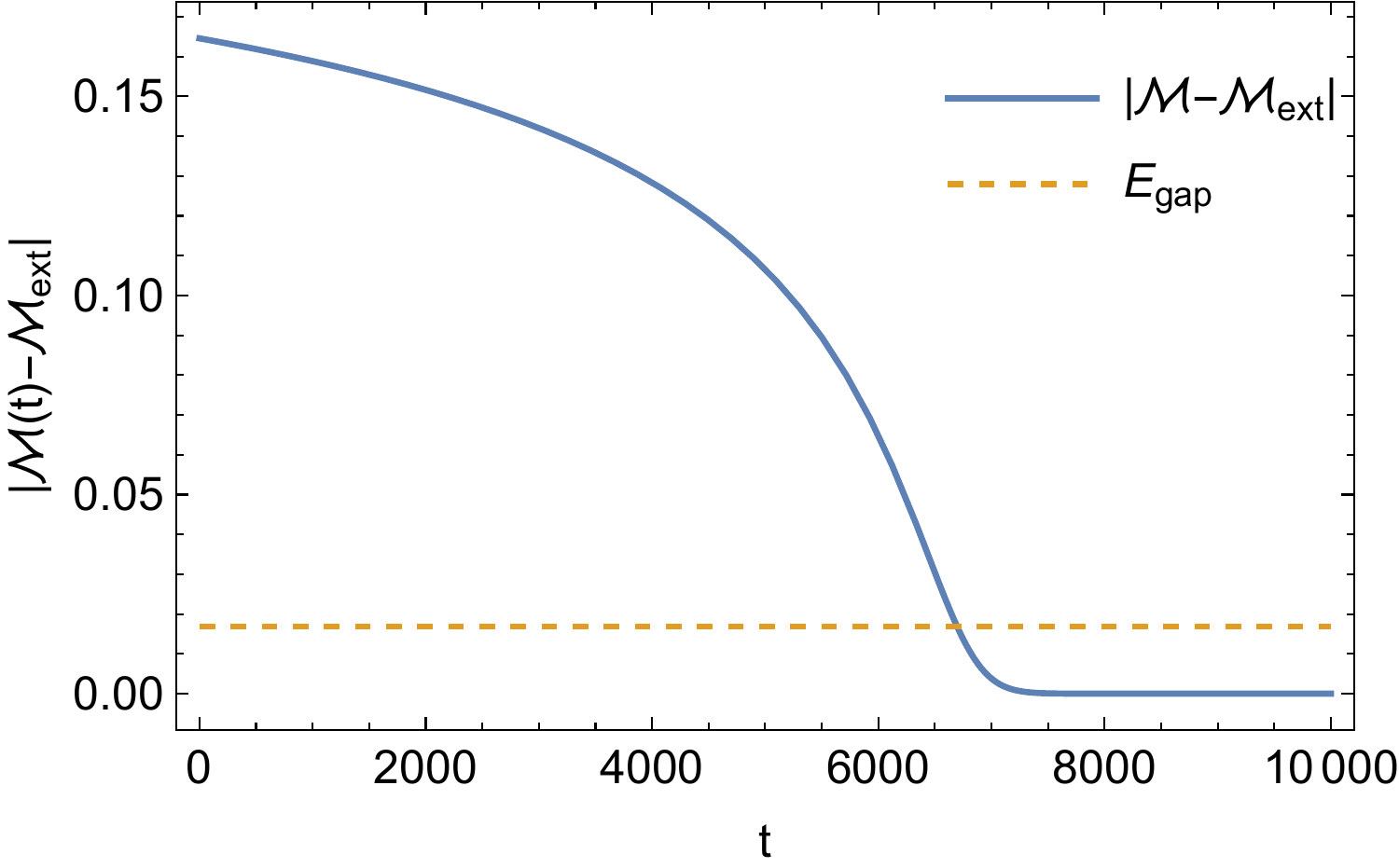}\label{fig:Egap}}
\hfill
\subfigure[]{\includegraphics[width= 8 cm, height = 8 cm,keepaspectratio]{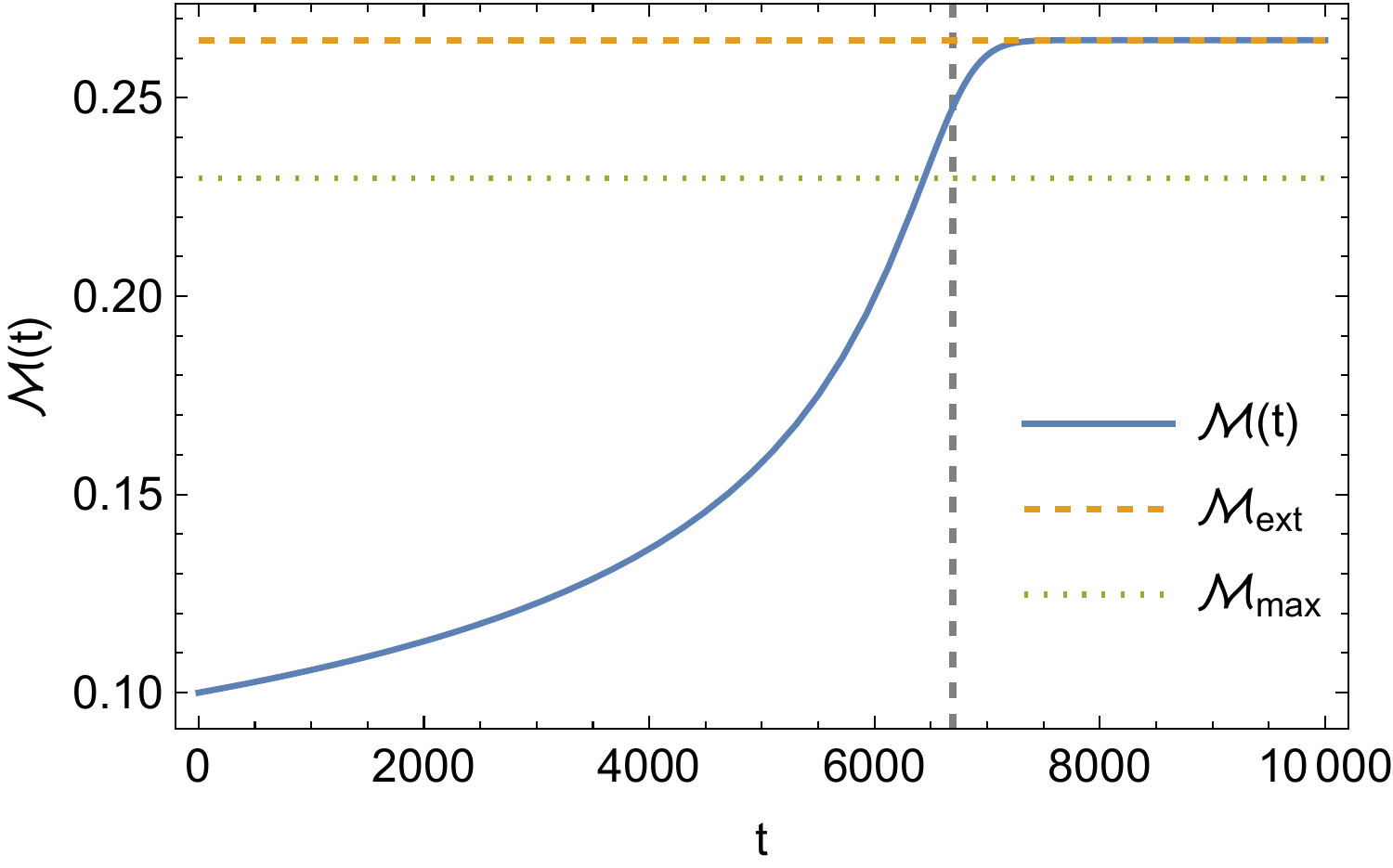}\label{Msemiclassicalbd}}
\caption{\textbf{Figure (a):} Plot of $\Delta E$ as a function of time (solid blue line). The horizontal dashed orange line corresponds to the value of the mass gap \eqref{Egap}. The intersection point between the two curves gives the  time where the semiclassical approximation breaks down. \textbf{Figure (b):} Mass evolution in time (solid blue line). The horizontal dashed orange line corresponds to the value of the mass of the extremal configuration. The horizontal dotted green line corresponds to the value of the mass at the temperature peak \eqref{Temppeak}. The vertical dashed line, instead, signals the instant of time where $|\M - \Mext|\sim E_\text{gap}$, with $E_\text{gap}$ given by \cref{Egap}. As it can be seen, the breakdown of the semiclassical approximation happens after the evaporation process reaches the temperature peak. \\
For both figures, we set $\M(t=0) = 0.1\, \lambda$ and $\ell = \lambda = 1$.}
\end{figure}

This mass gap can be determined by expanding \cref{masshorizon} near extremality, which yields
\begin{equation}
\Delta E \simeq \frac{\left(\rH - \sqrt[3]{2} \ell \right)^2}{6 \ell^3}=6 \pi^2 \lambda^2 \ell^3 \, \Th^2\, .
\end{equation}
From $\Delta E\simeq \Th$, we easily find the energy gap 
\begin{equation}\label{Egap}
E_\text{gap} \simeq \frac{1}{6\pi^2 \ell^3 \lambda^2}\, .
\end{equation}
This result is consistent with those obtained for $4\text{D}$ black holes with two horizons merging into a single one, in particular for charged black holes. For instance, in the Reissner-Nordstr\"om case, the energy gap behaves as $E_\text{gap} \propto Q^{-3}$, where $Q$ is the black-hole charge \cite{Maldacena:1998uz}. As one can expect, here the role of $Q$ is played by $\ell$ \footnote{An important difference is that, contrary to $Q$, $\ell$ is not related to any conserved quantity at infinity and thus it is not associated to any thermodynamic potential.}. \Cref{fig:Egap} shows the time variation of $\Delta E=\left|\M - \Mext \right|$: the intersection with the horizontal dashed line, corresponding to $\Delta E \sim E_\text{gap}$, identifies the time at which the semiclassical approximation breaks down.

It is interesting to notice that the limit of validity of the semiclassical approximation sets also the limit of validity of the quasistatic one, which cannot be valid when the former is broken. In fact, the quasistatic approximation is valid in the initial stages, when the black hole is macroscopic, which essentially evaporates as a GR one. It remains also valid for most of the evaporation process because the evaporation time is much larger than the typical black-hole time scale $1/\Delta E$ (see also, e.g., Refs. \cite{Bonanno:2000ep,Carballo-Rubio:2018pmi}). Only at extremality, when $\Delta E$ goes to zero, the two time scales become comparable.

Moreover, it should be considered that the time at which the semiclassical approximation breaks down (which can be read from \cref{fig:Egap}) is close to the time at which the evaporation process reaches the maximum of the temperature \eqref{Temppeak}, where  we have the onset of the second order phase transition (see \cref{Msemiclassicalbd}).

\section{Coupling to conformal matter}
\label{sec:CouplingConformalMatter}

In the previous section we have described the black-hole evaporation process in the  semiclassical and quasistatic approximations. Within these approximations, the backreaction effects of the geometry on the presence of Hawking radiation is completely   encoded in the change in time of the black-hole mass $\M$. As we have seen above, this may be a good approximation in the early stages of the evaporation, but it is expected to fail at later times. Another shortcoming of the approximation is that the backreaction is not fully dynamic, since it does not involve the full metric solution. Its role is simply encoded in the variation of the black-hole mass.   

In this section, we will give an exact semiclassical description of the evaporation process by studying the coupling of our $2\text{D}$ model to quantum conformal matter, in the form of $N$  massless scalar fields. This coupling is most easily analyzed in the conformal gauge, where the $2\text{D}$ metric reads
\begin{equation}\label{conformalgauge}
\dd s^2 = -e^{2\rho(x^+, \, x^-)}\dd x^+ \dd x^-\, ,
\end{equation}
where $e^{2\rho}$ is the conformal factor of the metric. The transition  from the metric in the Schwarzschild gauge \eqref{LDsolution1} to that in the form \cref{conformalgauge} is realized by using the  coordinates, $x^\pm = t \pm r_\ast$, where $\rst \equiv \int \dd r/f$ is  the  tortoise coordinate. The system of coordinates $x^\pm$ does not cover the interior of the black hole, but only the region outside the outer horizon. Indeed, $x^+-x^- \to -\infty$ corresponds to the horizon ($x^+ \to -\infty$ gives the past horizon, while $x^- \to \infty$  future  horizon), while $x^+ - x^- \to \infty$ corresponds to asymptotic infinity. 

The field equations stemming from \cref{actionpresentpaper} are now
\begin{subequations}
\begin{align}
&8e^{-2\rho} \partial_+ \partial_-\rho = -\frac{\dd\V}{\dd\phi}\, ; \label{EqScalarfieldConfv}\\
&\partial_+^2 \phi - 2\partial_+ \rho \partial_+ \phi = 0\, ; \label{EqEinsteinppv}\\
&\partial_-^2 \phi - 2\partial_- \rho \partial_-\phi = 0 \, ; \label{EqEinsteinmmv}\\
&\partial_+ \partial_- \phi + \frac{\V}{4} e^{2\rho} =0\, . \label{EqEinsteinmpv}
\end{align}
\label{conformaleomvacuum}
\end{subequations}
The solution reads
\begin{subequations}
\begin{align}
&e^{2\rho} = f = \frac{2\M}{\lambda} + \frac{1}{\lambda^2}\int^\phi \dd\psi \, \V \, \equiv \frac{2\M}{\lambda} + \J\, ; \label{metricconfgauge}\\
&\int^\phi \frac{\dd\psi}{e^{2\rho}} = \int^\phi \frac{\dd\psi}{\frac{2\M}{\lambda} + \J} = \lambda \rst = \frac{\lambda}{2}\left(x^+ - x^- \right)\, , \label{dilatonconfgauge}
\end{align}
\label{vacuumsolution}
\end{subequations}
where we defined $\J \equiv \frac{1}{\lambda^2}\int^\phi \dd\psi \, \V(\psi)$. 

\subsection{Coupling to matter: shock wave solution}
\label{subsec:shockwavesolution}

We now couple 2D  dilaton  gravity to the $N$ massless scalar fields describing conformal matter. The full action reads
\begin{equation}\label{actioncouplingmatter}
\mathcal{S} = \frac{1}{2}\int \dd^2 x \, \sqrt{-g} \, \left[\phi R + \V - \frac{1}{2}\sum_{i=1}^N \left(\nabla f_i \right)^2 \right]\, .
\end{equation}
The stress-energy tensor of matter fields reads
\begin{equation}
T_{\mu\nu} = -\frac{1}{4} g_{\mu\nu}\sum_{i=1}^N g^{\rho\sigma}\partial_\rho f_i \partial_\sigma f_i + \frac{1}{2}\sum_{i=1}^N\partial_\mu f_i \partial_\nu f_i \, .
\end{equation}
The field equations \eqref{conformaleomvacuum} now become
\begin{subequations}
\begin{align}
&8e^{-2\rho} \partial_+ \partial_-\rho = -\frac{\dd\V}{\dd\phi}\, ; \label{EqScalarfieldConfM}\\
&\partial_+^2 \phi - 2\partial_+ \rho \partial_+ \phi = -T_{++} =  -\frac{1}{2}\sum_{i=1}^N\partial_+ f_i \partial_+ f_i\, ; \label{EqEinsteinppM}\\
&\partial_-^2 \phi - 2\partial_- \rho \partial_-\phi = -T_{--} = -\frac{1}{2}\sum_{i=1}^N\partial_- f_i \partial_- f_i \, ; \label{EqEinsteinmmM}\\
&\partial_+ \partial_- \phi + \frac{\V}{4} e^{2\rho} =0\, ; \label{EqEinsteinmpM}\\
&\partial_+ \partial_- f_i = 0 \quad \Rightarrow \quad f_i = f_{i,+}(x^+) + f_{i, -}(x^-)\, .\label{matterfields}
\end{align}
\label{conformaleommatter}
\end{subequations}
The matter-coupled field equations above admit an exact solution if we consider an ingoing shock wave starting at $x^+ = x^+_0$ and propagating in the $x^-$ direction, while no energy flux is present in the $x^+$ direction
\begin{equation}\label{SETshockwave}
T_{++} = -\M \, \delta\left(x^+-x^+_0\right) = \frac{1}{2}\sum_{i=1}^N \partial_+ f_i \partial_+ f_i \, , \qquad T_{--} = 0\, .
\end{equation}
The minus sign in $T_{++}$ is again due to the normalization of the Killing vector of the metric. 

From Birkhoff's theorem, we can write the full solution by patching, on the infall line $x^+ = x^+_0$, the vacuum solution together with the one after the shock wave \cite{Cadoni:1995dd}. 

\subsubsection*{\fbox{$x^+\leq x_0^+$}}
At the end of \cref{subsec:potentials_firstprinciple}, we showed that, in the linear dilaton case, the GS of the theory, i.e., the state retaining the least internal energy, is the extremal black-hole configuration, characterized by a mass $\M_\text{ext}$  given by \cref{ellcritico}.

The vacuum solution (before the shock wave) therefore is equivalent to \cref{vacuumsolution} with $\M = \M_\text{ext}$

\begin{equation}\label{GSextremal}
e^{2\rho} = \frac{2\M_\text{ext}}{\lambda} + \J \, , \qquad \int^\phi \frac{\dd\psi}{\frac{2\M_\text{ext}}{\lambda} + \J} = \frac{\lambda}{2}\left(x^+-x^- \right) \, .
\end{equation}

\subsubsection*{\fbox{$x \geq x_0^+$}}
Since $T_{--} = 0$, now the solution is (see Ref.~\cite{Cadoni:1995dd})
\begin{equation}\begin{split}
&e^{2\rho} = \left(\frac{2\M}{\lambda} + \J \right) F'(x^-) \, ; \\
&\int^\phi \frac{\dd\psi}{\J + \frac{2\M}{\lambda}} = \frac{\lambda}{2} \left[x^+ - x^+_0 - F(x^-) \right]\, ; \\
&F'(x^-) \equiv \frac{\dd F(x^-)}{\dd x^-} = \frac{\J_{0} + \frac{2\Mext}{\lambda}}{\J_0 + \frac{2\M}{\lambda}}\, ,
\label{exactsolutionshockwave}
\end{split}\end{equation}
where $\J_0 \equiv \J|_{x^+ = x^+_0}$  and  $F(x^-)$  is a function needed to map the old coordinate $x^-$ of the observer in the GS solution into a new coordinate, which pertains to an observer in the excited solution. Its form is fixed by requiring continuity of the metric function across the shock wave at $x^+ = x^+_0$. This defines the function up to an integration constant, which is fixed by requiring the continuity of the dilaton across the shock wave. 

As usual, and as we will see in more details in \cref{subsec:Hawkingfluxmain}, the function $F(x^-)$ generates the Hawking flux of particles, which can be described in terms of the change of the  coordinate  $x^-$ defined by $F$. In fact, in $2\text{D}$, the flux of Hawking particles can be described in terms of the Schwarzian derivative of the function   $F(x^-)$ (see, e.g., Refs.~\cite{Cadoni:1995dd,Cadoni:1994uf}).  

\begin{figure}[h!]
\centering
\includegraphics[width= 9 cm, height = 9 cm,keepaspectratio]{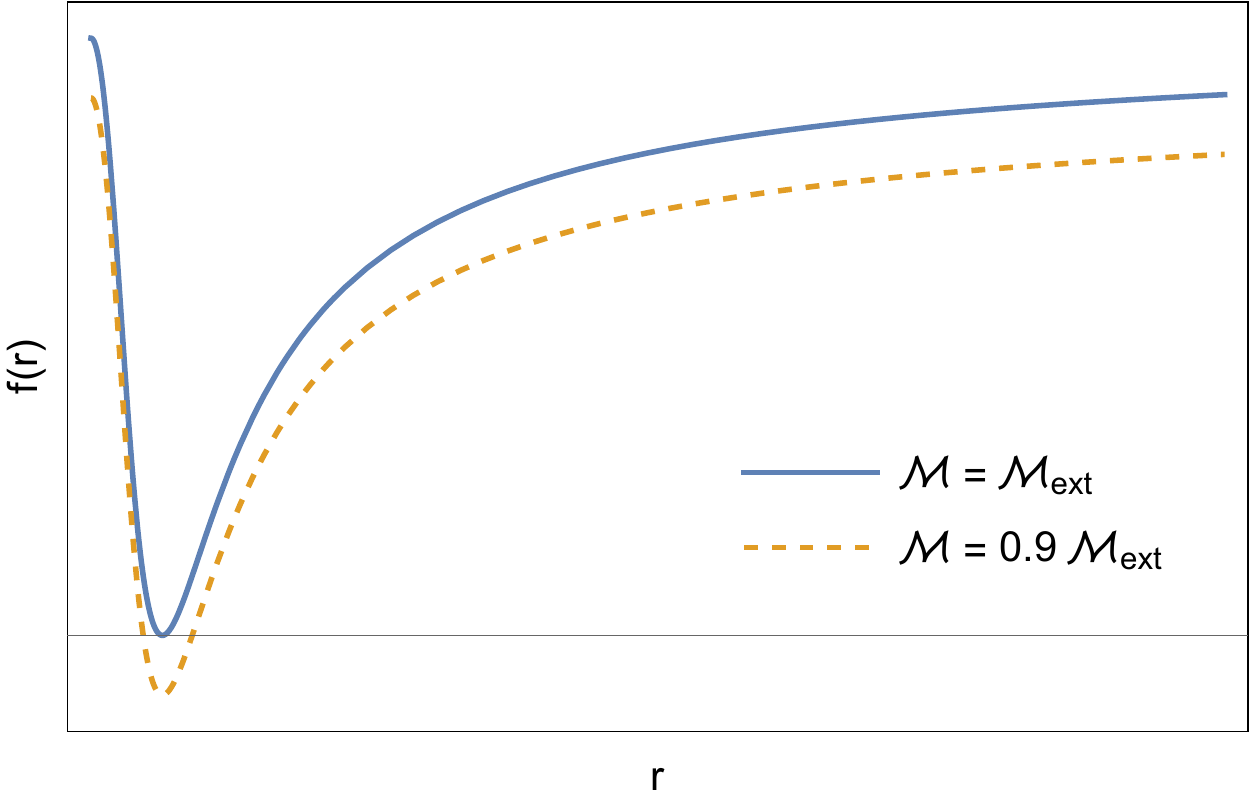}
\caption{Metric functions with varying mass  $\M$. The blue solid line refers to the extremal configuration with $\M = \Mext$. The orange dashed line, instead, refers to an excited state (after the shock wave) with $\M = 0.9 \, \Mext$, which presents two event horizons. }
	\label{ExtremalvsNonextremal}
\end{figure}

Due to the sign of the shock wave \eqref{SETshockwave} and to the minus sign in \cref{masshorizon}, the mass $\M$ of the excited states is \textit{less} than the extremal mass, as the shock wave increases the internal energy of the system. The physical picture we expect is therefore the following (see also \cref{ExtremalvsNonextremal}). The shock wave increases the internal energy of the initial system, i.e., the GS, extremal configuration: the degenerate horizon splits into two apparent horizons. According to the thermodynamic analysis and the results of \cref{subsubsec:massevolution}, we then expect that, when dynamically evolving the system, these two horizons will meet again at the end of evaporation, and merge to give again the extremal configuration. 

We will confirm this in the next section by keeping into account the dynamic contributions of backreaction effects of the radiation on the geometry and dynamically describing the evaporation process by numerically integrating the field equations. 

\section{Black hole evaporation and backreaction}
\label{sec:blackholeevaporationandbackreaction}

The evaporation and its backreaction effects on the spacetime geometry, are studied by quantizing the conformal matter on the curved $2\text{D}$ background. An important consequence of the curvature of the spacetime is that the otherwise classically traceless stress-energy tensor acquires a nonzero trace, proportional to the Ricci scalar, which is the so-called conformal anomaly \cite{Christensen:1977jc}
\begin{equation}\label{conformalanomaly}
\langle T^\mu_\mu \rangle = \frac{N}{24} R\, ,
\end{equation}
where $N$ is the number of matter fields. This can be accounted for by adding, to the classical action, the nonlocal Polyakov term
\begin{equation}\label{Polyakovactionsimple}
\mathcal{S}_\text{Pol} =-\frac{N}{96}\int \dd^2 x \, \sqrt{-g} \, R \Box^{-1} R\, ,
\end{equation}
where $\Box^{-1}$ is the scalar Green function. 

Using \cref{conformalanomaly}, we can derive the expectation value of $T_{+-}$, which in the conformal gauge is entirely local
\begin{equation}\label{confanomTpm}
\langle T_{+-} \rangle = -\frac{N}{12}\partial_+ \partial_- \rho \, . 
\end{equation}
The nonlocal effects stemming from the action \eqref{Polyakovactionsimple} are, instead, encoded in the other components of the stress-energy tensor $\langle T_{\pm\pm}\rangle$, which can be obtained from covariant conservation of the latter
\begin{subequations}
\begin{align}
&\langle T_{--} \rangle = -\frac{N}{12} \left[\partial_- \rho \partial_- \rho - \partial_-^2 \rho + t_-(x^-) \right] \, ; \label{confanomTmm}\\
&\langle T_{++} \rangle = -\frac{N}{12} \left[\partial_+ \rho \partial_+ \rho - \partial_+^2 \rho + t_+(x^+) \right] \, . \label{confanomTpp}
\end{align}
\end{subequations}
Here, $t_\pm(x^\pm)$ are integration functions, which depend on the boundary conditions and therefore encode the nonlocal effects of the Polyakov action \eqref{Polyakovactionsimple}. These functions are also sensitive to the choice of the coordinates. Indeed, under a conformal transformation of the coordinates $x^\pm \to y^\pm(x^\pm)$, the conformal factor transforms as
\begin{equation}
\rho(y^+, \, y^-) = \rho(x^+, \, x^-) - \frac{1}{2}\ln \frac{\dd y^+}{\dd x^+}\, \frac{\dd y^-}{\dd x^-}\, ,
\end{equation}
which, plugged into \cref{confanomTpm,confanomTmm,confanomTpp}, yields the anomalous transformation of the stress-energy tensor with the Schwarzian derivative
\begin{equation}
\begin{split}
\left(\frac{\dd y^{\pm}}{\dd x^\pm} \right)^2 T_{\pm\pm} (y^\pm) = T_{\pm\pm}(x^\pm) -\frac{N}{12} \left\{y^{\pm}, x^\pm \right\}\, , \qquad \left\{y , x \right\} = \frac{y'''}{y'}-\frac{3}{2} \frac{y''^2}{y'^2}\, ,
\end{split}
\label{anomaloustransformation}
\end{equation}
where the prime indicates a derivative with respect to $x$. The form of the stress-energy tensor in the new coordinate system is preserved if the $t_\pm$'s trasform as
\begin{equation}\label{anomaloustransftpm}
\left(\frac{\dd y^{\pm}}{\dd x^\pm} \right)^2 t_\pm (y^\pm) = t_\pm(x^\pm) + \left\{y^{\pm}, x^\pm \right\}\, .
\end{equation}
Including the conformal anomaly, the field equations \eqref{conformaleommatter} become
\begin{subequations}
\begin{align}
&8e^{-2\rho} \partial_+ \partial_-\rho = -\frac{\dd\V}{\dd\phi}\, ; \label{EqScalarfieldConfMAnom}\\
&\partial_+^2 \phi - 2\partial_+ \rho \partial_+ \phi = \M \, \delta\left(x^+-x^+_0\right) + \frac{N}{12} \left[\left(\partial_+ \rho \right)^2 - \partial_+^2 \rho + t_+(x^+) \right]\, ; \label{EqEinsteinppMAnom}\\
&\partial_-^2 \phi - 2\partial_- \rho \partial_-\phi =  \frac{N}{12} \left[\left(\partial_- \rho \right)^2 - \partial_-^2 \rho + t_-(x^-) \right]\, ; \label{EqEinsteinmmMAnom}\\
&\partial_+ \partial_- \phi + \frac{\V}{4} e^{2\rho} =-\frac{N}{12}\partial_+ \partial_- \rho\, ; \label{EqEinsteinmpMAnom}\\
&f_i = f_{i,+}(x^+) + f_{i, -}(x^-)\, .\label{matterfieldsAnom}
\end{align}
\label[plural]{conformaleommatterwithanomaly}
\end{subequations}
\Cref{conformaleommatterwithanomaly} can be solved once suitable initial conditions to fix the functions $t_\pm(x^\pm)$ are imposed. These can be determined assuming the GS as the initial state. In conventional AF models, like the CGHS one \cite{Callan:1992rs}, the GS is pure Minkowski spacetime. One can therefore define a global coordinate transformation in which the conformal metric \eqref{conformalgauge} is manifestly flat, i.e., we can define a system of coordinates in which $e^{2\rho} = \text{constant} = 1$. One can then assume that there is no incoming radiation (except from the classical shock wave) and that there is no net outcoming flux, so that
\begin{equation}\label{TmunuGS}
\langle T_{\mu\nu} \rangle_{\text{GS}} = 0
\end{equation}
identically, which implies $t_\pm = 0$ on the GS in this system of coordinates. One can then transform back to the original coordinates and exploit the anomalous transformation \eqref{anomaloustransftpm} to obtain their final form in the new coordinates.

In the case under consideration, however, we saw that the GS does not correspond to Minkowski spacetime (which is only reached asymptotically), but it is given by the extremal configuration \eqref{GSextremal}. This, of course, prevents from defining a global coordinate transformation which brings $e^{2\rho} \to \text{constant}$. Despite this difficulty, we can still use \cref{TmunuGS} as a boundary condition, similarly to  the CGHS model. 

Once the boundary conditions on the GS have been imposed, the solution before the shock wave ($x^+ < x^+_0$) is \cref{GSextremal}, the vacuum one, while after the shock wave ($x^+ > x_0^+$) it is given by an evaporating black-hole solution.

\subsection{Adding counterterms and fixing the boundary conditions}
\label{subsec:CountertermsandBCs}

In order to preserve the physically motivated boundary condition \eqref{TmunuGS}, we can follow Refs. \cite{Cadoni:1995dd, Cadoni:1995mi} and modify the usual Polyakov action \eqref{Polyakovactionsimple} by adding the most general local covariant counterterms with no second order derivatives
\begin{equation}\label{Polyakovactioncounterterms}
\mathcal{S}_\text{Pol} =-\frac{N}{96}\int \dd^2 x \, \sqrt{-g} \, \left[ R \Box^{-1} R - 4 \A(\phi) R + 4 \B(\phi) \left(\nabla \phi \right)^2 \right]\, ,
\end{equation}
where $\A$ and $\B$ are functions of the scalar field. The presence of these new terms, of course, does not alter the classical limit $N \to 0$. Also notice that the addition of new counterterms was already employed in, e.g., the CGHS model, in order to  make the theory exactly solvable \cite{Bilal:1992kv,Russo:1992ax}.

In Ref.~\cite{Cadoni:1995dd}, the addition of the counterterms was necessary to prevent $\langle T_{\mu\nu}\rangle_\text{GS}$ from diverging for $\phi \to \infty$. In the present case, it can be shown that divergences are absent due to the peculiar properties of the potential outlined in \cref{subsec:Nonsingularconditions} (see \cref{sec:nodivergencesTmunuGS}). Nevertheless, adding counterterms is needed to implement the boundary condition \eqref{TmunuGS} in a consistent way. With the new terms, the components of the stress-energy tensor \eqref{confanomTpm}, \eqref{confanomTmm} and \eqref{confanomTpp} become
\begin{subequations}
\begin{align}
&\langle T_{+-}\rangle = -\frac{N}{12}\left(\partial_+ \partial_-\rho + \partial_+ \partial_- \A \right)\, ; \label{TmpCounterPol}\\
&\langle T_{\pm\pm}\rangle =-\frac{N}{12} \left[\partial_\pm \rho \partial_\pm \rho - \partial^2_\pm \rho + 2 \partial_\pm \rho \partial_\pm \A - \partial_\pm^2 \A - \B \partial_\pm \phi \partial_\pm \phi + t_\pm(x^\pm) \right]\, . \label{TmmppCounterPol}
\end{align}
\end{subequations}
We now impose the boundary condition \eqref{TmunuGS}. Requiring $t_\pm(x^\pm) = 0$ on the GS completely fixes the two functions $\A$ and $\B$ (see also Ref.~\cite{Cadoni:1995dd})
\begin{subequations}
\begin{align}
\A(\phi) &= -\frac{1}{2}\ln \left(\frac{2\Mext}{\lambda} + \JGS\right)\, =-\rho_\text{GS}\, , \label{Acounter}\\
\B(\phi) &= -\left(\partial_\phi \rho_\text{GS} \right)^2 = -\frac{1}{4 \left(\frac{2\Mext}{\lambda} + \JGS \right)^2}\left(\frac{d\JGS}{d\phi} \right)^2\, , \label{Bcounter}
\end{align}
\end{subequations}
where the subscript GS indicates that $\J$ is computed  at extremality.

\Cref{TmpCounterPol,TmmppCounterPol} now read
\begin{equation}\begin{split}
\langle T_{+-}\rangle &=  -\frac{N}{12}\left(\partial_+ \partial_-\rho -\frac{\partial_+ \partial_- \JGS}{2\left(\frac{2\Mext}{\lambda} + \JGS\right)}  + \frac{\partial_-\JGS \partial_+\JGS}{2\left(\frac{2\Mext}{\lambda} + \JGS\right)^2}  \right)\, ;
\end{split}
\label{TmpCounterPolExpl}
\end{equation}

\begin{equation}
\begin{split}
\langle T_{\pm \pm}\rangle =-\frac{N}{12}\biggl[&\partial_\pm \rho \partial_\pm \rho - \partial_\pm^2 \rho - \frac{\partial_\pm \rho \partial_\pm \JGS}{\frac{2\Mext}{\lambda} + \JGS} + \frac{\partial_\pm^2 \JGS}{2\left( \frac{2\Mext}{\lambda} + \JGS \right)} +\\
&-\frac{\partial_\pm \JGS \partial_\pm \JGS}{2\left( \frac{2\Mext}{\lambda} + \JGS \right)^2} +\frac{\partial_\pm \phiGS \partial_\pm \phiGS}{4\left( \frac{2\Mext}{\lambda} + \JGS \right)^2} \left(\frac{d\JGS}{d\phi} \right)^2 \biggr]\, .
\end{split}
\label{TmmppCounterPolExpl}
\end{equation}

Since the scalar field is a function of $x^+$ and $x^-$, while we are treating $\J$ as a function of $\phi$, it is convenient to rewrite all derivatives of $\J$ with respect to the coordinates as derivatives with respect to $\phi$ (to lighten the notation, we indicate derivation with respect to $\phi$ with a subscript $_{,\phi}$). With the new components of the stress-energy tensor \eqref{TmpCounterPolExpl} and also using \cref{GSextremal}, the field equations \eqref{conformaleommatterwithanomaly} become
\begin{subequations}\label[plural]{EqEinsteinAnomaly}
\begin{align}
8e^{-2\rho} \partial_+ \partial_-\rho &= -\V_{,\phi}\, ; \label{EqScalarfieldConfMAnomC3}\\
\partial_+^2 \phi - 2\partial_+ \rho \partial_+ \phi &= \M \, \delta\left(x^+-x^+_0\right) +\frac{N}{12} \left[\partial_+ \rho \partial_+ \rho - \partial_+^2 \rho - \frac{\lambda}{2} \partial_+\rho \, \J_{\text{GS},\phi} + \right.\nonumber\\
&\qquad\qquad\qquad\qquad\qquad\quad\left.+\frac{\lambda^2}{8}\left(\frac{2\Mext}{\lambda}+\JGS \right) \J_{\text{GS},\phi\phi} + \frac{\lambda^2}{16}\left(\J_{\text{GS},\phi} \right)^2 \right]\, ; \label{EqEinsteinppMAnomC3}\\
\partial_-^2 \phi - 2\partial_- \rho \partial_- \phi &= \frac{N}{12}\biggl[\partial_- \rho \partial_- \rho - \partial_-^2 \rho +\frac{\lambda}{2} \partial_-\rho \, \J_{\text{GS},\phi} + \frac{\lambda^2}{8}\left(\frac{2\Mext}{\lambda}+\JGS \right) \J_{\text{GS},\phi\phi} + \frac{\lambda^2}{16}\left(\J_{\text{GS},\phi} \right)^2 \biggr]\, ; \label{EqEinsteinmmMAnomC3}\\
\partial_+ \partial_- \phi + \frac{\V}{4} e^{2\rho} &=-\frac{N}{12}\left[\partial_+ \partial_-\rho + \frac{\lambda^2}{8}\left(\frac{2\Mext}{\lambda}+\JGS \right) \J_{\text{GS},\phi\phi}\right]\, . \label{EqEinsteinmpMAnomC3}
\end{align}
\end{subequations}
\subsection{Hawking flux and apparent horizon trajectory}
\label{subsec:Hawkingfluxmain}

We now derive the asymptotic form of the Hawking flux in our model. This can be done by studying the behavior of $\langle T_{\mu\nu}\rangle$ at future null infinity $x^+ \to \infty$. In this region, we are considering $\phi\to \infty$. We are, therefore, in the decoupling regime, where the gravitational coupling is weak, so that the effects of backreaction can be approximately neglected. This means that the solution in the region of interest ($x^+ > x^+_0)$ corresponds to the classical one \eqref{exactsolutionshockwave}. In this limit, $\J \to 0$, $\J_{,\phi} \to 0$ and $\J_{,\phi\phi}\to 0$. We have thus
\begin{subequations}
\begin{align}
&\langle T_{+-}\rangle \to 0\, ;\\
&\langle T_{++}\rangle \to 0\, ;\\
&\langle T_{--}\rangle \to \frac{N}{24}\left\{F, x^- \right\}\, , \qquad \left\{F, x^- \right\} = \frac{F'''}{F'}-\frac{3}{2}\left(\frac{F''}{F'} \right)^2\, , \label{SchwarzianDerivativeF}
\end{align}
\end{subequations}
where now $'$ indicates differentiation with respect to $x^{-}$.
This result agrees with that of Ref.~\cite{Cadoni:1995dd}, as it is naturally expected. As it is noted there, this expression diverges once the (outer) event horizon is reached, due to the choice of coordinates adopted. One way to solve this problem is to redefine the $x^-$ coordinate as $\hat x^- \equiv F(x^-)$ and exploit the anomalous transformation \eqref{anomaloustransformation}. This leads to a well-behaved expression at the horizon, which reads
\begin{equation}
\langle \hat T_{--} \rangle = \frac{N}{24}\frac{\left\{F, x^- \right\}}{F'^2}\, . 
\end{equation}
Using the form of $F$ given by \cref{exactsolutionshockwave}, we obtain, approaching the horizon 
\begin{equation}\label{Hflux}
\langle \hat T_{--} \rangle = \frac{N}{192} \left[\V(\phi_\text{H})\right]^2\, \propto \Th^2\, .
\end{equation}
The proportionality relation is the same as the SB law \eqref{2DSBLawGeneral}, which confirms the Planckian nature of the emitted spectrum. Here, however, the flux is modified with respect to standard singular models due to the specific form of the potential \eqref{potentialHayward}.

Since the outgoing flux of Hawking radiation is positive, we expect the (outer) apparent event horizon to recede. To see this, we closely follow the approach adopted in Ref.~\cite{Russo:1992ht}. The apparent horizon trajectory $\hat x^- = \hat x^-(x^+)$ can be derived using the definition of apparent horizon, which satisfies $\partial_+ \phi = 0$. This implies
\begin{equation}\label{apphordef}
0 = \frac{\dd}{\dd x^+}\partial_+\phi \biggr|_{x^- = \hat x^-} = \partial_+^2 \phi + \left(\frac{\dd\hat x^-}{\dd x^+} \right) \partial_+ \partial_- \phi \, ,
\end{equation}
from which follows
\begin{equation}\label{dpdmphiapphor}
\partial_+\partial_-\phi = -\partial_+^2 \phi \,  \left(\frac{\dd\hat x^-}{\dd x^+} \right)^{-1}\,. 
\end{equation}
Combining \cref{EqEinsteinmpMAnomC3,EqEinsteinppMAnomC3,EqScalarfieldConfMAnomC3} into the above yields
\begin{equation}\label{dxminusdxplusfinal}
\frac{\dd\hat x^-}{\dd x^+} = \frac{N}{12}\frac{\partial_+\rho \partial_+\rho -\partial_+^2 \rho -\frac{\partial_+\rho}{2\lambda} \, \V_{\text{GS}} + \frac{e^{2\rho_\text{ext}}}{8} \, \V_{\text{GS},\phi} + \frac{\V_{\text{GS}}^2}{16 \lambda^2}}{\frac{\V}{4}e^{2\rho} + \frac{N}{96}\left(e^{2\rho_\text{ext}}\V_{\text{GS},\phi}-e^{2\rho} \V_{,\phi} \right)}.
\end{equation}
The qualitative behavior of the trajectory of the apparent horizon is determined by the sign of the right hand side of the expression above. In order to assess the latter, we would need the full solution of \cref{EqScalarfieldConfMAnomC3,EqEinsteinmpMAnomC3}, which however can only be solved numerically. This will be done in the next section. Here, as a first test, we exploit the fact that the full solution of the field equations approaches the classical one in the asymptotic region $\phi\to \infty$, where the coupling with matter fields and backreaction effects become negligible. In this region, the solution is then given by \cref{exactsolutionshockwave}. Using this solution into \cref{dxminusdxplusfinal}, it can be shown that $\dd\hat x^-/\dd x^+$ is indeed positive. This implies a receding outer apparent horizon, as expected. Moreover, it confirms the qualitative picture we expect from evaporation, outlined in \cref{subsec:semiclassicalevaporation} and at the end of \cref{subsec:shockwavesolution}: the outer horizon recedes and approaches the horizon of the extremal GS.  

Notice that, due to the limitations caused by the adopted system of coordinates, we are able to describe the behavior of the outer horizon only. 

\subsection{Numerical results}
We now numerically solve the equations of motion given in \Cref{EqEinsteinAnomaly}. This will allow us to capture the full dynamics of the evaporation process, keeping into account also backreaction effects. A numerical study of evaporating $2\text{D}$ models was performed in the past for the CGHS \cite{Piran:1993tq,Lowe:1992ed,Ashtekar:2010qz} and other (singular or regular) 2D models \cite{Lowe:1993zw,Diba:2002hb}. 

To numerically integrate the equations of motion, we construct a spacetime lattice by means of a grid of null lines and we impose two different sets of boundary conditions, one along $x^+$ and one along $x^-$:

\begin{itemize}
\item At the shock wave, i.e., at $x^+ = x^+_0$, we require the solution to coincide with the GS, given by \cref{GSextremal}; 
\item Above the shock wave, along $\mathcal{I}^-$, i.e., at $x^- \to -\infty$, where backreaction effects are expected to be negligible, we require the solution to match \cref{exactsolutionshockwave}, the classical one. Of course, we cannot numerically set a condition at infinity, so we choose a reasonably large negative value. Here we set $x^-_\infty = -220$. This value is found to be the minimal one for which the numerical solution coincides, in the classical limit $N \to 0$, with \cref{exactsolutionshockwave}, for every value of $x^+$ and $x^-$, within a reasonably small numerical error. For values of $x^-_\infty$ greater than $-220$, the numerical solution deviates from the expected analytical one, while for smaller values the results are the same as the one obtained with $x^-_\infty = -220$, but with a much higher integration time. 
\end{itemize}

Both the boundary conditions at $x^+ = x^+_0$ and at  $x^- \to -\infty$   are given  in implicit form. We therefore first need to integrate and invert the corresponding  expressions (the details of this computation are reported in \cref{sec:boundaryconditionsnumerical}).

To numerically integrate the field equations, we also need to select an appropriate integration interval along $x^+$, from the shock wave at $x_0^+$ up to a maximum value $x^+_\text{max}$. We chose the interval $x^+ \in [x^+_0, 5]$ (where we set $x^+_0 = 1$). We expect the general results to hold also for larger values of $x^+_\text{max}$. We chose it equal to $5$ to have a reasonable integration time interval. For larger values of  $x^+_\text{max}$, the time required to complete a computation increases considerably, given the high computational cost of the algorithm.

The interval on the $x^+$-axis is then discretized into a number $n_\text{steps}$ of small intervals, with length $\Delta x = (x^+_\text{max}-x^+_0)/n_\text{steps}$. The number of steps was set equal to $n_\text{steps} = 1000$. We checked that, for larger values  of $n_\text{steps}$, the results of the integration remain qualitatively the same, at the price of having, again, a much longer computational time.

Each point of the discretized $x^+$-axis is labeled by an index $i$. We choose to discretize the derivatives in the $x^+$ direction accordingly, 
\begin{align}\label{NumDer}
\partial_+ \phi &= \, \frac{\phi(x^+_{i+1}, x^-)-\phi(x^+_{i}, x^-)}{\Delta x} + \mathcal{O}(\Delta x)\, ,\\
\partial_+ \rho &= \, \frac{\rho(x^+_{i+1}, x^-)-\rho(x^+_{i}, x^-)}{\Delta x} + \mathcal{O}(\Delta x)\, .
\label{NumericalDerivatives}
\end{align}
Notice that, with this choice, our algorithm converges to the solution only at first order in $\Delta x$. However, since we are interested in the qualitative behavior of the solutions, \cref{NumDer,NumericalDerivatives}  represent a good approximation  for the derivatives   of $\phi$ and $\rho$.

\begin{figure}[h!]
\centering
\includegraphics[width= 8 cm, height = 8 cm,keepaspectratio]{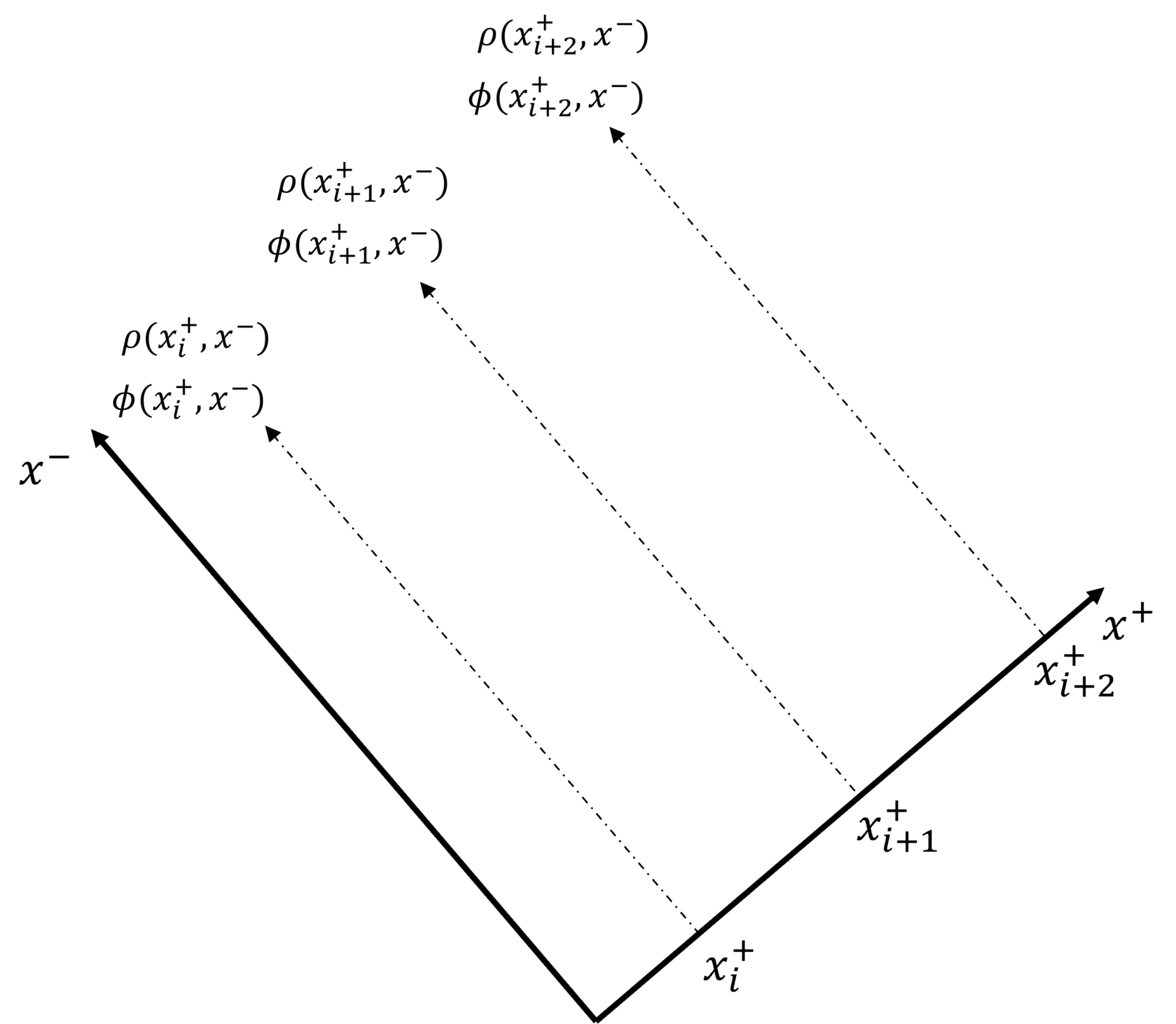}
\caption{Schematic representation of the numerical algorithm adopted to numerically integrate \cref{EqScalarfieldConfMAnomC3,EqEinsteinmpMAnomC3}. We discretize the $x^+$-axis, and for each interval on the latter, we numerically integrate the field equations along $x^-$ using a Runge-Kutta algorithm.}
	\label{schemanumerico}
\end{figure}

Along $x^-$ at fixed $x^+$, the field equations reduce to ordinary differential equations. Therefore, for each step in the $x^+$ direction, we numerically integrate the equations along $x^-$ by means of a 4th order Runge-Kutta algorithm. The outcome, thus, is a list of $x^-$-profiles of $\phi$ and $\rho$ for each point of the discretized interval on $x^+$ (see \cref{schemanumerico}). 

For all the cases considered here, the values of the parameters are set equal to $\lambda = 1$ and $\ell=1$. The mass of the evaporating solution is fixed equal to $\M = 0.1$ (in these units). 

\begin{figure}[h!]
\subfigure[$x^+ = x_0^+$]{\includegraphics[width=8cm]{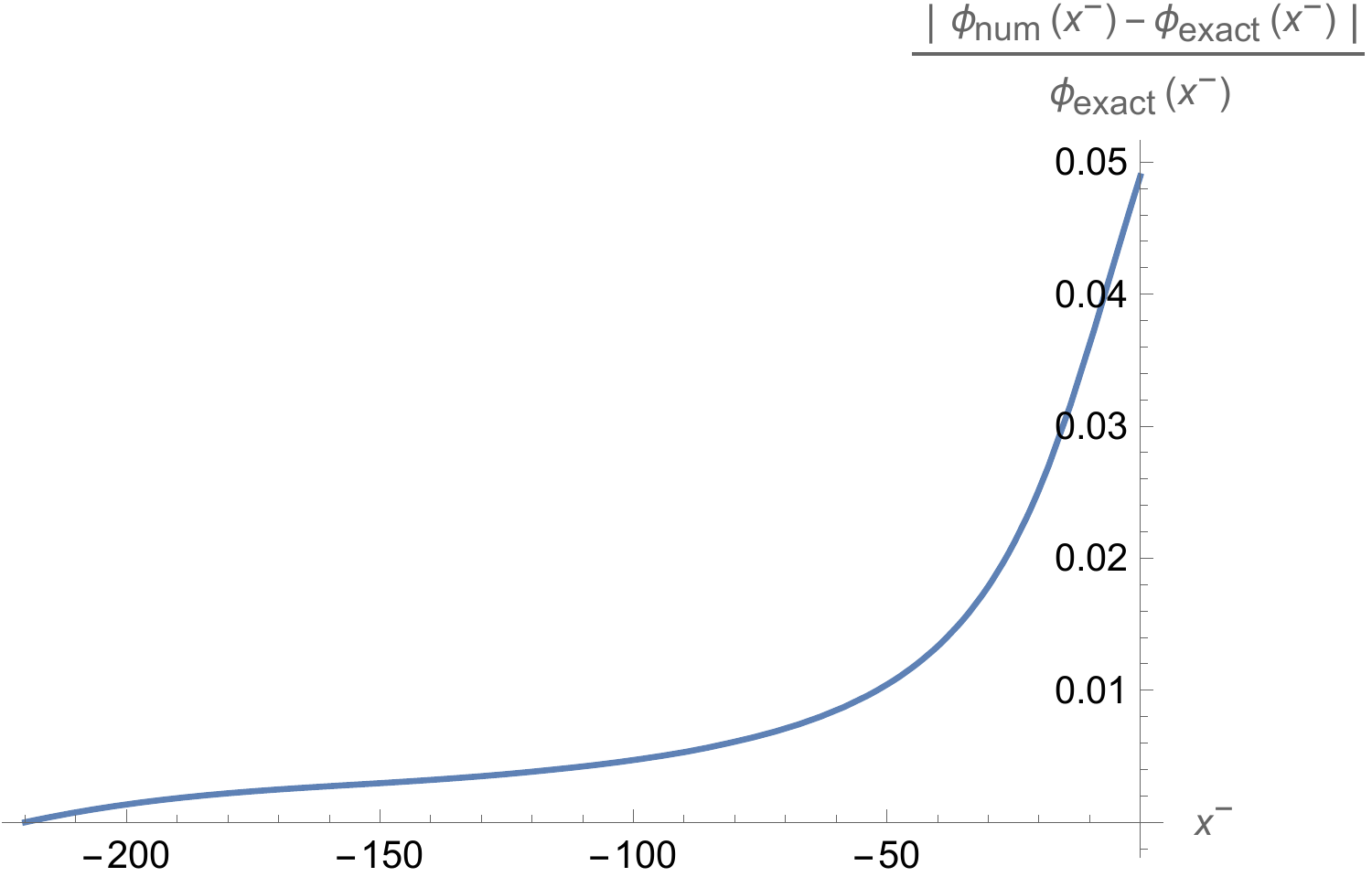}}
\hspace{0.7 cm}
\subfigure[$x^+ = 5$]{\includegraphics[width=8 cm]{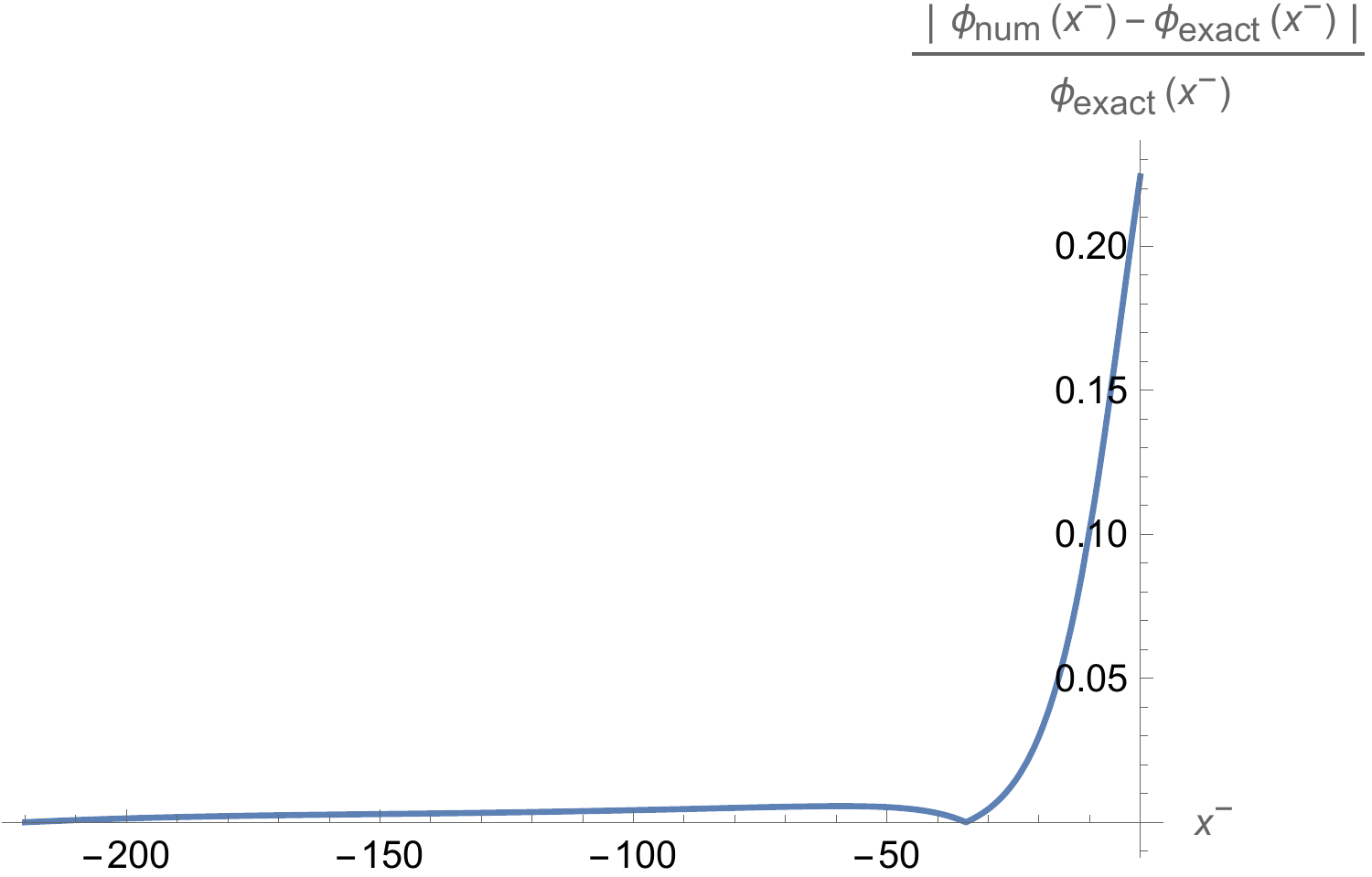}\label{fig:temp_d}}
\caption{Relative difference between the numerical (obtained by setting $N=0$) and the exact \eqref{exactsolutionshockwave} solutions at $x^+ = x^+_0$ and $x^+ = 5$. As it can be seen, the difference increases as we move towards small negative values of $x^-$. The numerical integration has been carried out on an $x^+$ interval of $[x_0^+, x^+_\text{max}]=[1,5]$, while on the $x^-$ direction, in the interval $[-220,0]$. The parameters of the integration are the following: $\lambda =\ell= 1$, $\M = 0.1 $ (in these units), $n_\text{steps} = 1000$, so that $\Delta x = 4 \cdot 10^{-3}$. We checked that, for higher values of $n_\text{steps}$, the relative differences decrease, without however altering the qualitative final results and with a much higher computational time.}
\label{CompNumAnalytical}
\end{figure}

As a first test, we have verified the accuracy of the integration algorithm in the absence of backreaction, i.e., for $N=0$, by comparing the numerical solution with  the analytical classical one \eqref{exactsolutionshockwave}. Overall, we find that the relative difference between the numerical and the analytical solutions is smaller than $1 \, \%$ as long as we consider large negative values of $x^-$, while it increases for $x^- \to 0$, staying however $\lesssim 20 \, \%$ (see \cref{CompNumAnalytical}). We checked that increasing $n_\text{steps}$ leads to an improvement in the accuracy of the integration algorithm (the relative differences decrease), without, however, altering the qualitative final results and at the price of having a much longer computational time. Although a relative discrepancy of $20 \, \%$ may seem quite important, one should consider that, in the presence of backreaction (see below), the extremal solution is reached at values of $x^-$ for which the relative discrepancy  stays always below $5 \, \%$. As in the following we are not interested in the exact details of the evaporation process, but rather in its qualitative evolution and outcome, we will adopt $n_\text{steps} = 1000$ anyway, favouring time efficiency over high precision.

After this preliminary test, we analyse three different cases of increasing $N$: $N = 0$, $N = 24$ and $N = 2400$, to study the backreaction effects in different regimes.

\begin{figure}[h!]
\subfigure[]{\includegraphics[width=8cm]{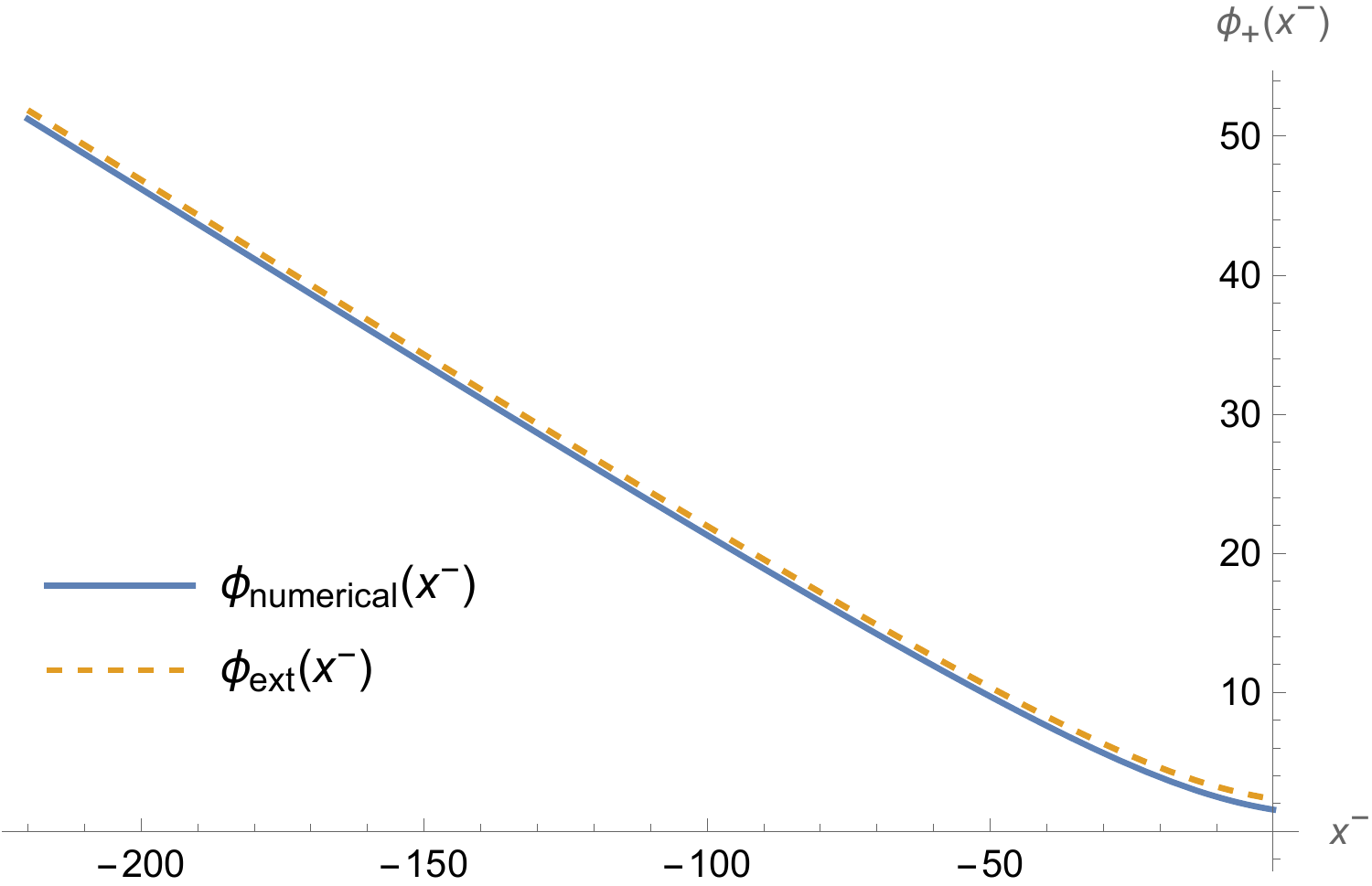}}
\hspace{0.7 cm}
\subfigure[]{\includegraphics[width=8cm]{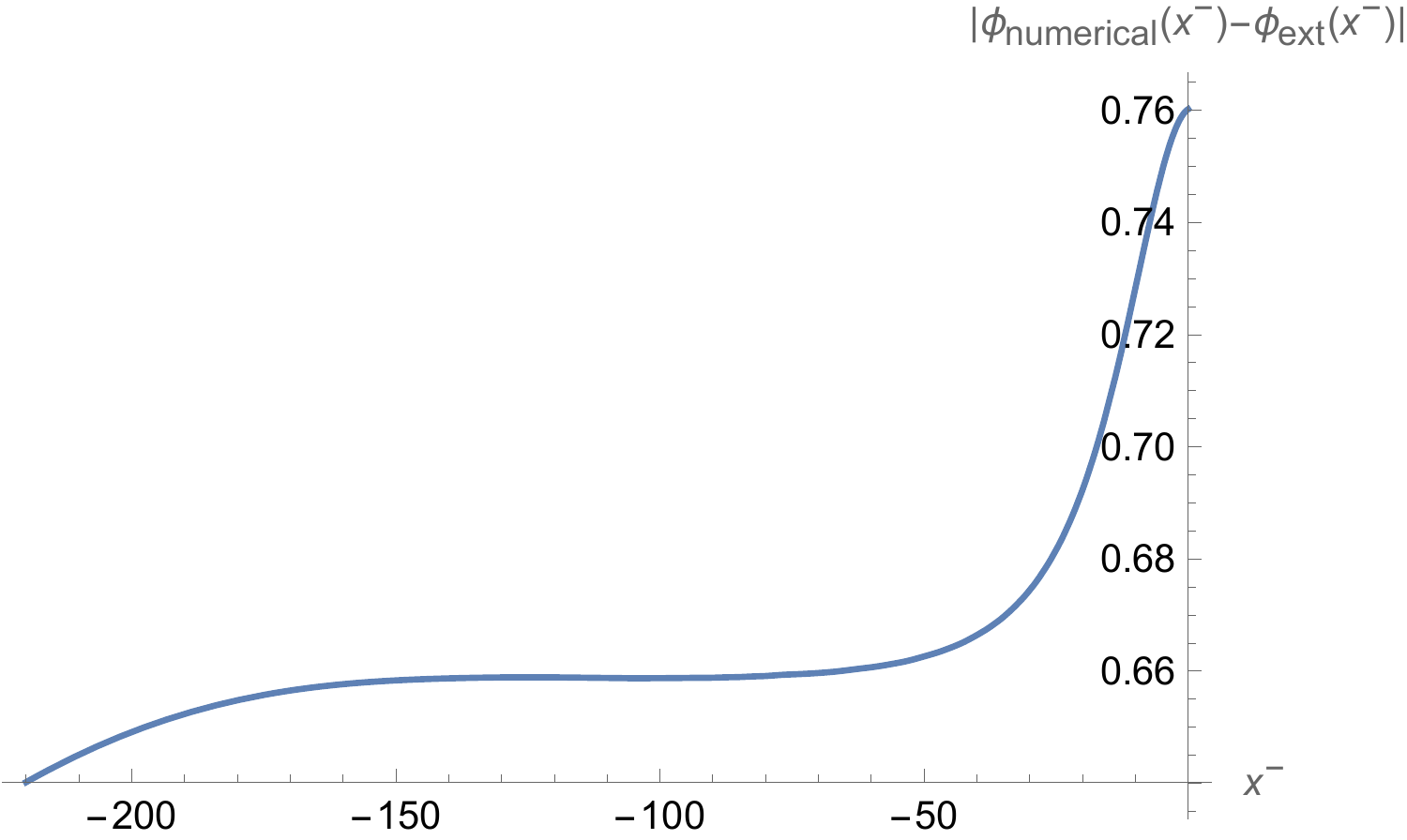}}
\vfill
\subfigure[]{\includegraphics[width=8cm]{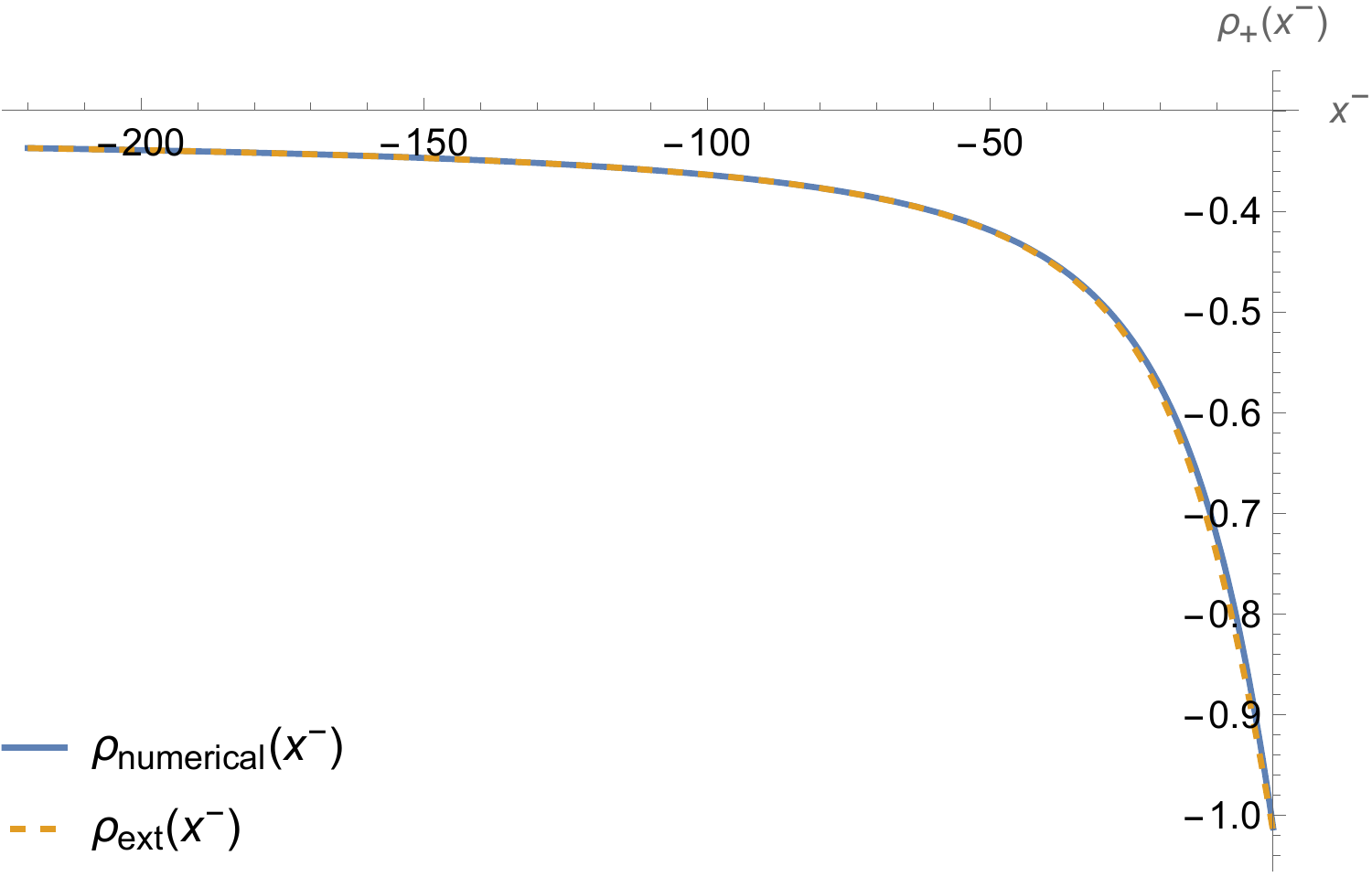}}
\hspace{0.7 cm}
\subfigure[]{\includegraphics[width=8 cm]{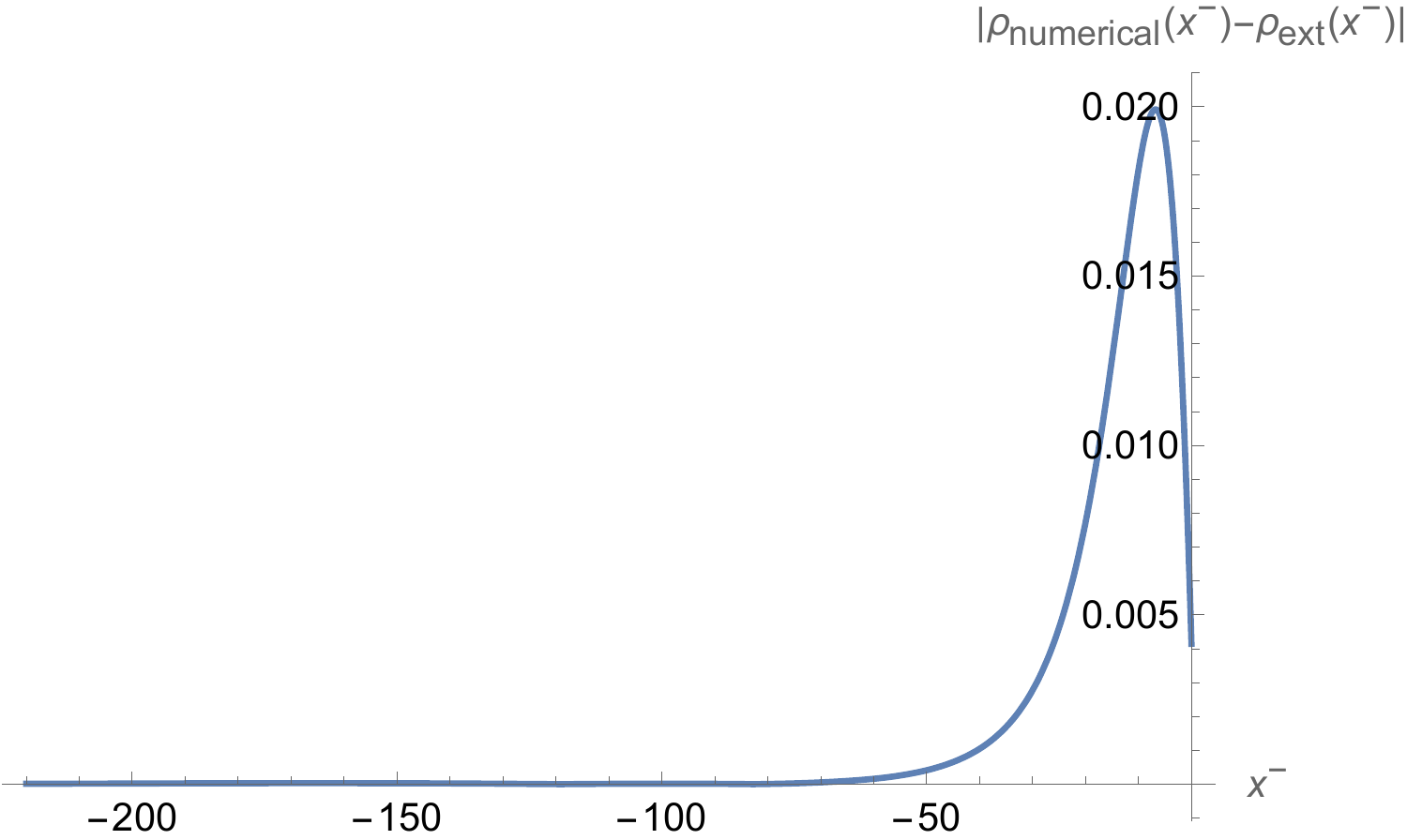}}
\caption{\textbf{Upper figures:} Comparison between the numerical (solid blue line) and analytical extremal (dashed orange line) dilaton solutions (left figure), and difference between the two (right figure), as functions of $x^-$.  \textbf{Lower figures:} Comparison between the numerical (solid blue line) and analytical extremal (dashed orange line) metric solution (left figure), and difference between the two (right figure), as functions of $x^-$. 
All figures are evaluated at $x^+ = 5$ and with $N=0$, in units where $\lambda = \ell = 1$. }
\label{fig:phirhoN0}
\end{figure}

\begin{figure}[h!]
\subfigure[]{\includegraphics[width=8cm]{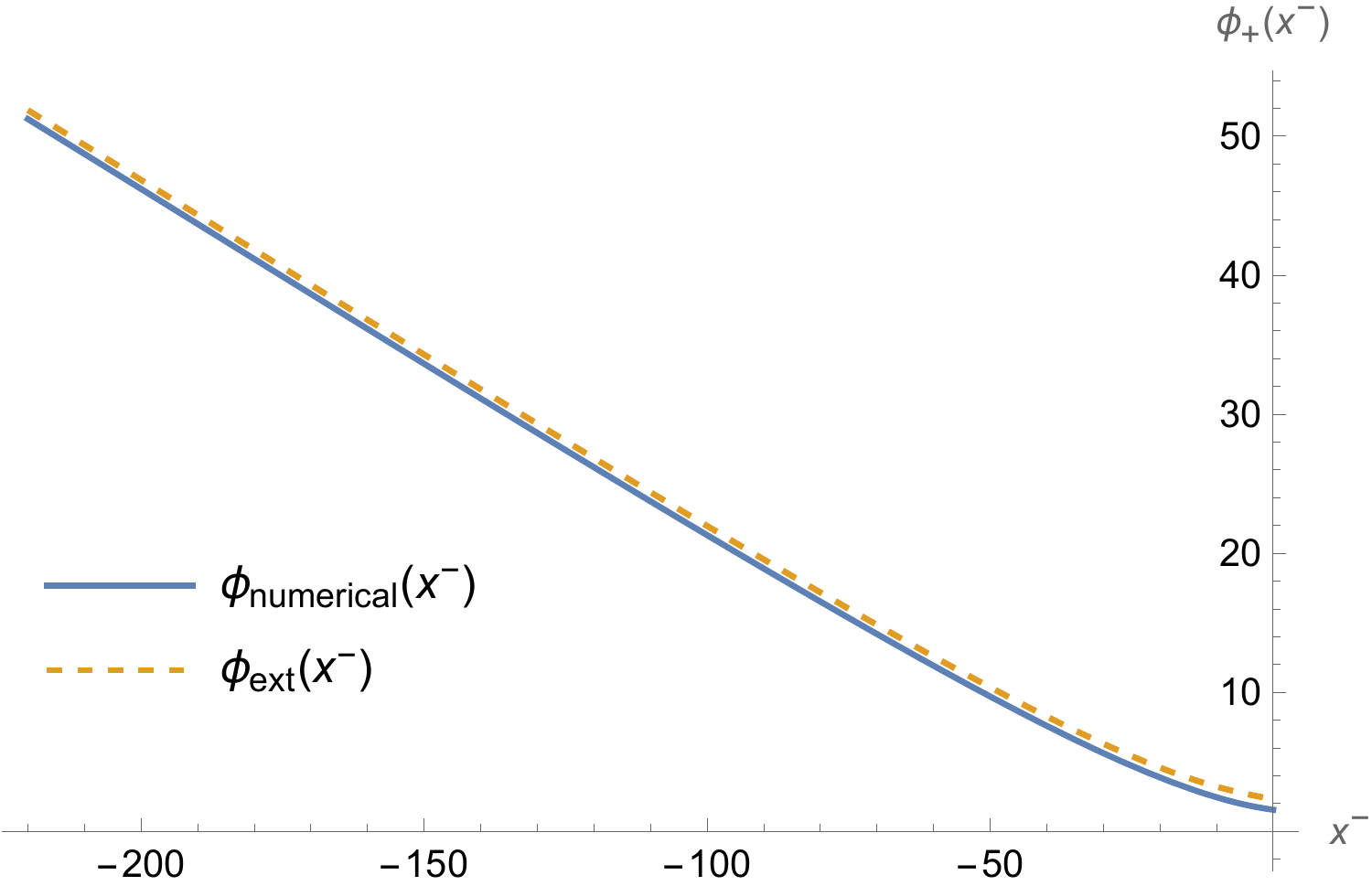}}
\hspace{0.7 cm}
\subfigure[]{\includegraphics[width=8cm]{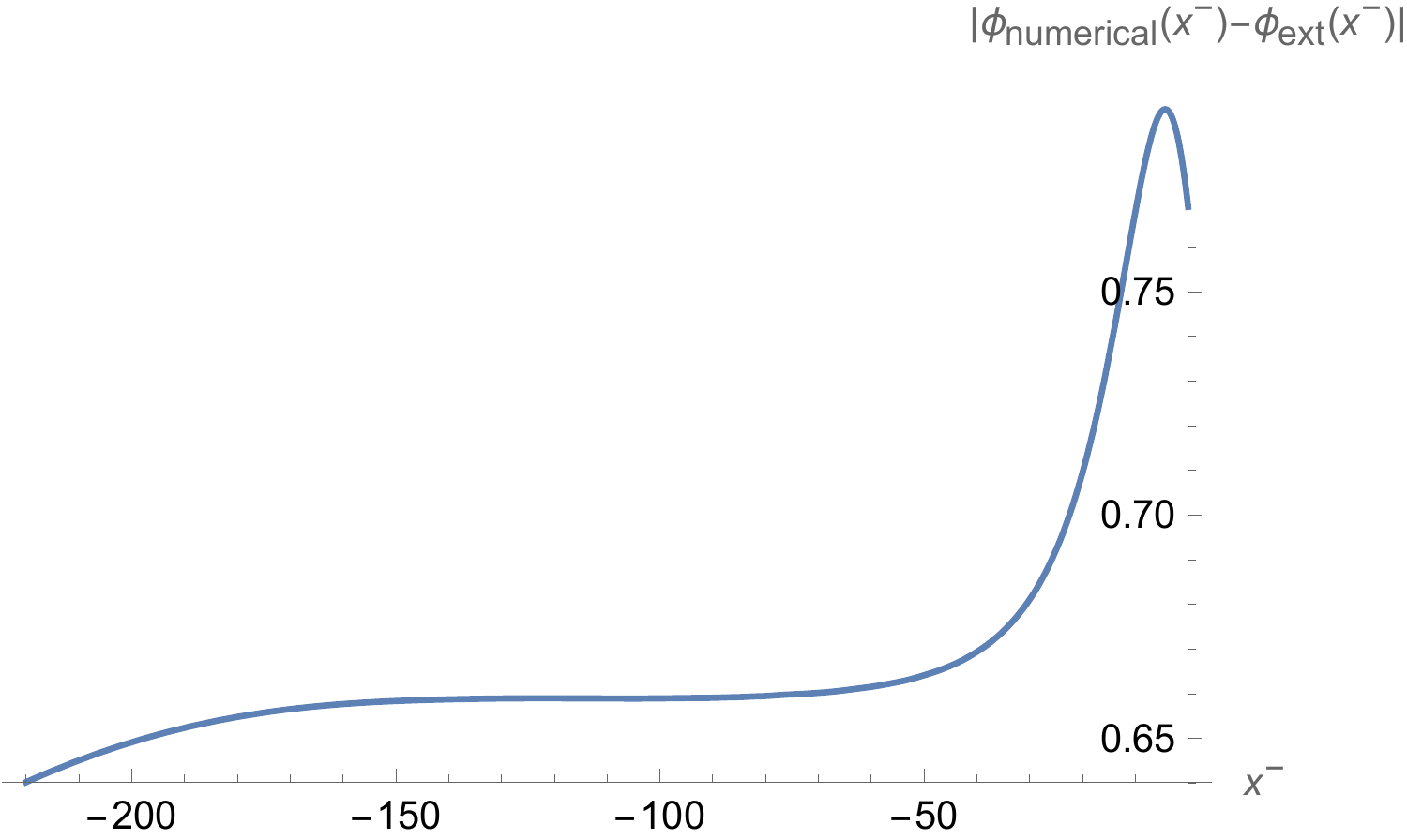}}
\vfill
\subfigure[]{\includegraphics[width=8cm]{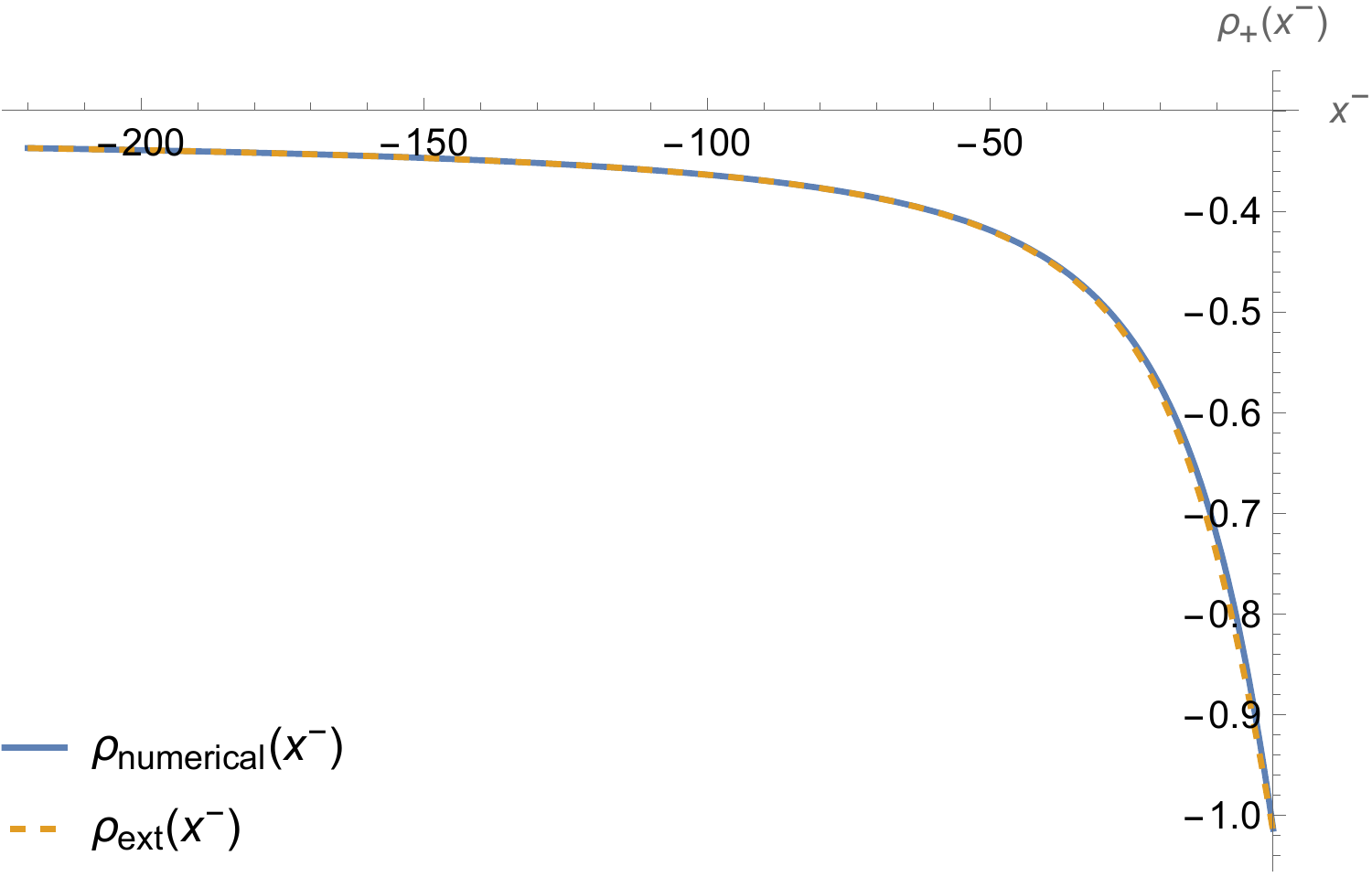}}
\hspace{0.7 cm}
\subfigure[]{\includegraphics[width=8 cm]{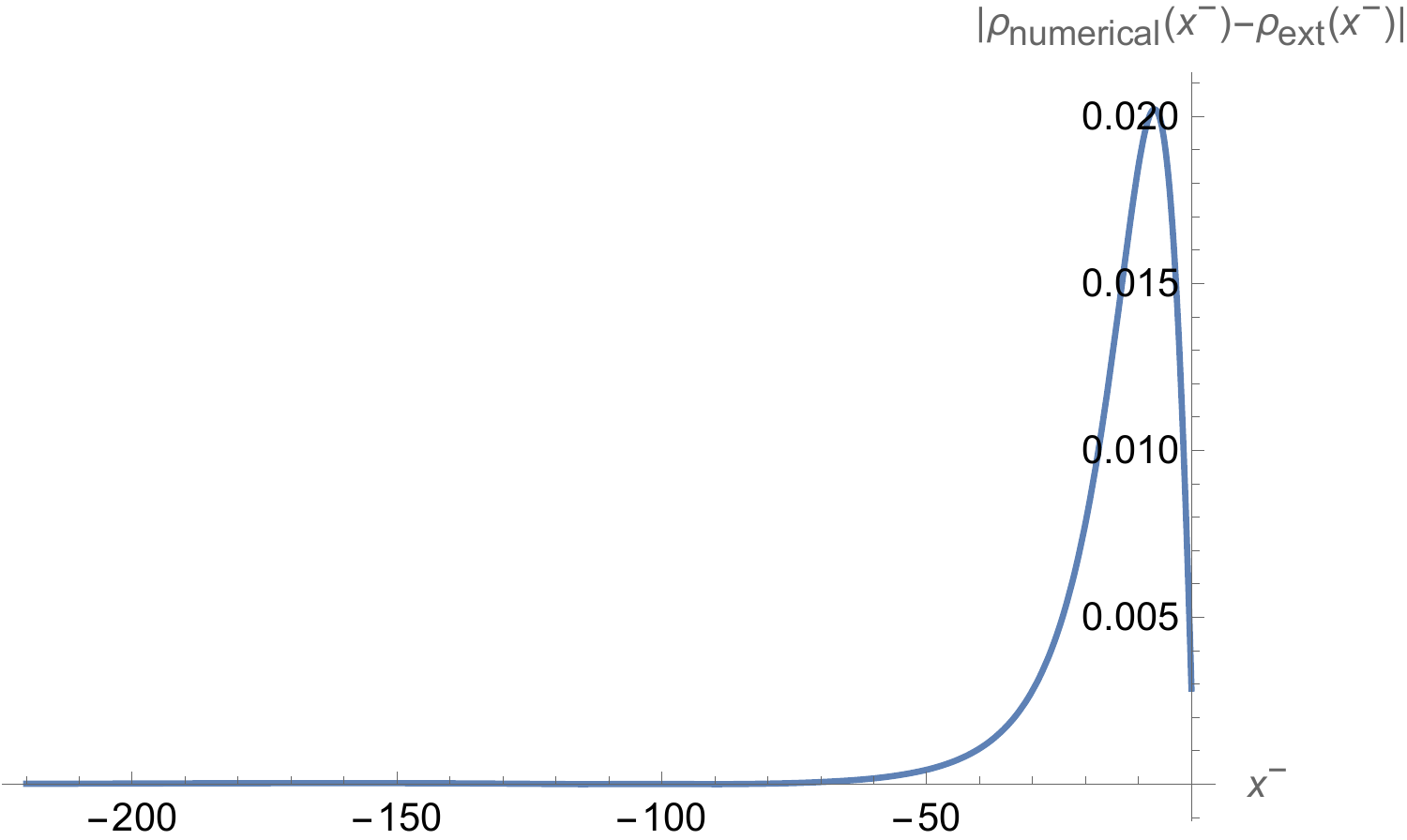}}
\caption{\textbf{Upper figures:} Comparison between the numerical (solid blue line) and analytical extremal (dashed orange line) dilaton solutions (left figure), and difference between the two (right figure), as functions of $x^-$.  \textbf{Lower figures:} Comparison between the numerical (solid blue line) and analytical extremal (dashed orange line) metric solution (left figure), and difference between the two (right figure), as functions of $x^-$.  
All figures are evaluated at $x^+ = 5$ and with $N=24$, in units where $\lambda = \ell = 1$.}
\label{fig:phirhoN24}
\end{figure}

\begin{figure}[h!]
\subfigure[]{\includegraphics[width=8cm]{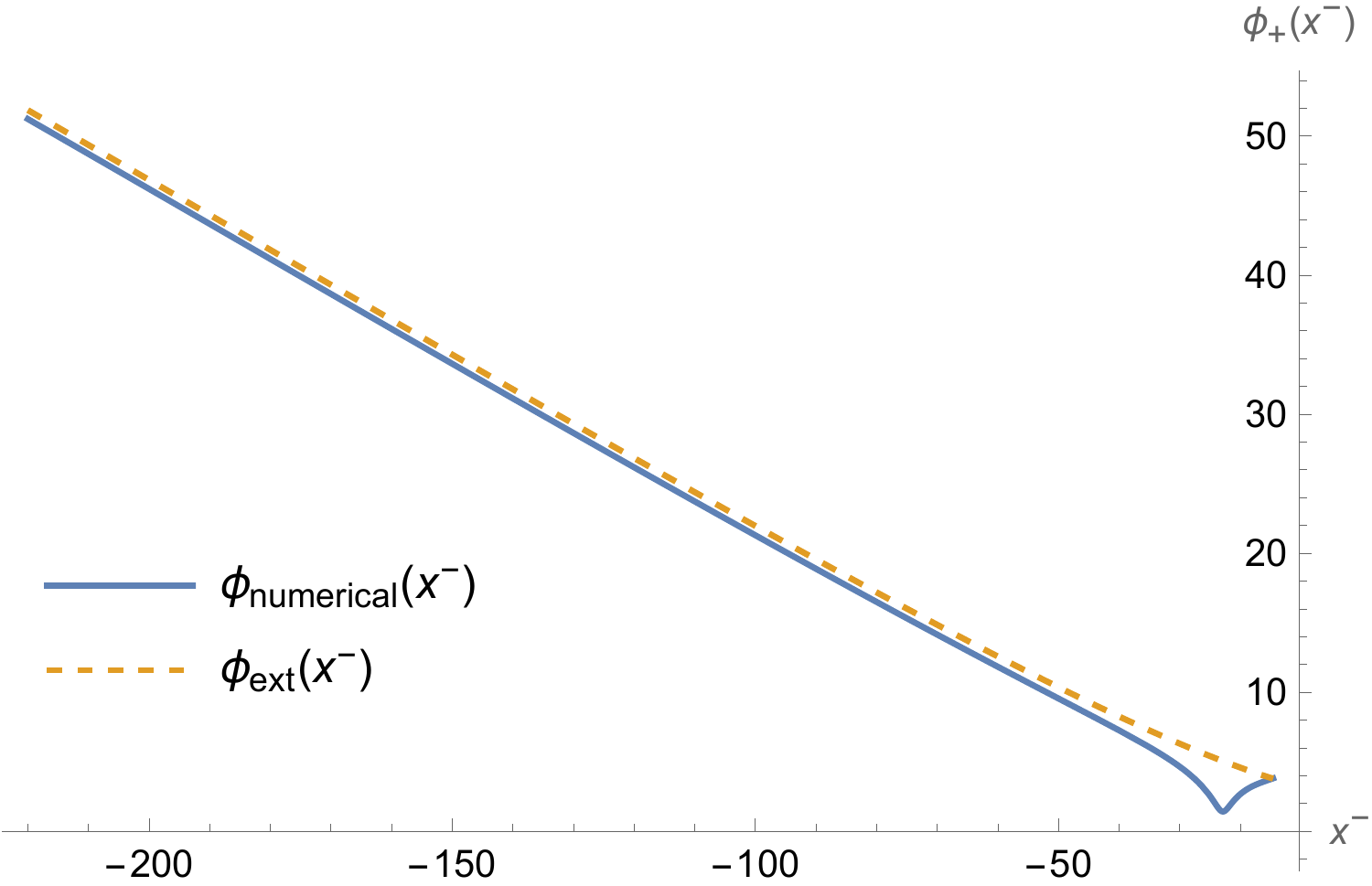}}
\hspace{0.7 cm}
\subfigure[]{\includegraphics[width=8cm]{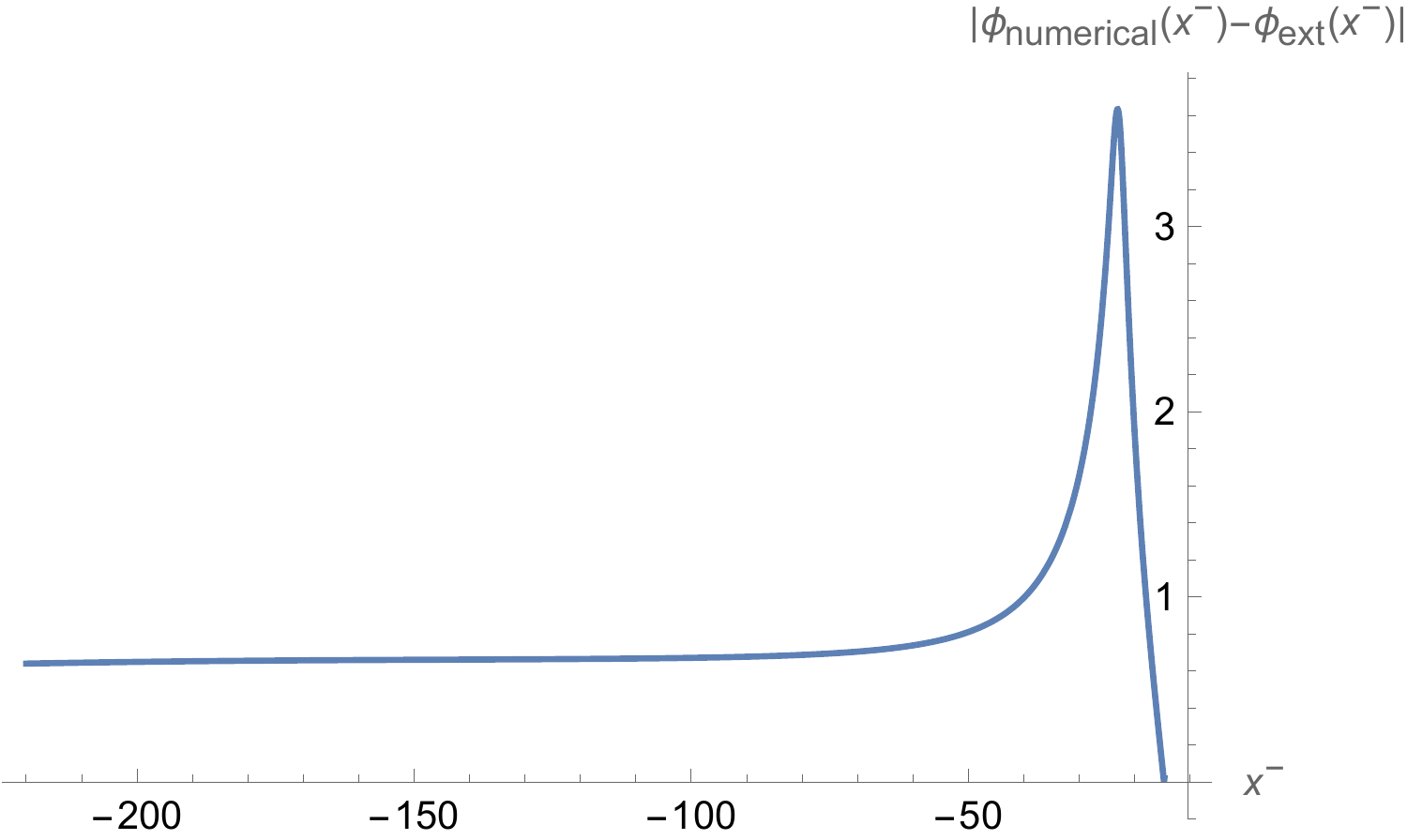}}
\vfill
\subfigure[]{\includegraphics[width=8cm]{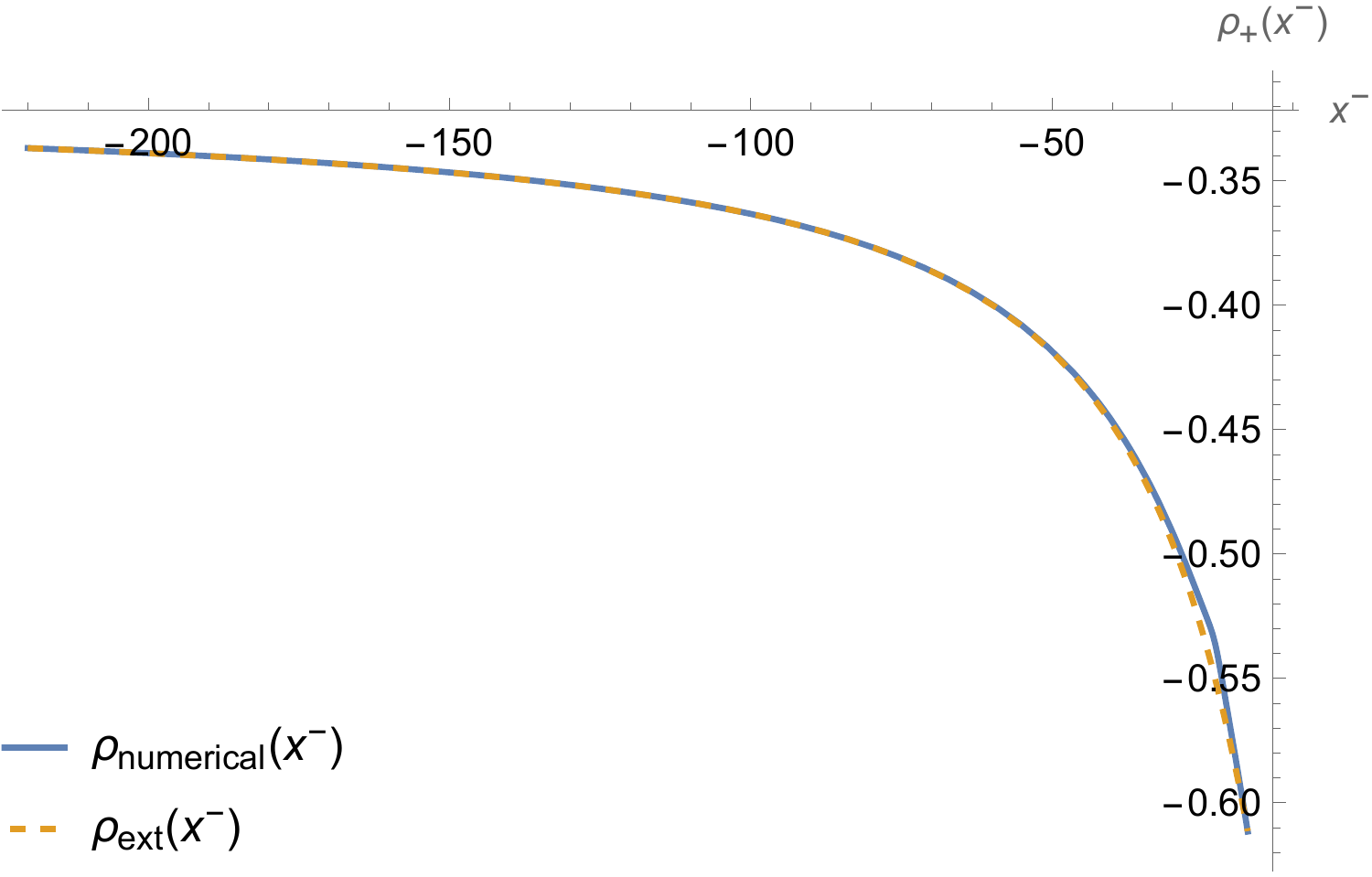}}
\hspace{0.7 cm}
\subfigure[]{\includegraphics[width=8 cm]{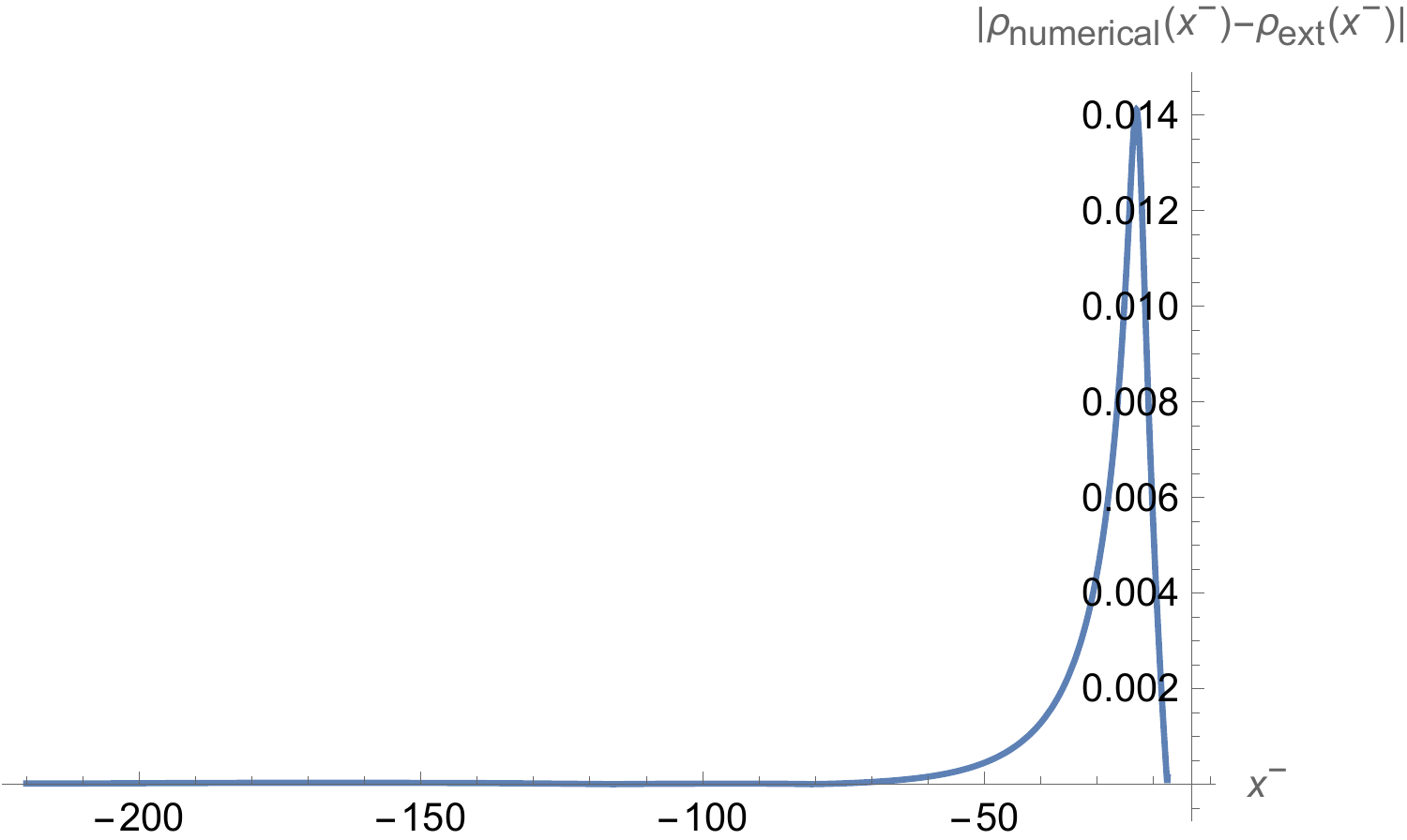}}
\caption{\textbf{Upper figures:} Comparison between the numerical (solid blue line) and analytical extremal (dashed orange line) dilaton solutions (left figure), and difference between the two (right figure), as functions of $x^-$.  \textbf{Lower figures:} Comparison between the numerical (solid blue line) and analytical extremal (dashed orange line) metric solution (left figure), and difference between the two (right figure), as functions of $x^-$. 
All figures are evaluated at $x^+ = 5$ and with $N=2400$. The $x^-$ axis has been cut at the point where the numerical solution matches the extremal one. }
\label{fig:phirhoN2400}
\end{figure}

The $x^-$-profiles of $\phi$ and $\rho$, together with their variations $\Delta\phi=|\phi(x^-)-\phi_\text{ext}(x^-)|$,  $\Delta\rho=|\rho(x^-)-\rho_\text{ext}(x^-)|$ with respect to the extremal configurations, are computed numerically for several values of $x^+$ in the range  $x^+ \in [x^+_0, 5]$. At $x^+ = x_0^+$, the numerical solutions match exactly the extremal ones, as it should be according to the boundary condition imposed at the shock wave. For simplicity, in \cref{fig:phirhoN0,fig:phirhoN24,fig:phirhoN2400}, we only show the plots for $x^+=5$ and $N=0$, $24$, $2400$. The plots for the other values of $x^+$ in the range considered here have the same qualitative behavior. Moreover, given the increase in computational errors near $x^- \sim 0$, we performed the integration in the range $x^-\in [-220,0]$ for convenience. We have checked that the results do not differ from those shown, even if we extend the $x^-$ axis to positive values: the convergence to the extremal solution either does not occur in the entire axis or always occurs  in the $x^-<0$ region.

For  $N=0$, i.e., in the absence of backreaction, we see, as expected, that the black-hole solution remains different from the extremal one for every value of the coordinate $x^-$. 

For $N=24$, namely when backreaction effects begin to become relevant, we see that, although $\Delta \phi$ and $\Delta\rho$ remain different from zero for all values of $x^-$, they begin to decrease towards zero after reaching a maximum. 

For $N\gg 24$, i.e., $N \sim 2400$, when backreaction effects become stronger, we see that $\Delta \phi$ and $\Delta\rho$  become always zero at some finite (negative) value of $x^-$. In general, the larger $N$, i.e., the stronger backreaction effects, the faster the evaporating configuration reaches the extremal GS. As remarked above, the convergence to the extremal configuration occurs at values of $x^-$ for which relative numerical errors are less than $5 \, \%$.

It is very important to notice that the convergence of the excited, evaporating solution towards the extremal one is non-monotonic. As one can see clearly from the plots shown (but the same happens also for other values of $N$ not shown here), $\Delta \phi$ and $\Delta\rho$ stay almost flat in the region of large $\phi$ (corresponding to $x^-\ll 0$). Then, they reach a sharp maximum at relatively large values of $x^-$ before falling rapidly towards zero. This behavior cannot be traced back to backreaction effects, since it is present also in the $N=0$ case. The sharp maximum seems to be related to the presence of the maximum in the potential $\V$ at (relatively) small values of the dilaton (see \cref{Potential}), thus to a self-interaction effect of the dilaton. On the other hand, this maximum in $\V$ is also responsible for both the presence of two horizons (instead  of only one) and for the phase transition small/large black holes (see \cref{sec:FirstLawThermo,subsec:Nonsingularconditions}). We will come back to this intriguing point in \cref{sec:conclusions}. 

Summarizing, the numerical integration of \cref{EqEinsteinAnomaly} clearly shows that, differently from what obtained in the rough quasistatic description, the effect of the backreaction is to bring the excited, evaporating solution back to the extremal state after a \emph{finite} time, when $N$ is chosen to be sufficiently large, i.e., at least $N \sim \mathcal{O}(10^2 - 10^3)$.
 
\clearpage 
 
\section{Entanglement Entropy and the Page curve}
\label{sec:EEandPagecurve}

In this section, we compute the entanglement entropy (EE) of Hawking radiation, described here as a collection of $N$ massless scalar fields, in the $2\text{D}$ nonsingular black-hole geometry. By assuming that the evaporation process is quasistatic, we also determine the time variation of the EE and construct the related Page curve.

\begin{figure}[h!]
\centering
\includegraphics[width= 9 cm, height = 9 cm,keepaspectratio]{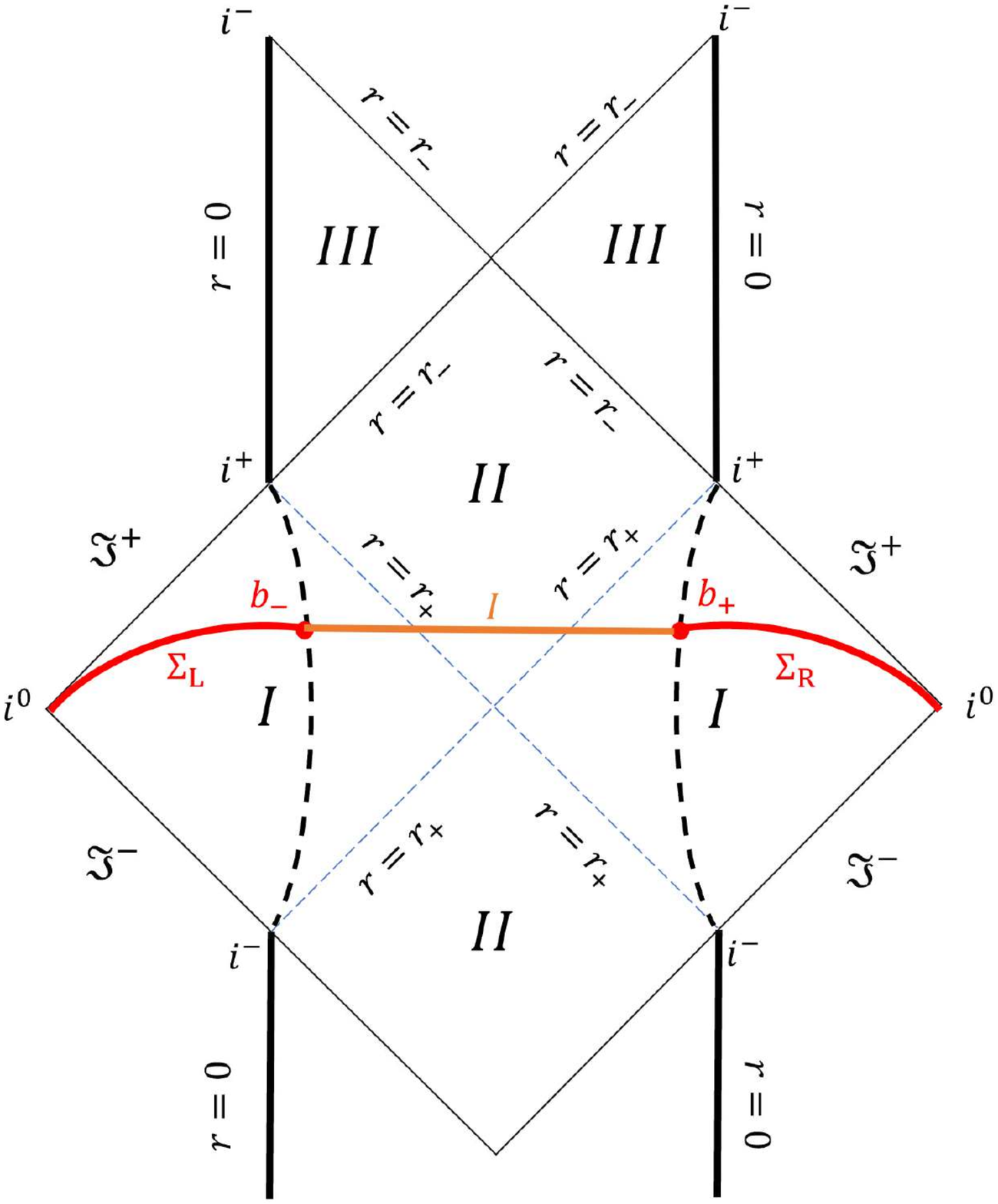}
\caption{Penrose diagram of the maximally extended spacetime of the nonextremal configuration. The two points $b_+$ and $b_-$, belonging to the right and left wedges, respectively, are highlighted, and represents points anchored to two timelike curves (dashed black lines in the two wedges). The union between the three hypersurfaces $\Sigma_\text{L} \cup I \cup \Sigma_\text{R}$ (the red and orange curves) is a spacelike surface and the state defined on it is pure. The radiation is defined on $\Sigma_\text{L}$ and $\Sigma_\text{R}$ and its state is mixed. }
	\label{PenroseEntE}
\end{figure}

The EE of the radiation can be computed by using Kruskal coordinates, covering the region outside the outer event horizon of the black hole,
\begin{equation}\label{Kruskalcoordouter}
\kappa X^{\pm} = \pm e^{\pm \kappa \, x^{\pm}} \, \longleftrightarrow x^{\pm} = \pm \frac{1}{\kappa}\ln \left(\pm \kappa \, X^{\pm} \right)\, ,
\end{equation}
where $\kappa$ is the surface gravity at the outer event horizon. In these coordinates, the conformal factor of the metric \eqref{conformalgauge} can be written as 
\begin{equation}
e^{2\rho} = \frac{f(r)}{-\kappa^2 X^+ X^-}\, .
\end{equation}
The entanglement entropy of $N$ massless scalar fields in two spacetime dimensions on a line can be evaluated by tracing out the degrees of freedom in a spacelike slice $[x, y]$ connecting two points. The resulting expression is \footnote{This expression should be dependent also on ultraviolet cutoffs, which are here considered as addittive constants.} (see, e.g., Refs.~\cite{Fiola:1994ir, Almheiri:2019yqk,Penington:2019kki,Almheiri:2019qdq})
\begin{equation}\label{ententropygeneral}
S_\text{matter} = \frac{N}{6}\ln d^2(x, y)\, ,
\end{equation}
where $d(x, y)$ is the geodesic distance between $x$ and $y$. In principle, \cref{ententropygeneral} is valid for a QFT on a flat spacetime \cite{Calabrese:2009qy}, but it has been generalized to \emph{static} curved spacetime \cite{Fiola:1994ir}, where $d^2(x, y)$ reads 
\begin{equation}\label{distancecurved}
d^2(x, y) = -\left[X^+(x) - X^+(y) \right] \left[X^-(x) - X^-(y) \right] e^{\rho(x)}e^{\rho(y)}\, .
\end{equation} 
To compute the entanglement entropy, we construct a spacelike surface encompassing different regions of the black hole. In \cref{PenroseEntE}, this surface is $\Sigma_\text{L} \cup I \cup \Sigma_\text{R}$, where $\Sigma_\text{L}$ and $\Sigma_\text{R}$ are two hypersurfaces on the \emph{outside} regions of the two copies of the black hole, where an observer collects Hawking radiation. They are the portion of the hypersurface, where the radiation degrees of freedom are defined.  They are anchored to two timelike surfaces (dashed black lines in \cref{PenroseEntE}) at the points $b_+ = (t_b, b)$ (right wedge) and $b_- = (-t_b + \ii \beta/2, b)$ (left wedge). The surface $\mathcal{J}$ defines, instead, the interior region of the black hole. The radiation quantum state over the whole hypersurface $\Sigma_\text{L} \cup I \cup \Sigma_\text{R}$ is pure. When tracing out the interior degrees of freedom in $I$, we obtain the mixed state of the radiation   described by the density matrix $\rho_\text{rad}$, which can therefore be used to compute the entanglement entropy. This is reminiscent of the thermofield double state of the black hole \cite{Maldacena:2001kr,Maldacena:2013xja}: the entanglement entropy takes into account the correlations between the two disjointed copies of the black hole (right and left wedges). Since we have radiation outside the black hole, there will be two copies of this thermal bath (the two regions $\Sigma_\text{L}$ and $\Sigma_\text{R}$). 

In our case, \cref{distancecurved} reads
\begin{equation}\begin{split}
d^2(b_+, b_-) &= -\frac{1}{\kappa^2} \left[e^{\kappa t_b + \kappa b} - e^{-\kappa t_b + \kappa b} e^{i\kappa \beta/2} \right] \left[-e^{-\kappa t_b + \kappa b}+ e^{\kappa t_b + \kappa b} e^{-i\kappa \beta/2} \right] \frac{f(b)}{e^{2\kappa b}}\, .
\end{split}
\end{equation}
This is valid off-shell. On shell $\kappa \to \kappaH$, we have $\frac{\beta \kappaH}{2} = \pi$, and thus
\begin{equation}\begin{split}
d^2(b_+, b_-) = \frac{4 f(b)}{\kappaH^2} \cosh^2(\kappaH t_b)\, .
\end{split}
\end{equation}
Finally, the EE of the matter fields is
\begin{equation}\label{EntanglementEntropyMatter}
S_\text{matter} = \frac{N}{6}\ln \left[\frac{4 f(b)}{\kappaH^2} \cosh^2(\kappaH t_b) \right]\, .
\end{equation}
As stressed above, this is valid as long as we consider the static case. However, $\kappaH$ varies due to the evaporation process. To get a qualitative picture of the behavior of the entropy in time, we can assume that the evaporation process happens in an adiabatic way, so that we can use a quasistatic approximation. The evaporation is thus again modelled in terms of a sequence of static states with decreasing mass. As we have seen explicitly in \cref{subsubsec:approachingextandsemiclassicalapprox}, the quasistatic approximation is reliable as long as the semiclassical one is valid. In a first approximation, therefore, we can use the time variation of the event horizon $\rH = \rH(t)$, computed as a solution of the SB law \eqref{drHdtSB}, and plug it into the expression of the surface gravity
\begin{equation}
\kappaH(\rH)= \frac{f'(\rH)}{2} = \frac{\rH^4-2 \ell^3 \rH}{2 \lambda  \left(\ell^3+\rH^3\right)^2}\, .
\end{equation}
\begin{figure}[h!]
\centering
\includegraphics[width= 8 cm, height = 8 cm,keepaspectratio]{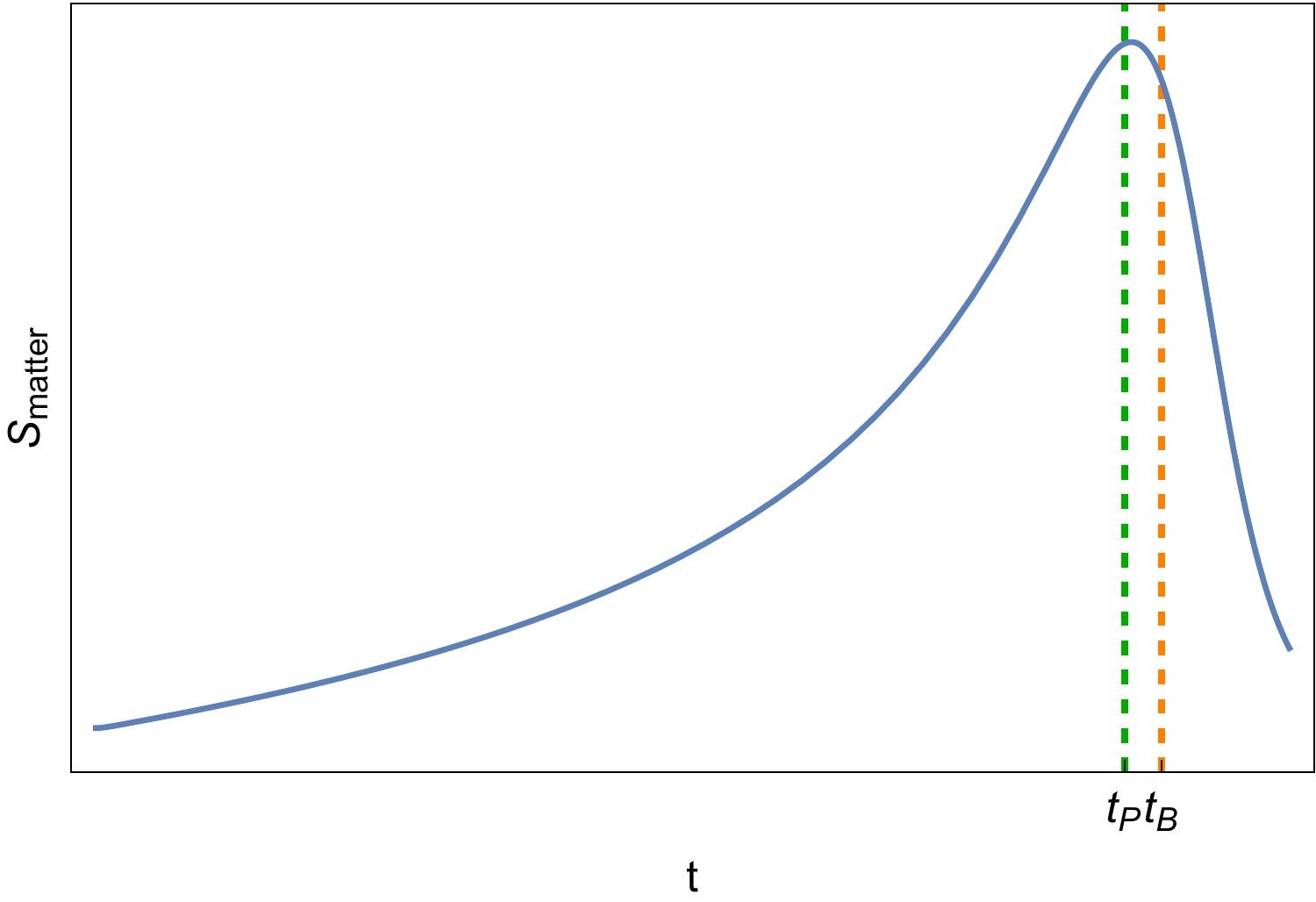}
\caption{Qualitative time variation of the entanglement entropy of matter fields, according to \cref{EntanglementEntropyMatter}, where we considered the time variation of the surface gravity, calculated using the solution of \cref{drHdtSB}. The vertical, dashed green line corresponds to the maximum of the curve (the Page time $t_\text{P}$), while the vertical dashed orange one indicates the time $t_\text{B}$ at which the semiclassical approximation should break down.} 
	\label{EntEntropyMatter}
\end{figure}

The qualitative result (obtained neglecting the irrelevant constants $\frac{N}{6}\ln\left[4 f(b) \right]$) is plotted in \cref{EntEntropyMatter}. As in singular black-hole models, initially the entanglement entropy of the radiation grows. However, this growth reaches a maximum at the ``Page time'' $t_\text{P}$ and then the entropy starts decreasing, due to the peculiar form of the surface gravity, which is related to the absence of a singularity. 

It is interesting to note that $t_\text{P}$ physically coincides with the onset time of the second order phase transition (graphically, the intersection point between the solid blue and the horizontal dotted lines in \cref{Msemiclassicalbd}). This feature was found before for nonsingular black holes in Ref.~\cite{Bianchi:2014bma}, where it was noted that the presence of the dS core traps Hawking modes, which cause a decrease in entropy once freed from the trapping region. This indeed happens as we get closer to the extremal configuration, right after the onset of the second order phase transition, as the role of the inner horizon becomes increasingly important. This release of information could also be related to the peculiarities of the latter, which has negative surface gravity, causing an outburst of energy in the final stages of the evaporation \cite{Frolov:2017rjz}, a process similar to the mass inflation. 

The mechanism described above is qualitative similar to that taking place in the island proposal \cite{Penington:2019npb,Penington:2019kki,Almheiri:2019qdq,Almheiri:2020cfm} (for an application to two-horizon models, both singular and regular, see, e.g., Refs.~\cite{Wang:2021woy,Kim:2021gzd,Goswami:2022ylc}; for an application to dS spacetime, see, e.g., Refs.~\cite{Svesko:2022txo,Ageev:2023mzu}), where the resolution of the information paradox is traced back to a transition in the behavior of the entanglement entropy functional. At late times (right after the Page time), this functional starts receiving contributions from nontrivial configurations in the  black hole interior (the ``islands''). This allows to correctly keep track of the entanglement structure of both the black hole and the radiation subsystems and to reconstruct the Page curve. Our approach provides, at qualitative level, a physical explanation for the appearance of such configurations in the black-hole interior. In our description, the islands correspond to the inner horizon, while the transition in the entropy functional is physically realized by the second order phase transition \footnote{As it was noted in Ref.~\cite{Kim:2021gzd}, the expression for the entanglement entropy in the island, when applied to singular two-horizon models, becomes mathematically ill-defined for the extremal configurations, as the boundary of the islands necessarily ends at the singularity at $r=0$. This problem is naturally avoided when dealing with regular models, as we do in this paper. }. In this paper we have only considered nonsingular black holes with two horizons, characterized therefore by a timelike singularity. It is, therefore, currently unclear if and how the results of the present paper compare with the island rule applied to black holes possessing spacelike singularities (see, e.g., Refs.~\cite{Hashimoto:2020cas,Gautason:2020tmk,Hartman:2020swn}) \footnote{It was shown in Ref.~\cite{Hartman:2020swn} that the quantum extremal surface, identifying the position of the island, could meet the singularity during the evaporation, leading to the impossibility of following the evaporation process until completion, at least in this setup. The absence of the singularity could possibly lead to a resolution of this problem.}. In general, we expect the  single-horizon case to be qualitatively different from  the two-horizon one. On the other hand, at least in some particular cases, the behavior of regular models with a \emph{single} event horizon could be not so drastically different from that of two-horizon black holes (see, e.g., Refs.~\cite{Trodden:1993dm,Fitkevich:2022ior,Lobo:2020ffi,Akil:2022coa}).

The assumptions used so far are the validity of the quasistatic and the semiclassical approximations. As we have seen in \cref{subsubsec:approachingextandsemiclassicalapprox}, the semiclassical approximation (and hence also the quasistatic one) breaks down near extremality, when we reach the energy gap \eqref{Egap}. This happens at the time corresponding to the vertical dashed orange line in \cref{EntEntropyMatter}. Therefore, we have to cut the Page curve when the semiclassical approximation breaks down. What happens beyond this point cannot be inferred from our semiclassical description of the dynamics. In particular, we cannot assess whether the decrease in EE continues until it becomes zero at extremality, as it would be expected for an evaporation process that leaves behind a quantum pure state. Results from $\text{AdS}_2$ quantum gravity  indicate the occurrence of a quantum phase transition from the LDS vacuum to the $\text{AdS}_2$ CDV \cite{Cadoni:2017dma}. Similarly to what happens in the case of extremal charged black holes, the near-extremal, near-horizon state of $4\text{D}$ nonsingular Hayward black holes, described by the AdS$_2 \times S^2$ spacetime, could have a  purely topological entropy content, explained in terms of AdS$_2$ fragmentation \cite{Maldacena:1998uz}.  

\section{Conclusions}
\label{sec:conclusions}
In this paper we have investigated the thermodynamics and the classical and semiclassical dynamics of $2\text{D}$, AF, nonsingular dilatonic black holes with a dS core. The aim has been trying to understand both the end point of  the black-hole evaporation and  the related information flow. Our analytic and numerical results provide some evidence that the latter leads to a regular extremal configuration in a finite amount of time. This conclusion is supported both by a thermodynamic analysis, showing that extremal configurations are energetically preferred, and by the numerical integration of the semiclassical field equations, which allows to take into account the full dynamics of the backreaction. Concerning the information flow during the evaporation process, the Page curve we have constructed clearly shows the presence of a maximum at the Page time, followed by a descent, in which information is recovered from the hole. These features have a nice physical explanation in terms of the trapping/flow of Hawking modes in the region between the inner and outer horizons. Additionally, they may provide some physical insight on the mechanism underlying the recently advanced ``island'' proposal to address the information paradox.

On the other hand, the intrinsic limitation of our approach, the validity of the semiclassical approximation, prevents us from putting a definitive final word on the issue. Near extremality, the semiclassical approximation breaks down and a mass gap is generated, which is a well-known feature of $\text{AdS}_2$ quantum gravity. This gap separates the extremal configuration, endowed with a linear dilaton, from another vacuum of the theory, the $\text{AdS}_2$ CDV. There are strong indications that, once the former is reached during evaporation, a quantum phase transition occurs that causes a transition to the latter. A similar phase transition between the extremal LDS and the $\text{AdS}_2$ CDV seems to occur also in the case under consideration. If this could be independently confirmed, it would imply that the end point of the evaporation process is the full $\text{AdS}_2$ spacetime endowed with a constant dilaton, which is a perfectly regular configuration. This would be a clear confirmation that the evaporation process is unitary.   

In terms of the $4\text{D}$ solution, which we are here modelling with the $2\text{D}$ dilaton gravity models, this corresponds to a (topology changing) phase transition between an extremal solution, which is an AF spacetime with an infinitely long throat, and an AdS$_2\times S^2$ spacetime, describing instead the near-horizon region.  

Our investigation has also revealed some intriguing peculiar features of the semiclassical dynamics of regular black holes with a dS core. The approach to extremality of excited black-hole solutions is not monotonic, but presents a sharp maximum of $\Delta\rho$ and $\Delta \phi$ near extremality, at relatively small values of the dilaton. The presence of this maximum can be explained in terms of the self-interaction of the dilaton, i.e., the presence of a maximum in the dilaton potential $\V(\phi)$. This, in turn, is what determines the presence of two horizons, instead of a single one, the presence of a large/small-black-hole phase transition and also the decreasing tail in the Page curve for the EE of Hawking radiation. Since the behavior of the dilatonic potential is essentially determined by the presence of a dS core, all the previous features are a consequence of the absence of singularities, which is achieved exactly thanks to this dS behavior in the black-hole interior. This strongly supports the hypothesis that the nonunitary evolution of the evaporation process has to be traced back to the presence  of a singularity in the black hole interior.  

\section{Acknowledgements}
We thank D. Grumiller for valuable comments and for having pointed our attention to Ref.~\cite{Bagchi:2014ava}. 
    
\begin{appendix}

\section{No divergences in the ground state stress-energy tensor}
\label{sec:nodivergencesTmunuGS}

In Ref.~\cite{Cadoni:1995dd}, counterterms are added to the Polyakov action (as in \cref{Polyakovactioncounterterms}) to eliminate a divergence in the stress-energy tensor of the GS $\langle T_{\mu\nu}\rangle \sim \phi^{2a}$, which, for $a>0$ diverges at asymptotic infinity $\phi \to \infty$. We show that in our case the rapid fall of the potential at infinity allows to avoid this divergent behavior. 

We consider, as an example, $\langle T_{+-}\rangle_{\text{GS}}$, but similar considerations also hold for the other components of $T_{\mu\nu}$. Using \cref{confanomTpm}, we have
\begin{equation}
\langle T_{+-}\rangle_{\text{GS}}  = -\frac{N}{12}\partial_+ \partial_- \rho \biggr|_{\text{GS}}\, ,
\end{equation}
where $\rho$, computed at the GS, is given by \cref{GSextremal}
\begin{equation}
\rho_\text{GS} = \frac{1}{2}\ln\left(\frac{2\M_\text{ext}}{\lambda} + \JGS \right) \, , \qquad \J \equiv \frac{1}{\lambda^2}\int^\phi \dd\psi \, \V(\psi) \, .
\end{equation}
Therefore 
\begin{equation}
\partial_- \rho_\text{GS} = \frac{1}{2}\frac{\partial_- \JGS}{\frac{2\M_\text{ext}}{\lambda} + \JGS} = \frac{1}{2}\frac{\partial_-\phi_\text{GS} \, \J_{\text{GS},\phi}}{\frac{2\M_\text{ext}}{\lambda} + \JGS} = -\frac{1}{4\lambda} \V_\text{GS} \, ,
\end{equation}
where $\J_{,\phi} = \lambda^{-2} \V$ and we used the vacuum solution \cref{dilatonconfgauge}. Differentiating with respect to $x^+$ yields
\begin{equation}
\partial_+\partial_- \rho_\text{GS} = -\frac{1}{4\lambda} \partial_+ \phi \, \V_{\text{GS},\phi} = -\frac{1}{8} e^{2\rho} \V_{\text{GS},\phi}\, .
\end{equation}
For $\phi \to \infty$, $e^{2\rho} \to \text{constant}$, while $\V_{,\phi} \sim -\phi^{-3} \to 0$. So we do not have divergences. 

\section{Boundary conditions for numerical integration}
\label{sec:boundaryconditionsnumerical}

\subsection{Boundary condition at $x^+ = x^+_0$}
For our numerical integration, we set $\lambda =1$ and $\ell = 1$. 

When the shock wave is turned on, the solution is set equal to the extremal configuration \eqref{GSextremal}. Our procedure to implement this boundary condition is the following:

\begin{enumerate}
\item We first integrate the equation for the dilaton (expression on the right of \cref{GSextremal}), leading to an implicit relation between the dilaton and the coordinates;
\item We numerically invert it to have explicitly $\phi = \phi(x^+, x^-)$;
\item We plug the result into the equation for $e^{2\rho}$.
\end{enumerate}

The first point is achieved by solving the differential equation 
\begin{equation}\label{diffeqrstarextremal}
\frac{\dd r_\ast}{\dd\psi} = \frac{1}{\frac{2\Mext}{\lambda} + \J(\psi)}\, .
\end{equation}
This integral can be done analytically. With $\Mext = \sqrt[3]{2} \, \ell / 3$ and $\ell =\lambda=1$ it reads
\begin{equation}
r_{\ast, \text{ext}} = 2^{5/3} \ln \left(\left| \sqrt[3]{2}-\phi \right| \right)+\frac{3 \phi }{2^{2/3}}+\frac{3}{\sqrt[3]{2}-\phi }+\frac{\ln \left(2 \phi +\sqrt[3]{2}\right)}{2 \sqrt[3]{2}}. 
\end{equation}

We then numerically invert the result to get $\phi = \phi(x^+, x^-)$ 
and plug the result into \cref{GSextremal} to get $\rho = \rho(x^+, x^-)$, both  evaluated  at extremality. 

\subsection{Boundary condition at $x^- \to -\infty$}
The procedure is exactly the same as before, with $\Mext$ replaced by a different value of the mass. Here, we choose $\M = 0.1 \, \lambda$. 

The coordinate $r_\ast$ as a function of $\phi$ in this case reads
\begin{equation}
r_{\ast} \simeq -1.37 \ln (|0.47 -\phi |)+25.42 \ln (| 4.96 - \phi| )+5 \phi +0.95 \ln (\phi +0.43)\, .
\end{equation}

Again, we numerically invert this expression, to get $\phi = \phi(x^+, x^-)$. In this case, however, the quantity we are (improperly) calling $r_\ast$ contains the function $F(x^-)$, which has to be computed to fully obtain $\phi$ as a function of the coordinates. Starting from 
\begin{equation}
\mathscr{F}[\phi(x^+, x^-)] \equiv \int^\phi \,\frac{\dd\psi}{\frac{2\M}{\lambda}+\J} = \frac{\lambda}{2}\left[x^+ - x^+_0 - F(x^-) \right]\, ,
\end{equation}
we note that, when evaluated at $x^+ =x_0^+$, it reads 
\begin{equation}
\mathscr{F}[\phi(x_0^+, x^-)]  = -\frac{\lambda}{2} F(x^-) \, ,
\end{equation}

from which we get 
\begin{equation}
\phi(x_0^+, x^-) \equiv \phi_0(x^-) = \mathscr{F}^{-1}\left[-\frac{\lambda}{2} F(x^-) \right]\, .
\end{equation}

This expresssion of $\phi(x^+_0, x^-)$ can be used in \cref{exactsolutionshockwave} to compute $\J_0$. This gives us a differential equation in terms of $F$ which reads
\begin{equation}
F' (x^- ) = \frac{\frac{2\Mext}{\lambda}+ \J_{0,\text{ext}}(x^-)}{\frac{2\M}{\lambda}+ \J_0 \left[\phi_{0}(x^-)\right]} \equiv  \frac{\frac{2\Mext}{\lambda}+ \J_{0,\text{ext}}(x^-)}{\frac{2\M}{\lambda}+ \J_0 \left(\mathscr{F}^{-1}\left[-\frac{\lambda}{2} F(x^-) \right]\right)}\, .
\end{equation}
This equation is solved numerically. The integration constant is chosen so that, once the solution of $F$ is plugged into the dilaton solution \eqref{exactsolutionshockwave}, at $x^+ = x^+_0$, the dilaton is equal to the extremal one computed at $x^+_0$. This guarantees the continuity of the scalar field across the shock wave.

Once $F(x^-)$ is computed, we plug it into $\phi = \phi(x^+, x^-)$. With this and \cref{exactsolutionshockwave}, we obtain also $\rho =\rho(x^+, x^-)$ above extremality. 

\end{appendix}

\newpage

\bibliography{NonSingular2DBHFinal}
\bibliographystyle{ieeetr}

\end{document}